\documentclass[useAMS,usenatbib]{mn2e}

\usepackage{natbib}
\usepackage{color}
\usepackage{amsmath, amssymb}
\usepackage{bm}
\usepackage{pifont}
\usepackage{float}
\usepackage{graphicx}
\usepackage[usenames,dvipsnames]{xcolor}
 \usepackage{relsize}
\usepackage{subfigure}  
\usepackage[singlelinecheck=false 
]{caption}

\usepackage{soul} 
\usepackage{epsfig}
\usepackage{enumerate}
\usepackage[shortlabels]{enumitem}
\usepackage[version=3]{mhchem}
\usepackage{caption}
\usepackage{varwidth}
\newsavebox\CBox
\usepackage{color}
\usepackage{lineno}
\usepackage{epsf,epsfig}
\usepackage{epstopdf} 
\usepackage{float}
\usepackage{rotating} 
\usepackage{epstopdf} 

\title[Reference study to characterise plasma  and magnetic properties of ultra-cool atmospheres]{Reference study to characterise plasma  and magnetic properties of ultra-cool atmospheres\\ }
\author[M.\ I.\ Rodr\'iguez-Barrera, Ch.\ Helling, C.\ R.\ Stark and A.\ M.\ Rice]{M.\ I.\ Rodr\'iguez-Barrera$^{1}$\thanks{E-mail:
mirb@st-andrews.ac.uk}, Ch.\ Helling$^{1}$, C.\ R.\ Stark$^{1}$ and A.\ M.\ Rice$^{1}$\footnotemark[1]\thanks{This file has been amended to
highlight the proper use of \LaTeXe\ code with the class file.
These changes are for illustrative purposes and do not reflect the
original paper by M.\ I.\ Rodr\'iguez-Barrera.}\\
$^{1}$SUPA, School of Physics \& Astronomy, University of St.\ Andrews, St.\ Andrews KY16 9SS, UK}

\begin{document}
\maketitle

\label{firstpage}

\begin{abstract}
Radio and X-ray emission from brown dwarfs suggest that an ionised gas
and a magnetic field with a sufficient flux density must be present.
We perform a reference study for late M-dwarfs, brown dwarfs and giant
gas planet to identify which ultra-cool objects are most susceptible
to plasma and magnetic processes. Only thermal ionisation is
considered.  We utilise the {\sc Drift-Phoenix} model grid where the
local atmospheric structure is determined by the global parameters
T$_{\rm eff}$, $\log(g)$ and [M/H].
 
Our results show that it is not unreasonable to expect H$_{\alpha}$ or
radio emission to origin from Brown Dwarf atmospheres as in particular
the rarefied upper parts of the atmospheres can be magnetically
coupled despite having low degrees of thermal gas ionisation.  Such
ultra-cool atmospheres could therefore drive auroral emission without
the need for a companion's wind or an outgassing moon.  The minimum
threshold for the magnetic flux density required for electrons and
ions to be magnetised is well above typical values of the global
magnetic field of a brown dwarf and a giant gas planet. Na$^{+}$,
K$^{+}$ and Ca$^{+}$ are the dominating electron donors in low-density
atmospheres (low log(g), solar metallicity) independent of T$_{\rm
eff}$. Mg$^{+}$ and Fe$^{+}$ dominate the thermal ionisation in the
inner parts of M-dwarf atmospheres. Molecules remain unimportant for
thermal ionisation.  Chemical processes (e.g. cloud formation)
affecting the most abundant electron donors, Mg and Fe, will
have a direct impact on the state of ionisation in ultra-cool
atmospheres.

\end{abstract}

\begin{keywords}
brown dwarfs, planets and satellites: atmospheres, stars: atmospheres, plasma, radio continuum: planetary systems, radio lines: planetary systems
\end{keywords}

\section{Introduction}

Ultra-cool objects like brown dwarfs and giant gas planets have masses
below the hydrogen-burning limit of $\sim0.08$~M$_{\odot}$
(e.g. \citealt{Burrows2001}).  Brown dwarfs are born like stars by
gravitational collapse, however, they have not sufficient mass to
achieve the required core temperature to provide a steady rate of
nuclear hydrogen fusion. As a consequence, gravitational collapse
provides the only energy source for most of the brown dwarf's
lifetime. Cooling and contracting during their entire life, brown
dwarfs cannot compensate the radiative losses by thermonuclear
processes.  Brown dwarfs evolve from a young objects with an effective
temperature T$_{\rm eff}\approx$ 3000 K that is comparable to  late M-dwarfs
and surface gravities like giant gas planets to high-gravity objects
of $\log(g)$=5.0 with an effective temperature lower than T$_{\rm
  eff}\approx$ 500 K. Brown dwarfs are fully convective objects,
  and they can be fast rotators. Brown dwarfs and giant gas planets form clouds in their
atmospheres which have strong feedback onto the atmospheric structure
due to element depletion and due to a high opacity
(\citealt{HellingCasewell2014} and references there in). Observations by \citet{Biller2013},
\citet{Buenzli2014} and \citet{Crossfield2014}, for example, suggest
that brown dwarf atmospheres show a patchy cloud coverage.  Transit
spectroscopy from extrasolar planets suggest that giant gas planets
are covered in hazes and clouds too (\citealt{Pont2008};
\citealt{Gibson2012}; \citealt{Sing2014}). Model simulations suggest
that cloud formation prevails for a large range of effective
temperatures up to 2800 K, and for metallicities as low as
[M/H]$=-5.0$ \citep{Witte2009}.  A significant volume of these clouds
is susceptible to local discharge events \citep{Helling2013}:
large-scale discharges in the upper cloud regions and corona-like
small-scale discharges in the inner, denser part of the cloud.  These
local discharge events may generate atmospheric electrical storms. At
the same time, storms may ionise the local gas. \cite{Luque2014}
modelled the ionosphere of Saturn and Jupiter finding that the
atmospheric electrical storms may produce up to 10$^{3}$ cm$^{-3}$ of
free electrons below the ionosphere.



Radio and X-ray emission from ultra-cool dwarfs are now well
established (e.g. 2MASSJ10475385+2124234 by \citealt{Route2012}; 2MASS
J13153094-2649513AB by \citealt{Burgasser2013}). \citet{Berger2002}
observed twelve sources in radio, X-ray and H$_\alpha$ emission
between the spectral types M8 and L3.5. \citet{Sorahana2014} present
mid-IR AKARI observations which suggest the presence of chromospheric
activity in brown dwarfs.  \cite{Schmidt2015} derive a rise in
magnetic activity in form of H$\alpha$ emission from SDSS spectra from
2\% for early M-dwarfs (M0) to 88\% for early L-dwarfs (L0). 39\% of
the L dwarfs in their sample observed multiple times are suggested to
be variable.  Such observations indicate that an appropriately ionised
gas is present in the atmospheres of such cool objects that allows chromospheric heating. The prevailing question is how
much of such a cool, cloud-forming atmosphere needs to be ionized that
a chromosphere could form.

 It is suggested that the magnetic field strength of brown dwarfs is
 as high as $10^{3}$G (e.g. \citealt{Shulyak2011}). \cite{Lynch2015}
 suggest field strengths of $\sim 2.5\,\ldots\,2.5$kG for the surface
 field on TVLM 0513-46 and 2M 0746+20.  A correlation between radio
 activity and rotation has not been settled for brown dwarfs given the
 wide parameter range occupied by these objects (\citealt{McLean2012,
   Antonova2013}).  \citet{Williams2014} demonstrate that ultra-cool
 objects do not follow the classical G\"udel-Benz relation where the
 X-ray and the radio luminosity from F\,$-$\,M stars correlate
 \citep{Gudel1993}. This relation was shown to persist for solar
 flares and active rotating binaries \citep{Benz1994}. The deviation
 of ultra-cool stars from the G\"udel-Benz relation beyond than approximately M5 may suggest a change
 in the dynamo mechanism that produces the magnetic field in such
 ultra-cool objects \citep{Cook2014}.
If the atmospheric gas can couple to the strong magnetic field in
brown dwarfs, the kinetic energy carried by large-scale convective
motions may be transported to the top of the atmosphere and released
in form of flares, quiescent or quasi-quiescent emissions.  The flares
are a sudden release of magnetic energy from the deeper convective
layers. The quiescent emission is a continuous emission but at lower
energy levels than flares.
The most likely mechanism to
produce this emission is electron cyclotron maser emission from highly
relativistic electrons or plasma emission alternatively, incoherent
synchrotron or gyrosynchrotron emission from relativistic electrons
\citep{Hallinan2006}. Flares and quiescent emission were reported for
example by \cite{Burgasser2005} for two late-type M-dwarfs,
with a magnetic field strength of 1 kG. \citet{Tanaka2014} study the
mass loss of hot Jupiters through MHD waves.  They suggest 
the formation of a chromosphere by Alfv\'en wave
heating. \citet{Tanaka2014}, however, observe, that the gas does not
need to be fully ionised, i.e. degree of ionisation f$_{\rm e}<1$, for
the mechanism to work. \citet{Mohanty2002} analysed magnetic Reynolds
number, R$_{\rm m}\propto f_{\rm e}$, for a set of model atmospheres
to show why the chromospheric $H_{\alpha}$ activity should be low in
rapid rotating brown dwarfs with T$_{\rm eff}\geq 1500$K and solar
metallicity. The works by \cite{Schmidt2015} and \citet{Sorahana2014}
suggest that L-dwarfs should have a chromosphere (but with a reduced
filling factor) despite having low magnetic Reynolds numbers.

Our paper presents a theoretical framework using a set of fundamental
parameters to analyse the ionisation and magnetic coupling state of
objects with ultra-cool atmospheres. This paper focuses on late
M-dwarfs, brown dwarfs and giant gas planets spanning T$_{\rm
  eff}=1000\,\ldots\,3000$K. The approach is also applicable to, for
example, protoplanetary disks. Only thermal ionisation is considered
for this reference study against which the effect of additional
processes (e.g. dust-dust collisions, cosmic ray ionisation, Alfv\'en
ionisation, lighting, photoionisation) can be tested in future works.
Our investigations utilise results from {\sc Drift-Phoenix} 1D model
atmosphere simulations (\citealt{Helling2008b, Witte2009, Witte2011}).
This allows us to perform an extensive reference study across the 
  late M-dwarf, brown dwarf and planetary regime on the basis of a
consistently calculated model atmosphere grid for a large set of
global parameters (T$_{\rm eff}$, $\log(g)$, [M/H]). The aim of this
study is to identify ultra-cool objects that are most susceptible to
plasma processes by itself or that lead to instabilities that trigger
the emergence of a strong plasma.  This study does not include
  any multi-dimensional atmospheric flows and resulting mulit-D
  radiative transfer effects.

Section~\ref{s:DF} describes our approach and the use of the {\sc
  Drift-Phoenix} model atmosphere results. Sections~\ref{s:plasma}
and~\ref{s:magnetic} introduce our theoretical frame work in form of
basic plasma and magnetic parameters, respectively. Our study shows
that ultra-cool atmospheres are composed of an ionised and magnetised gas. The local
degree of ionisation varies largely amongst the objects and throughout
the atmospheres. While a late M-dwarf has a considerable degree of
ionisation throughout the whole atmosphere, the degree of thermal
ionisation for a L-dwarf is rather low but may well be enough to seed
other ionisation processes like for example due to Alfv\'en ionisation
\citep{Stark2013}.  In Sec.~\ref{sss:fp} we demonstrate that
electromagnetic interactions can dominate over electron-neutral
interactions also in regions of a very low degree of ionisation.  In
Sec.~\ref{sss:lD} we investigate the relevant length scales effected
by electrostatic interactions in the gas phase.     Section~\ref{ions} (with additional
material in Appendix~\ref{chem}) contains our assessment of the local
equilibrium chemistry abundances with respect to electron donors
species across the late M-dwarfs, brown dwarfs and giant planets regime.

In Sec.~\ref{s:magnetic} we demonstrate that 30\%-50\% of the
atmospheric volume can be magnetically coupled in L-dwarfs for a
magnetic field strength of 10$^{3}$G. The atmospheric volume with a
degree of thermal ionisation above a plasma threshold value of
$>10^{-7}$ is, however, considerably lower.  Our results show that it
is not unreasonable to expect ultra-cool atmospheres to emit
H$_{\alpha}$ or even in radio wavelength as in particular the rarefied
upper parts of the atmospheres are affected by electromagnetic
interactions over many pressure scale heights despite having low
degrees of ionisation.  Section~\ref{discussion} contains the
discussion of our results in view of previous
publications. Section~\ref{s:con} presents our conclusions.

\begin{table}
\caption{{\sc Drift-Phoenix}  atmosphere structures used: \\{\bf Group\,1}:\,giant gas planets and young brown dwarfs.\\ {\bf Group\,2}:\,dependence on log(g). \\{\bf Group\,3}:\,dependence on metallicity.}
\centering
\scriptsize\addtolength{\tabcolsep}{-1.5pt}
\hspace*{-1cm}\begin{tabular}{|c|c|c|c|c|} 
\hline\hline
       & $T_{\rm eff}$ [K]					  & \bf $\log(g)$	 & \tiny [M/H]	 \\ [0.5ex]
 \hline\hline
 \rule{0pt}{4ex} \bf Group\,1   & 1000-3000 & 3.0    & 0.0  \\
\hline
\rule{0pt}{4ex}    & 1000 &  & \\ 
					\bf Group\,2        & 2000  &   3.0, 4.0, 5.0   & 0.0  \\  
					       & 2800 \\
\hline
\rule{0pt}{4ex}     &1000  &  & \\ 
					\bf Group\,3       & 2000  &   3.0 & -0.6, -0.3, 0.0, +0.3  \\  
					       & 2800 \\
 [1ex] \hline
\end{tabular}
\label{tab}
\end{table}

\section{Approach} \label{s:DF}

We aim to assemble a theoretical framework that allows us to assess
the plasma and magnetic character in atmospheres of objects across the
stellar-planetary boundary, namely for late M-dwarfs, brown dwarfs and
planets. Our approach is not limited to these objects as the plasma
parameters used are fundamental properties of a gas rather than of a
particular object.  We utilise the grid of {\sc Drift-Phoenix} model
atmosphere structures in order to quantify the plasma and magnetic
characteristics.  {\sc Drift-Phoenix} is a combination of two
complementary codes, {\sc Drift} and {\sc Phoenix}
(\citealt{Helling2008b}; \citealt{Witte2009,Witte2011}). The {\sc
  Drift} code (\citealt{Woitke2004}; \citealt{Helling2008b}) solves a
system of equations that describes the stationary dust formation
process of mineral clouds (seed formation, growth, evaporation,
sedimentation, element depletion) and interaction between the dust
grains and gas (\citealt{Woitke2003, Woitke2004};
\citealt{Helling2006,Helling2008}). {\sc Phoenix} is a hydrostatic
radiative transfer model atmosphere code \citep{Hauschildt1999} that
determines the resultant thermodynamic structure of the atmosphere
(local gas temperature T$_{\rm gas}$ [K], local gas pressure p$_{\rm
  gas}$ [bar] and local electronic pressure p$_{\rm e}$ [bar]) from
fundamental stellar parameters (effective temperature T$_{\rm eff}$
[K], surface gravity $\log(g)$ [cm\,s$^{-2}$] and metallicity [M/H]).
When combined with {\sc Drift} it provides a self-consistent
atmospheric model that takes into account cloud formation and its
impact on the thermodynamic structure and the resulting spectral
energy distribution.

The {\sc Drift-Phoenix} atmosphere simulations model the kinetic
  formation of mixed mineral cloud particle made of TiO$_2$[s], Al$_2$O$_3$[s], Fe[s],
  SiO$_2$[s], MgO[s], MgSiO$_3$[s], Mg$_2$SiO$_4$[s] which effects 6
  elements (O, Mg, Si, Fe, Al, Ti). Mg, Si, and Fe are the most
  abundant elements after O and C in a gas with solar element abundances.

\begin{figure}
\centering
\hspace*{-0.9cm}\includegraphics[angle=0,width=0.6\textwidth]{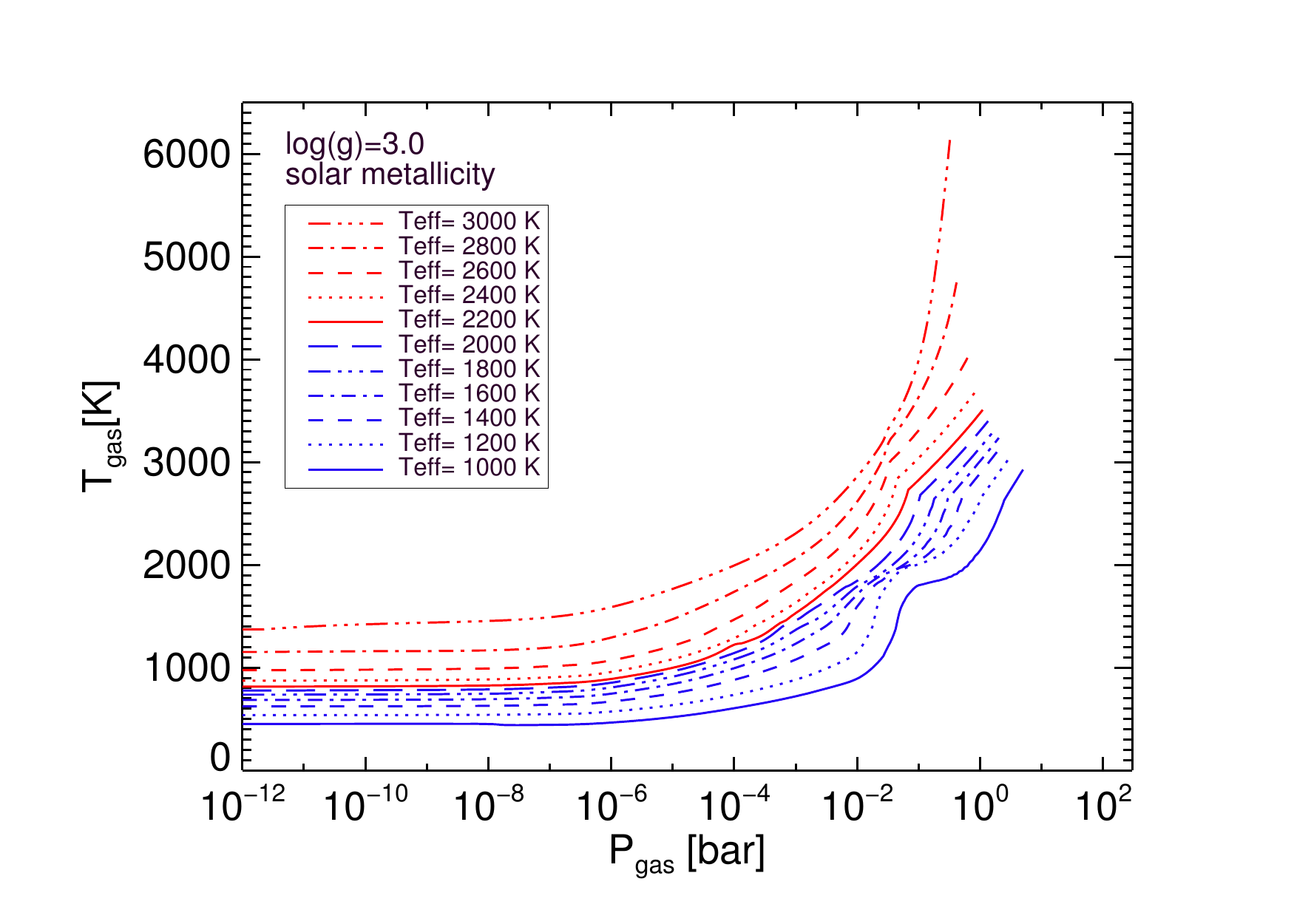}\\*[-0.8cm]
\hspace*{-0.9cm}\includegraphics[angle=0,width=0.60\textwidth]{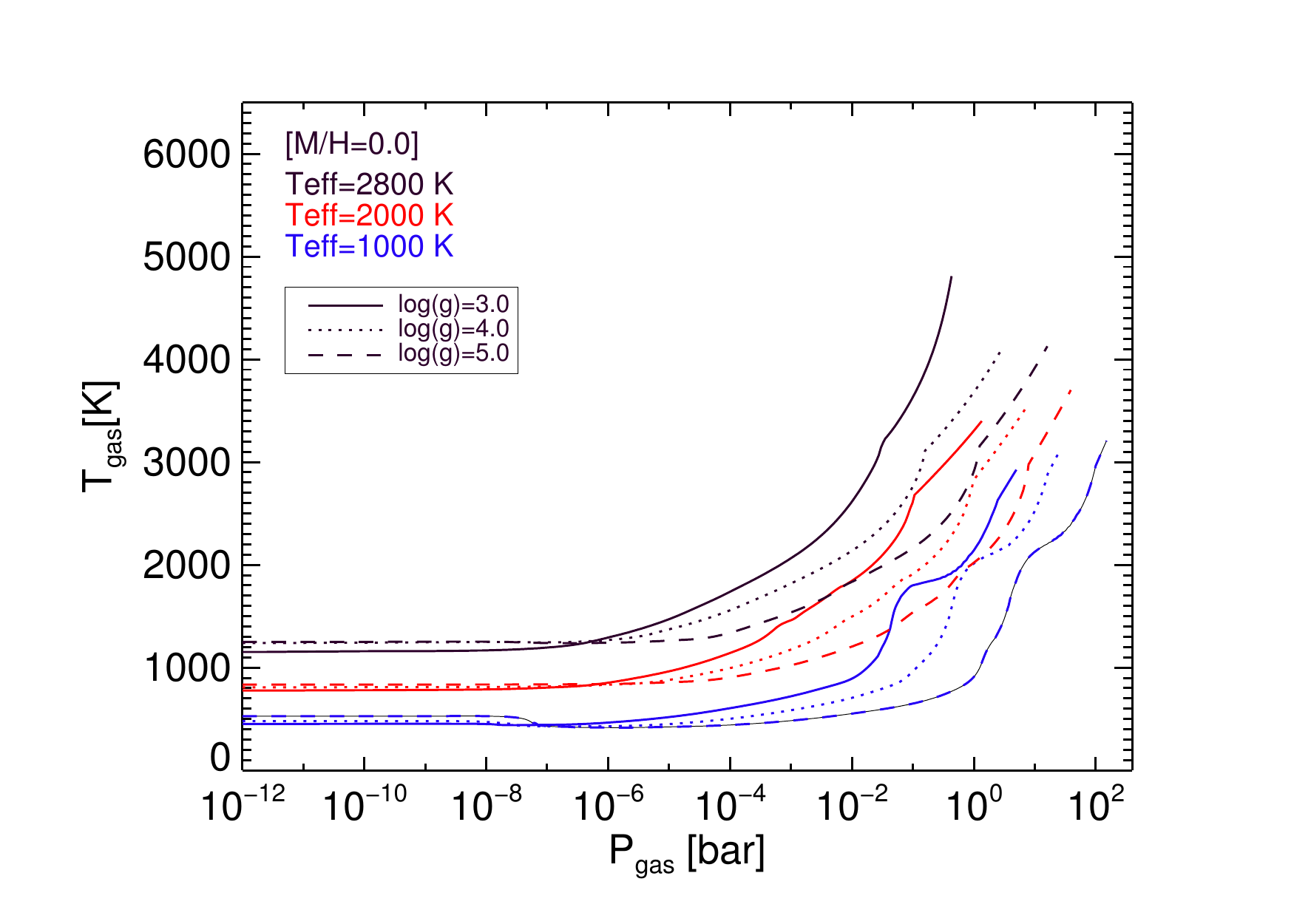}\\*[-0.8cm]
\hspace*{-0.9cm}\includegraphics[angle=0,width=0.60\textwidth]{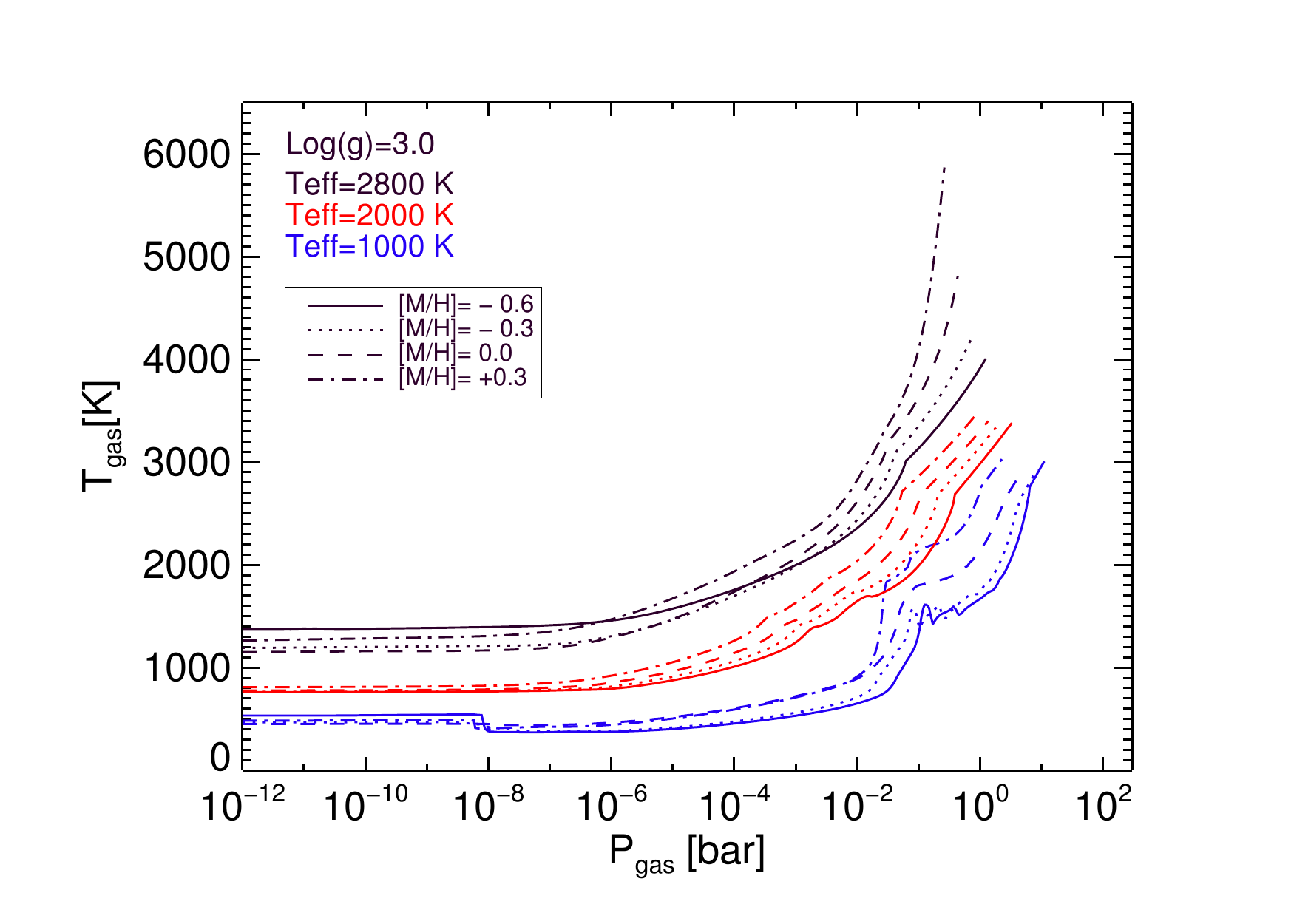}\\*[-0.8cm]
\hspace*{-1cm}
\caption{(T$_{\rm gas}$-p$_{\rm gas}$) input structure from the {\sc Drift-Phoenix} model atmosphere grid. The hottest models represent late M-dwarf atmospheres or atmospheres of young brown dwarfs.
The coolest models represent atmospheres in the planetary regime.
 {\bf Top:} Group 1.
{\bf Middle:} Group 2.
 {\bf Bottom:} Group 3. 
}
 \label{fig:Tpgas}
\end{figure}
  Using {\sc Drift-Phoenix} atmospheric models with a range of
  effective temperatures (T$_{\rm eff}$ = 1000 K - 3000 K), surface
  gravity ($\log(g)$ = 3.0, 4.0, 5.0) and metallicity ({\small [M/H]}=
  -0.6, -0.3, 0.0 +0.3), we evaluate the degree of thermal gas
  ionisation, the plasma parameter (Sec.~\ref{s:plasma}) and magnetic
  parameters (Sec.~\ref{s:magnetic}). Applying a separate chemical
  equilibrium code, we evaluate the gas-phase composition of various
  atmosphere models to assess if the dominating electron donating
  species changes or remains the same (Sec.~\ref{ss:chem1}). \par We
  apply a chemical equilibrium routine to calculate the chemical
  composition in more detailed than available from the standard {\sc
    Drift-Phoenix} output. Our main interest is to evaluate the
  results regarding the ions in the gas phase.  A combination of $155$
  gas-phase molecules (including $33$ complex hydrocarbon molecules),
  $16$ atoms, and various ionic species were used under the assumption
  of local thermodynamic equilibrium (LTE).  The \cite{Grevesse2007}
  solar composition is used for calculating the gas-phase chemistry
  outside the metal depleted cloud layers. No solid particles were
  included in the chemical equilibrium calculations but their presence
  influences the gas phase by the reduced element abundances due to
  cloud formation and the cloud opacity impact on the radiation field,
  both accounted for in the {\sc Drift-phoenix} model simulations.

We utilize {\sc Drift-Phoenix} model atmosphere
  (T$_{\rm gas}$, p$_{\rm gas}$, p$_{\rm e}$) structures and the dust element
  depleted abundances as input for our calculations.  We group the
  {\sc Drift-Phoenix} atmosphere structures into three groups for an
  easier presentation of our results (Table~\ref{tab}). These groups,
  defined by a range of global parameters, cover young and old giant gas planets,
  young and old brown dwarfs and late M-dwarfs. Figure~\ref{fig:Tpgas} shows the thermodynamic profiles
  (T$_{\rm gas}$, p$_{\rm gas}$) for each of these groups. As the
  effective temperature  increases, the thermodynamic gas
  temperature of the atmosphere increases as well for a given
  $\log(g)$ and [M/H]. The effect of dust on the local gas temperature
  in the atmosphere appears as a step-like temperature change
  (backwarming) in the atmosphere models.

\section{Basic plasma parameters}\label{s:plasma}


In the following section, we lay out the theoretical framework that we
use to characterise the plasma of an atmospheric environment with
respect to its electrostatic and magnetostatic behaviour.
This is inspired by the wealth of radio observations of substellar
objects (\citealt{Berger2002,Berger2010}; \citealt{Route2012};
\citealt{Burgasser2013}; \citealt{Williams2014}). To understand these
observations, radio wavelength (quiescent emission) and X-ray emission
(flares), the atmospheric gas must couple with the background magnetic
field. 
 As a consequence, free charged particles produced in the
atmosphere would be accelerated along magnetic field lines and released
into the upper parts of the atmosphere. A magnetic coupling of the
local gas would also be required for Alfv\'en waves to develop and
potentially contributing to an acoustic heating of a chromosphere
also on such ultra-cool objects (e.g. \citealt{Testa2015}).  We note
that ideal and non-ideal MHD simulations require a certain degree of
ionisation to allow Alfv\'en wave heating to develop as possible
mechanisms for chromospheric heating.


For a plasma to exist, the gas needs to be ionised. The degree of
ionisation, $f_{\rm e}$ measures the extent to which a gas is ionised,
and it is defined as
\begin{equation} \label{eq:fe}
f_{\rm e}=\frac{p_{\rm e}}{p_{\rm e}+p_{\rm gas}} \,\,, 
\end{equation}
where $p_{\rm gas}$ and $p_{\rm e}$ are the gas and electron pressure
respectively, both in $[\rm bar ]$.  Once we have determined the
degree of ionisation of the atmospheric gas depending on the global
parameters (T$_{\rm eff}$, $\log(g)$, [M/H]) we evaluate the plasma
frequency to investigate where in the atmosphere electromagnetic
interactions dominate over kinetic collisions between electrons and
neutrals,
\begin{equation}
\omega_{\rm pe}\gg\nu_{\rm ne}.
\label{eq:wpe}
\end{equation}
$\omega_{\rm pe}$ [$\rm rad\,s^{-1}$] is the plasma frequency
(i.e. the frequency at which the plasma reacts to an imposed or perturbed electric fields),
$\nu_{\rm ne}$ [$\rm s^{-1}$] is the electron-neutral collision
frequency. Only if Eq.~\ref{eq:wpe} is fulfilled, we
can expect the ionised gas to undergo electromagnetic interactions that
could lead, for example, to discharge processes.
A more refined insight about electrostatic interactions influencing the atmospheric gas can be gained by determining the length scales beyond which the Coulomb force of a charge does not any
more effects its surrounding. On length-scale larger than the Debye
length, a gas will be quasi-neutral and no electrostatic forces will
effect the gas behaviour. Hence,

\begin{equation}
\lambda_{\rm D}\ll L   
\label{eq:lD}
\end{equation}
with $\lambda_{\rm D}$ the Debye length and $L$ the typical length
 scale of the considered plasma, both in [$\rm m$]. Ideally, this
 would be associated with the atmospheric volume where the ionised gas can couple
 to an external magnetic field.

In the following subsections we define each of the plasma criteria
and evaluate them for our model atmosphere grid. All equations and
natural constants are given in SI units. All results, however, have
been converted into cgs unit for an easier representation in the
astrophysical context.

\begin{figure}
\centering
\hspace*{-0.9cm}\includegraphics[angle=0,width=0.6\textwidth]{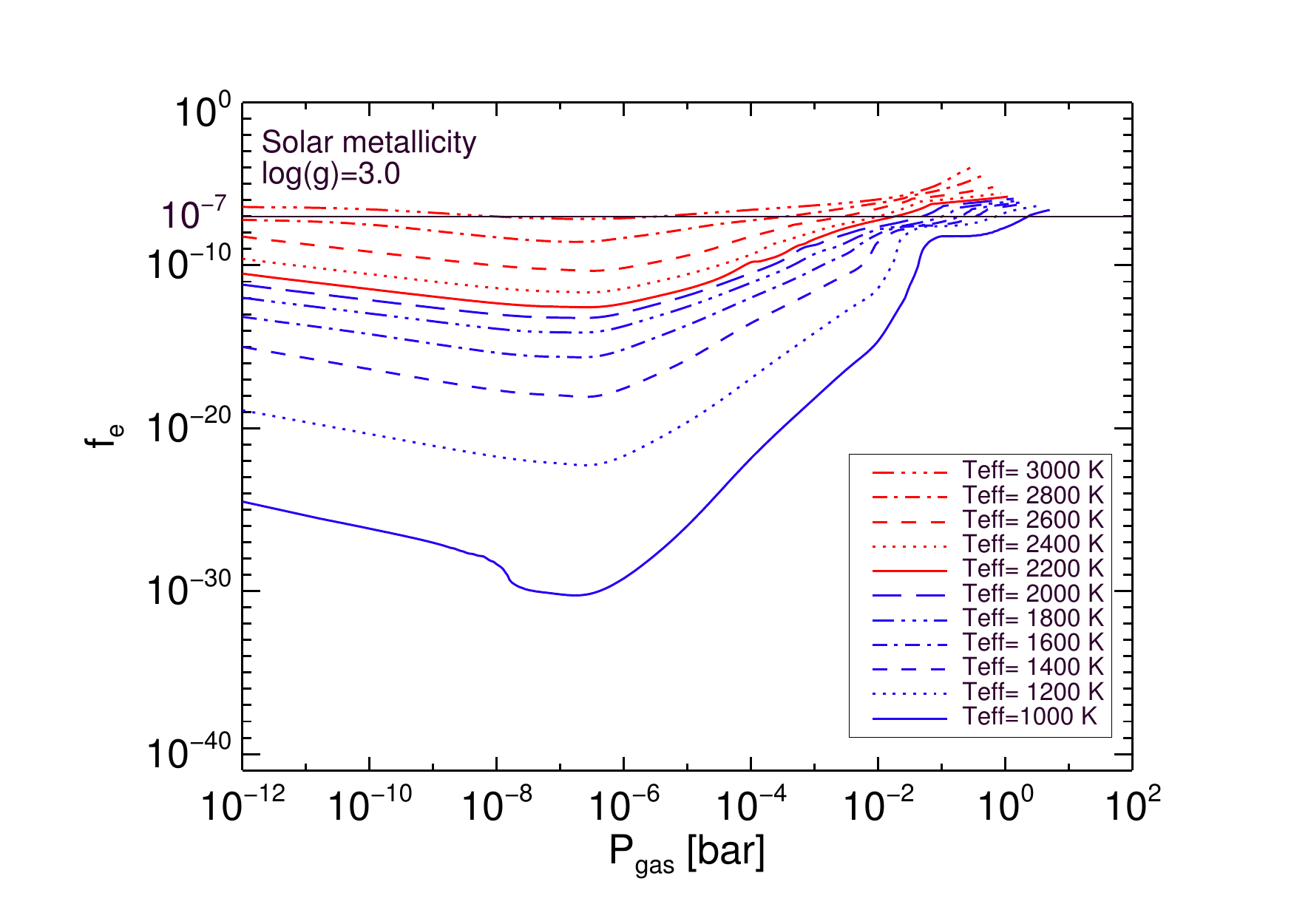}\\*[-0.8cm]
\hspace*{-0.9cm}\includegraphics[angle=0,width=0.6\textwidth]{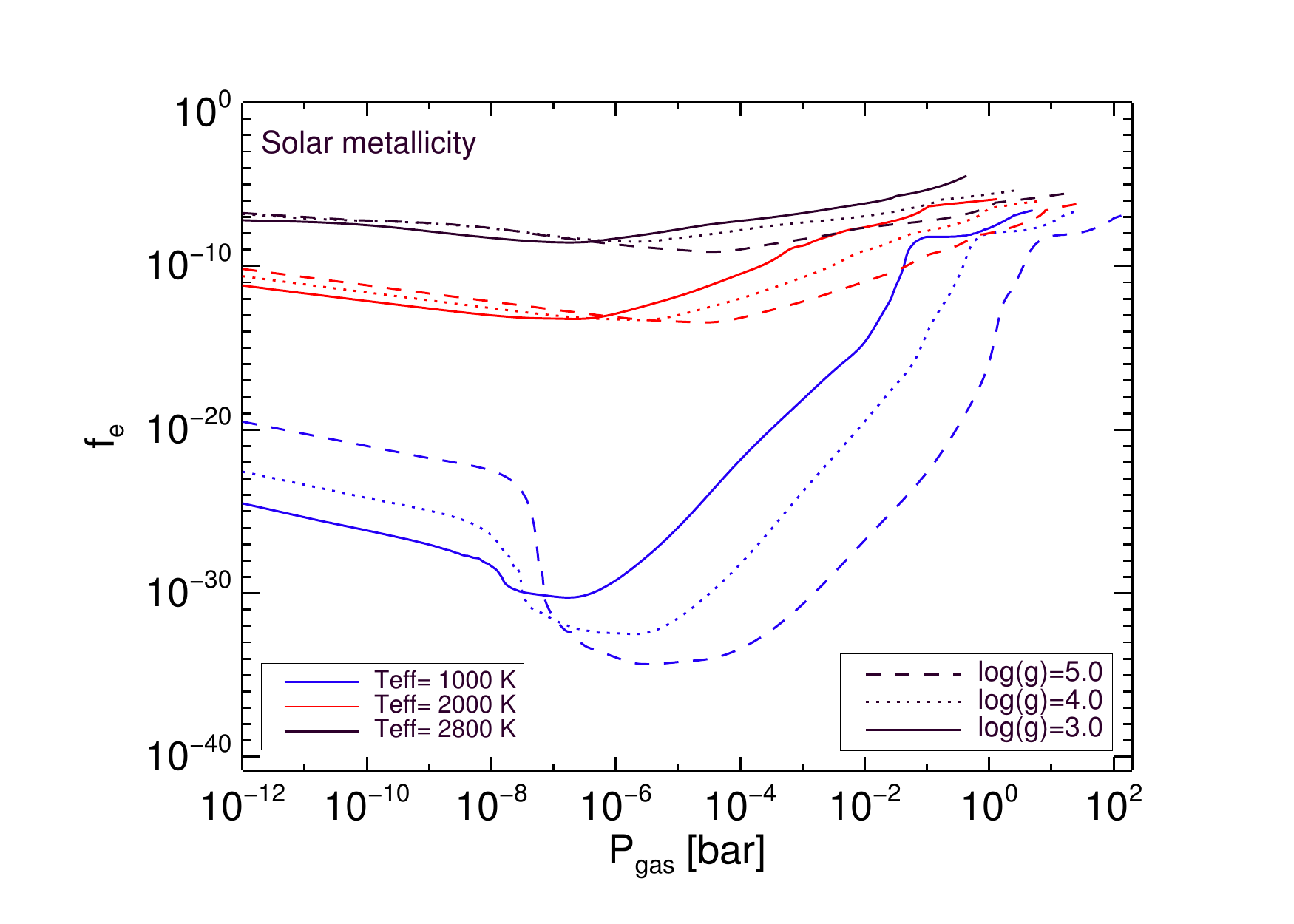}\\*[-0.8cm]
\hspace*{-0.9cm}\includegraphics[angle=0,width=0.6\textwidth]{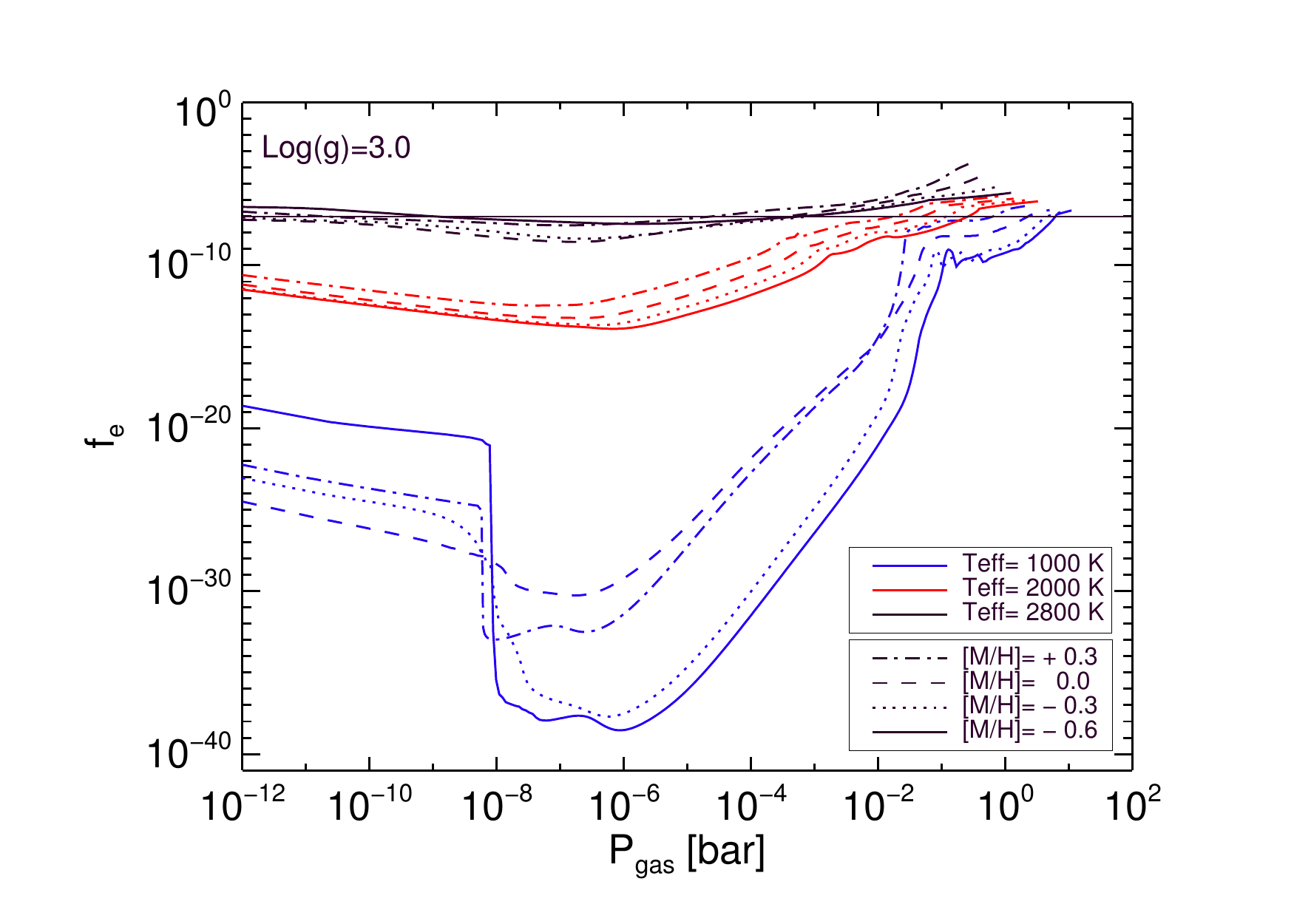}
\\*[-0.6cm]
\vspace{+0.1cm}\caption{The degree of thermal ionisation, $f_{\rm
    e}=p_{\rm e}/(p_{\rm e} +p_{\rm gas})$ as a measure of free
  charged particles for M-dwarf, brown dwarf and giant gas planet
  atmospheres. The M-dwarf atmosphere is easily ionised by thermal
  processes. Atmosphere of brown dwarfs can only be thermally ionised
  in deeper layers.
  {\bf Top:} Group 1.
{\bf Middle:} Group 2.
 {\bf Bottom:} Group 3. 
}
 \label{fig:fe}
\end{figure}

\subsection {Plasma Frequency: $\bm { \omega_{\rm pe}\gg\nu_{\rm ne}}$}\label{ss:fp}
 
\begin{figure}
\centering
\hspace*{-0.9cm}\includegraphics[angle=0,width=0.6\textwidth]{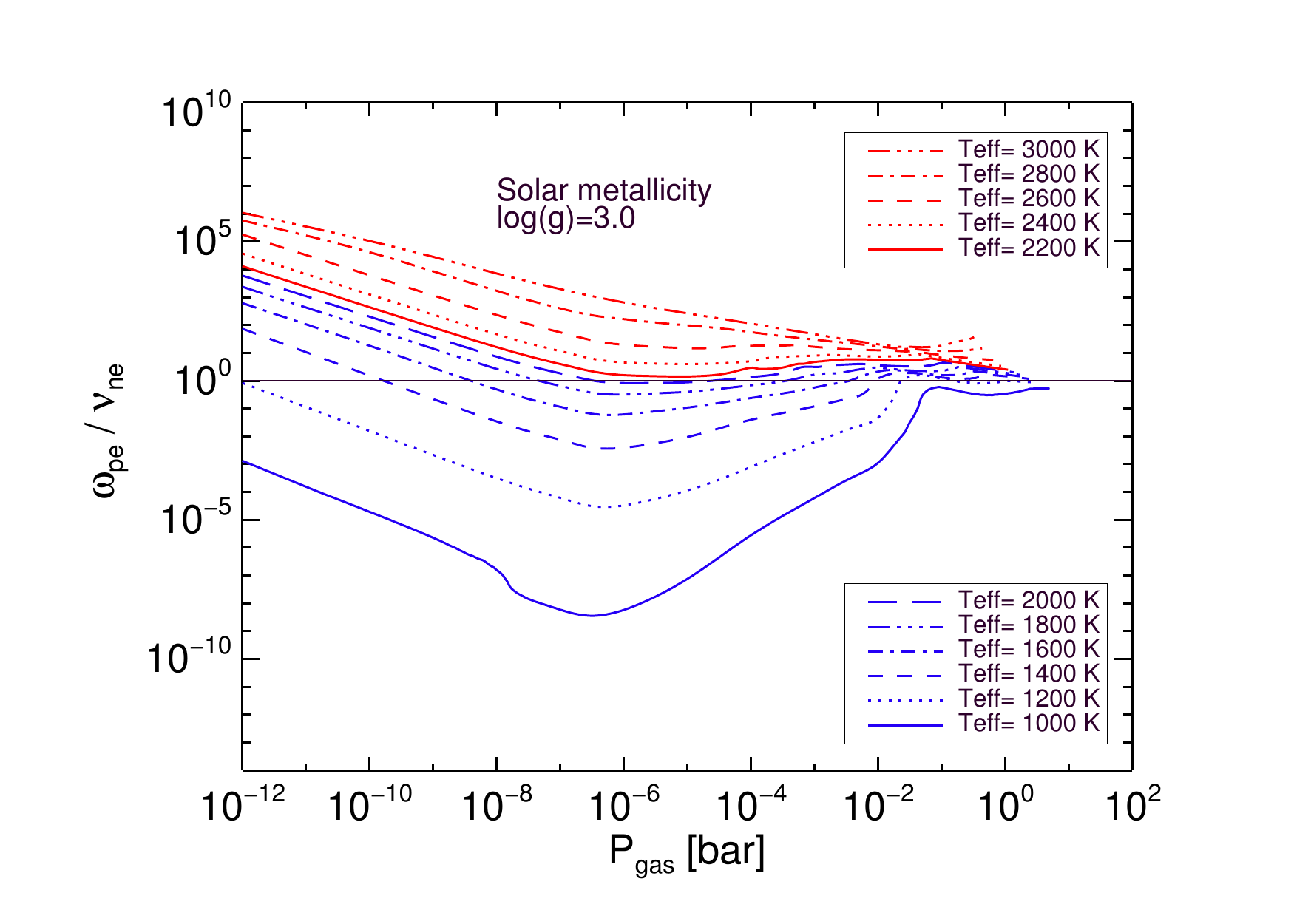}\\*[-0.8cm]
\hspace*{-0.9cm}\includegraphics[angle=0,width=0.6\textwidth]{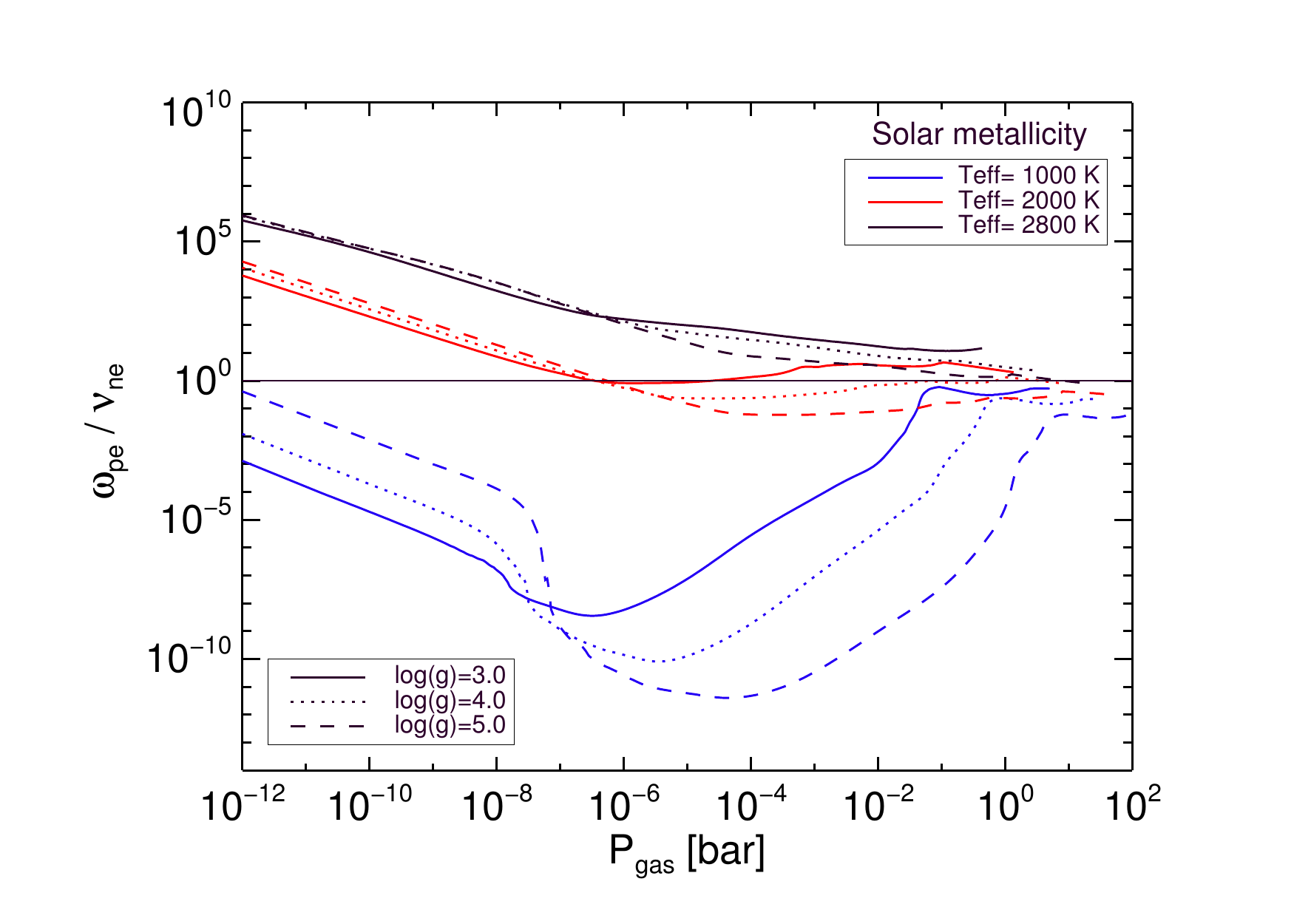}\\*[-0.8cm]
\hspace*{-0.9cm}\includegraphics[angle=0,width=0.6\textwidth]{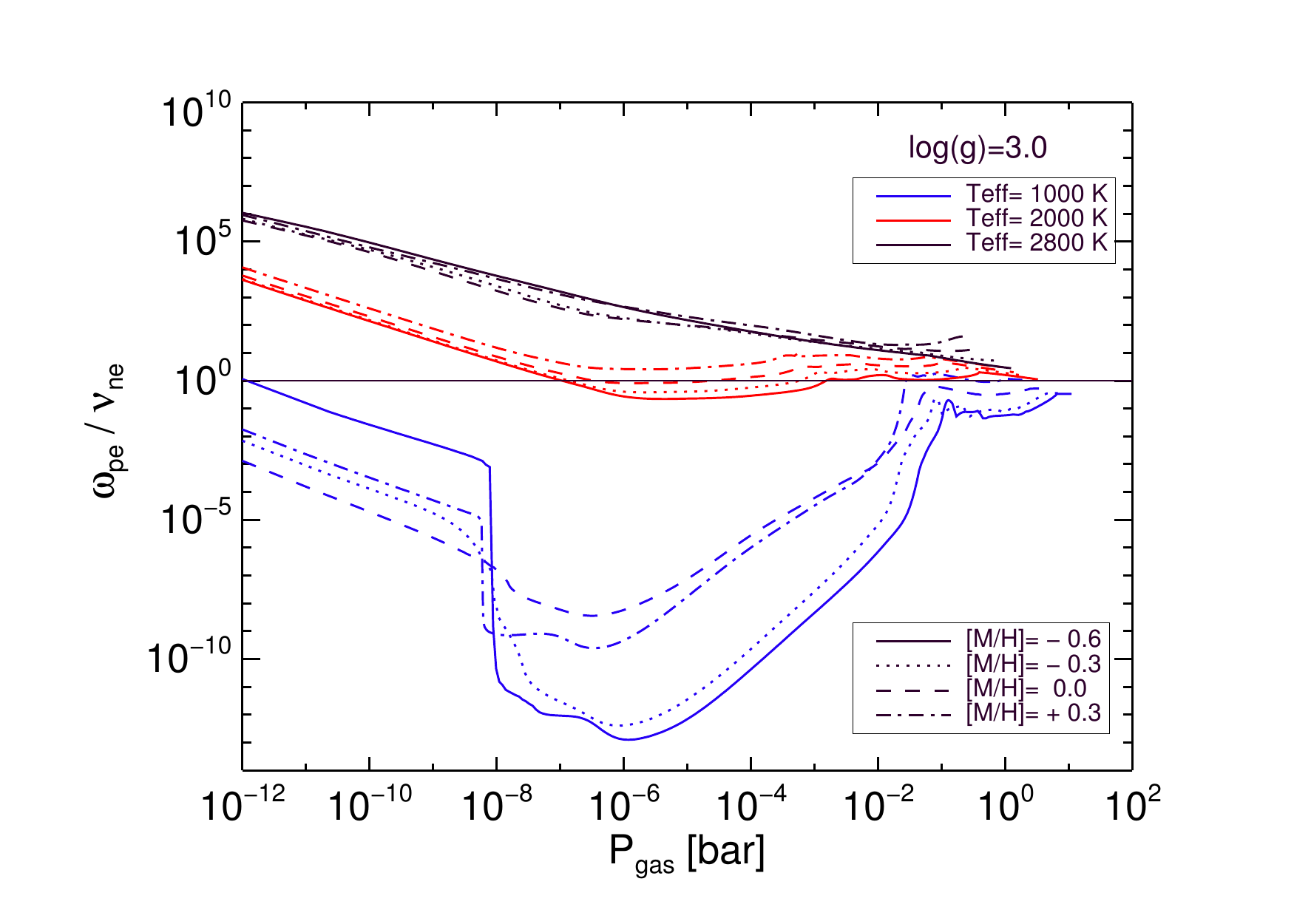}\\*[0.05cm]
\vspace*{-0.5cm}\caption{Ratio of plasma frequency of the electrons and the frequency of collisions between neutral particles and electrons. Electromagnetic interactions dominate over electron-neutral interactions if $\omega_{\rm pe}/\nu_{\rm ne}\gg 1$. {\bf Top:} Group 1.
{\bf Middle:} Group 2.
 {\bf Bottom:} Group 3.}
 \label{fig:fp}
\end{figure}

In a plasma, if electrons are displaced from their equilibrium position (assuming a
uniform, stationary ionic background), a charge imbalance is imposed
on the plasma, creating a local, restoring electric field. Consequently, the electrons try to re-establish charge
neutrality resulting in them oscillating around their equilibrium
position with a particular frequency called plasma frequency. The
plasma frequency is defined as,

\begin{equation}
\omega_{\rm pe}=\left(\frac{n_{\rm e}e^{2}}{\epsilon_{0}m_{\rm e}}\right)^{1/2},\label{eq:fp}
\end{equation}
with  $n_{\rm e}$ the electron number density [$\rm m^{-3}$], $e$ the electron charge [C], $m_{\rm e}$ the electron mass [$\rm kg$].
If the plasma frequency, $\omega_{\rm pe}$,  is greater
than the frequency of  collisions between the electrons and neutral
particles, $\nu_{\rm ne}$ $[\rm s^{-1}]$ then, long-range electromagnetic collective effects
dominant over short-range binary interactions (see Fig.~\ref{fig:fp}). 
The collision frequency for
neutral particles with electrons is given by $\nu_{\rm ne}=\sigma _{\rm gas}n_{\rm  gas}v_{\rm th,e }$, where $v_{\rm th,e }$ is the thermal velocity of electrons given by $v_{\rm th,e}$=$(k_{\rm B}T_{\rm s}/m_{\rm s })^{ 1/2 }$ [m s$^{-1}]$, $n_{\rm gas}$ the ambient gas density and $\sigma_{\rm gas}$ the collision, or scattering, cross section of particles. The latter is assumed to be  $\sigma_{\rm gas}= \pi\cdot r_{\rm gas}^2$ with $r_{\rm gas}=r_{\rm H_2}$ as the
 atmospheric gas in late M-dwarfs, brown dwarfs and most likely in
 giant gas planets is composed mostly of molecular hydrogen, H$_2$. Therefore,
 the collision cross section is approximated by $\sigma_{\rm
   gas}\approx\sigma_{\rm H_{2}}\approx\pi\cdot {r_{\rm
     H_{2}}}^{2}= 5.81\cdot10^{-20}$ m$^{2}$ (r$_{\rm
   H_{2}}=1.36\cdot10^{-10}$ m).
 If the charged particles collide frequently with
the ambient neutral gas ($\omega_{\rm pe}/\nu_{\rm ne}<1$), their
motion will be determined by nearest neighbour interactions and not by
collective, long-range electromagnetic interactions.

 \subsection{Debye length: $\bm{\lambda_{\rm D}\ll L$}}\label{ss:lD}
  
The Debye length, $\lambda_{\rm D}$, is the spatial length scale 
beyond which a plasma can be considered quasi-neutral ($n_{\rm
  e}\approx n_{\rm i}\approx n_{\rm gas}$).
For length scales less than the Debye length, a test charge will
experience the influence of the charge imbalance inside the Debye
sphere. The Debye length, $\lambda_{\rm D}$ [m], resulting from the
solution of the Poisson equation for a non-zero charge density near
test charge, is defined as
\begin{equation}
\lambda_{\rm D}=\left(\frac{\epsilon_{0}k_{B}T_{\rm e}}{n_{\rm e}e^{2}}\right)^{1/2}. \label{debye}
\end{equation} 
with $k_{B}= 1.38\times10^{-23}$ [J\,K$^{-1}$] and $\epsilon_{0}= 8.85\times10^{-12}$ [F\,m$^{-1}$].
A plasma is quasi-neutral if   
\begin{equation}
\lambda_{\rm D}\ll  L.
\end{equation} 
 For an ionised gas region to exhibit plasma behaviour, it is required that over the length scale of the region, the electron number density is high enough that L is greater than the Debye length.
The typical length scale of the plasma, L [m], considered in the literature
(e.g. \citealt{Mohanty2002, Tanaka2014}) is the pressure scale height
which depends on the local gas properties and varies with
$1/g$. Typical values for the pressure scale height are
10$^{5}\ldots\,10{^6}$ cm for a brown dwarf with
$\log(g)=5$. \citet{Tanaka2014} base their length scale on the
definition of the Alfv\'en speed that is of the order of the velocity
of sound (their Eq.12).  Also their approach results in a typical length scale of the order of the pressure scale height. Associate
with the Debye length is the number of charges inside a Debye sphere,
N$_{\rm D}$ ({\it plasma parameter}, App. \ref{ss:Nd}). If N$_{\rm
  D}\gg 1$, the ionised gas exhibits plasma behaviour.

\begin{figure}
\centering
\hspace*{-0.9cm}\includegraphics[angle=0,width=0.6\textwidth]{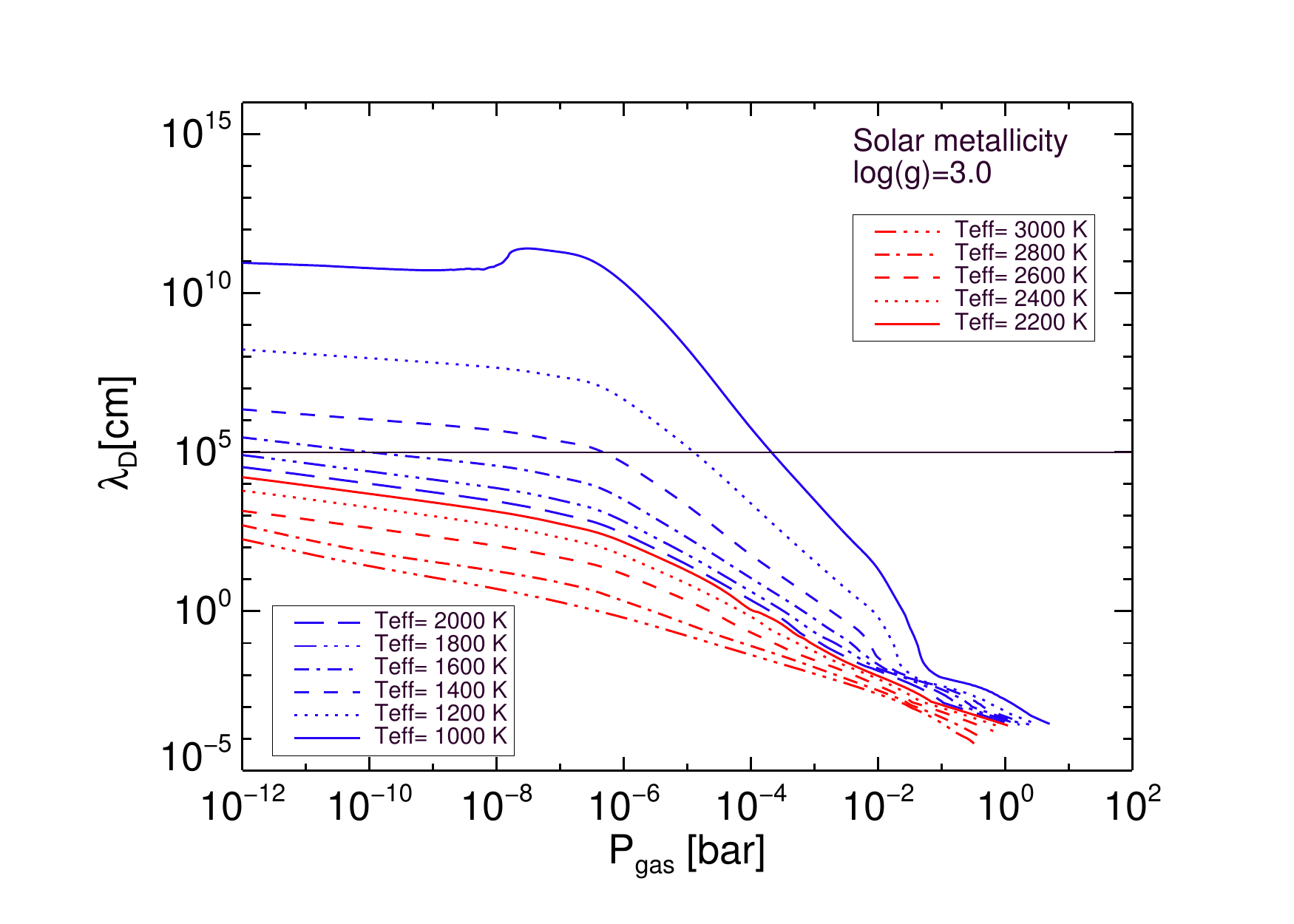}\\*[-0.8cm]
\hspace*{-0.9cm}\includegraphics[angle=0,width=0.6\textwidth]{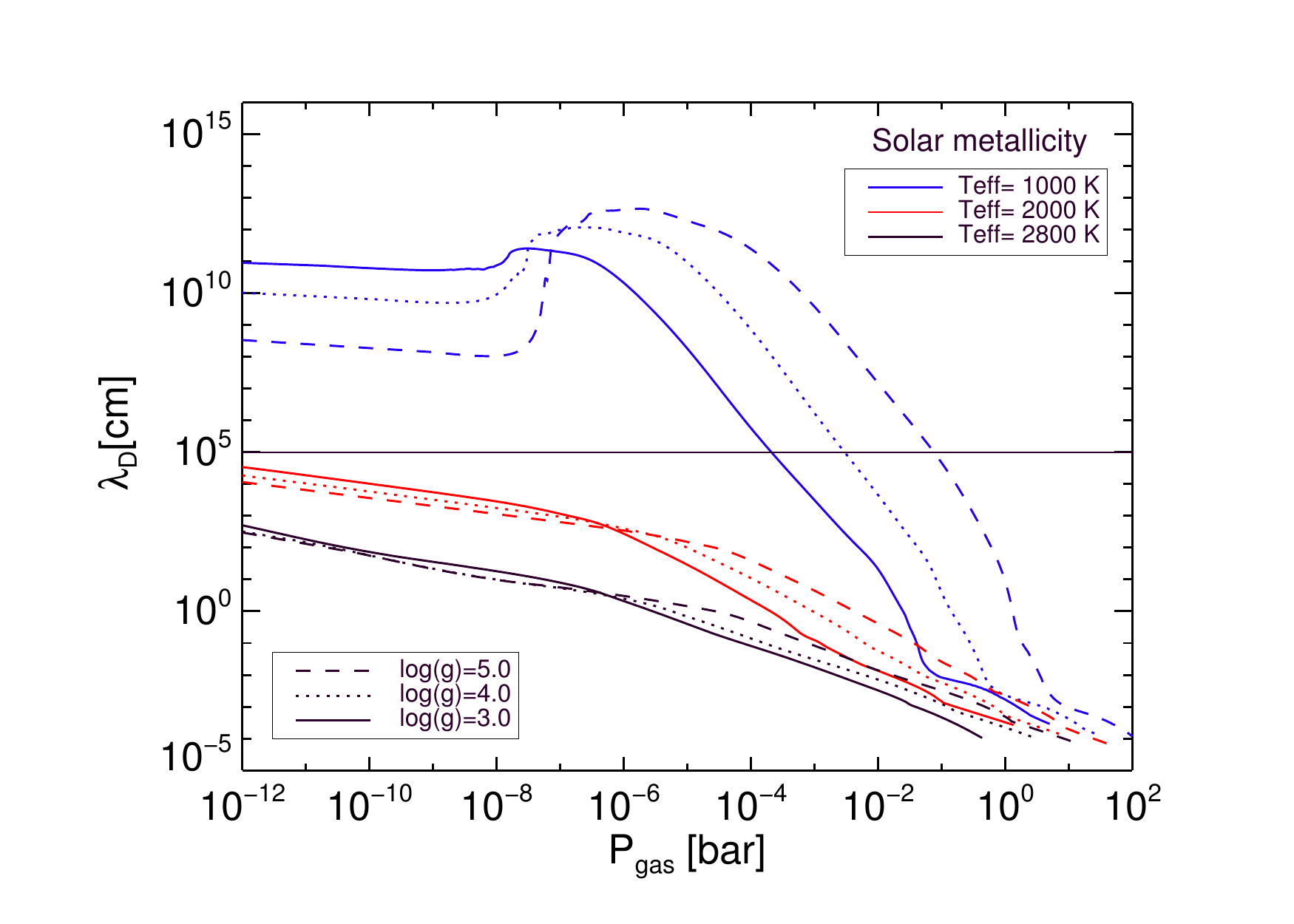}\\*[-0.8cm]
\hspace*{-0.9cm}\includegraphics[angle=0,width=0.6\textwidth]{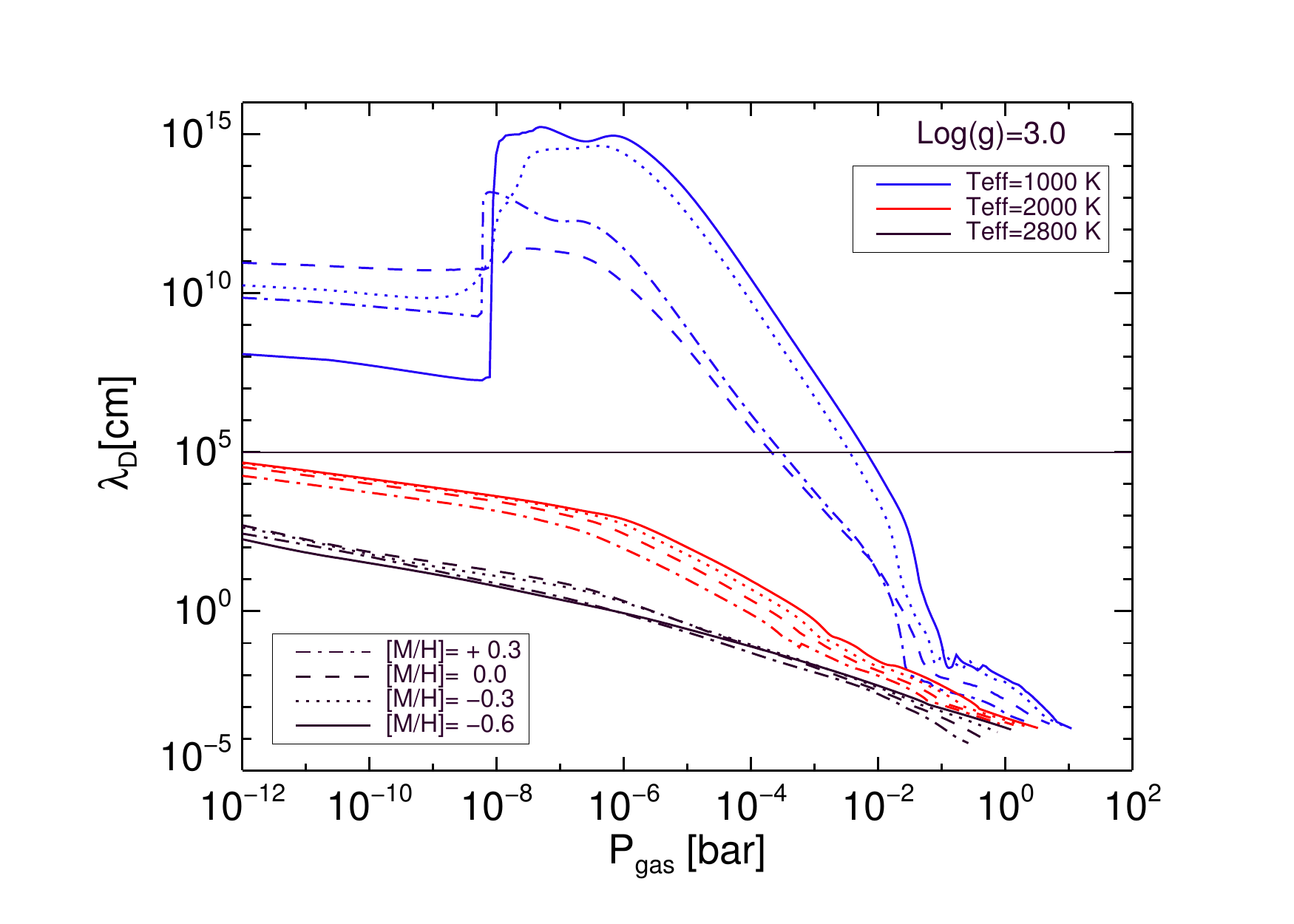}
\\*[-0.6cm]
\vspace{+0.1cm}\caption{Debye length versus the local gas pressure assuming $T_{\rm e}=T_{\rm gas}$. For length scales less than the Debye length, a test charge will experience the influence of the charge imbalance inside the Debye sphere. The horizontal line indicates the typical length scale of the plasma L=10$^3$m consider in this study \citep{Helling2011}.
{\bf Top:} Group 1.
{\bf Middle:} Group 2.
 {\bf Bottom:} Group 3. 
}
\label{fig:lD}
\end{figure}

\begin{figure}
\centering
\hspace*{-1.cm}\includegraphics[width=0.6\textwidth]{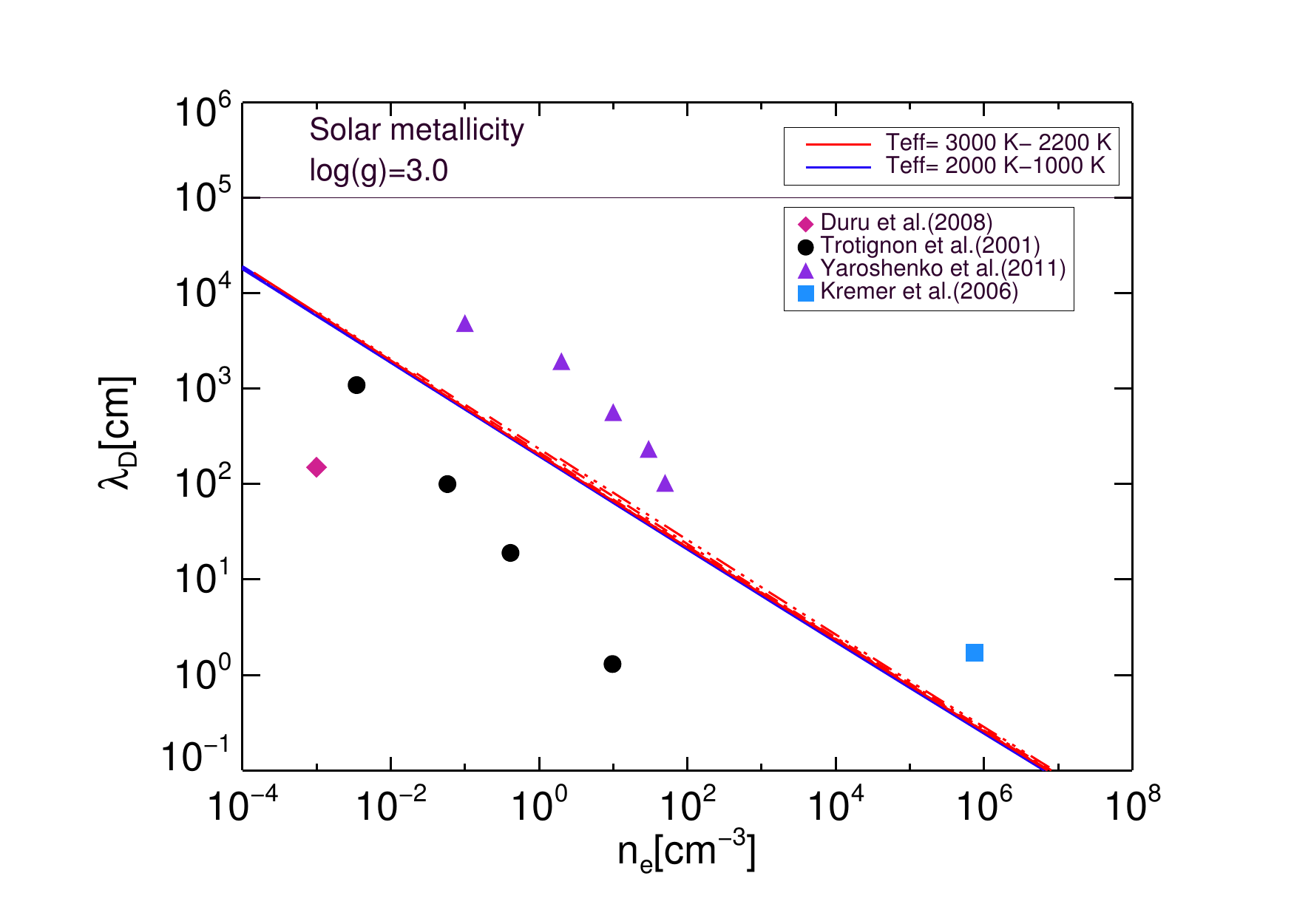}
\vspace{-1cm}\caption{Comparison Debye length of different astrophysical environments Group1 (Table\,\ref{tab}) is plotted here. The red lines represent T$_{\rm eff}=3000\,$K- 2200$ \,$K covering M-dwarfs and young L-dwarfs. Blue lines represent T$_{\rm eff}=2000$\,K-1000 $\,$K covering late L-dwarfs and giant gas planets regime. The Debye lengths for Martian's atmosphere are plotted as red diamond \citep{Duru2008} and black circle \citep{Trotignon2001}, for plasma composed of electrons, water group ions and protons as violet triangle \citep{Yaroshenko2011} and a pure electron plasma as light blue square symbols \citep{Kremer2006}. The black line represents the typical
atmospheric length scale, L=10$^{3}$ m \citep{Helling2011}  }
\label{fig:lDvsne}
\end{figure}

\subsection{Plasma parameters across the star-planet regime} \label{results}

In the following, we evaluate the plasma criteria for late M-dwarfs,
brown dwarfs and giant gas planet atmospheres. All results have been
calculated considering thermal ionisation only and compose our
reference study against which the need for additional ionisation
processes can be derived. First we examine for which global parameters
and $p_{\rm gas}$ values, the gas is ionised above the threshold value
of $f_{\rm e}>10^{-7}$ (Sec.~\ref{sss:fe}). In Sec.~\ref{sss:fp} we
demonstrate that long-range electromagnetic collective interactions of
many charged particles can dominate over short-range binary
interactions also in regions of a very low degree of
ionisation. Recent brown dwarf atmospheric investigations have focused
on the degree of ionization to characterize plasma behaviour, in this
paper we consider multiple parameters to gain a more detailed
characterization.  (e.g. \citealt{Osten2015}). In Sec.~\ref{sss:lD} we
demonstrate for which length scale ultra-cool atmospheres will be
effected by electrostatic processes and that it is not unreasonable to
expect ultra-cool atmospheres to emit H$_{\alpha}$ or even in radio
wavelength as in particular the rarefied upper parts of the
atmospheres fulfill plasma criteria easily despite having low degrees
of ionisation.

\subsubsection{Degree of ionisation by thermal processes, ${f_{\rm e}}$}\label{sss:fe}

  Figure~\ref{fig:fe} shows the degree of thermal ionisation evaluated
  for the same models represented in Fig.~\ref{fig:Tpgas}
  (Table~\ref{tab}). Guided by these results we consider $f_{\rm
    e}>10^{-7}$ to be a threshold above which the gas is
  partially ionised and it may exhibit plasma behaviour.
  The above choice of a threshold value is supported by results from
  laboratory experiments and laboratory plasma devices (e.g. Tokamak;
  \citealt{Diver2001}; \citealt{Fridman2008}).  For a fluorescent
  tube, the degree of ionisation is $f_{\rm e}\approx10^{-5 }$
  according to \citet{Inan2011}.  \citet{Christophorou2004} showed
  that at low temperature (T$_{\rm gas}\approx300-600$~K) and low
  density ($10^{13}-10^{16}$~cm$^{-3 }$) the gas is weakly ionised
  with $f_{\rm e}\approx10 ^{ -6 }-10^{-1}$.  If the density of the
  charged particles increases towards $f_{\rm e}\rightarrow1$ the gas
  will be fully ionised. For example, a fully ionised gas is assumed
  in ideal MHD calculations. This threshold, f$_{\rm e}> 10^{-7}$,
  allows us to derive the atmospheric volume that can be considered as
  an ionised gas (Sect.\,\ref{s:comp}, Fig.~\ref{fig:vfe}).  Deriving
  such atmospheric volume fractions will enable us to compare the
  results from different plasma criteria
  (Eqs.\,\ref{eq:wpe},\,\ref{eq:lD}) and to demonstrate that a gas
  does not need to be fully ionised in order to exhibit collective
  plasma effects.

\begin{figure*}
\hspace*{-0.0cm}\includegraphics[width=0.83\textwidth]{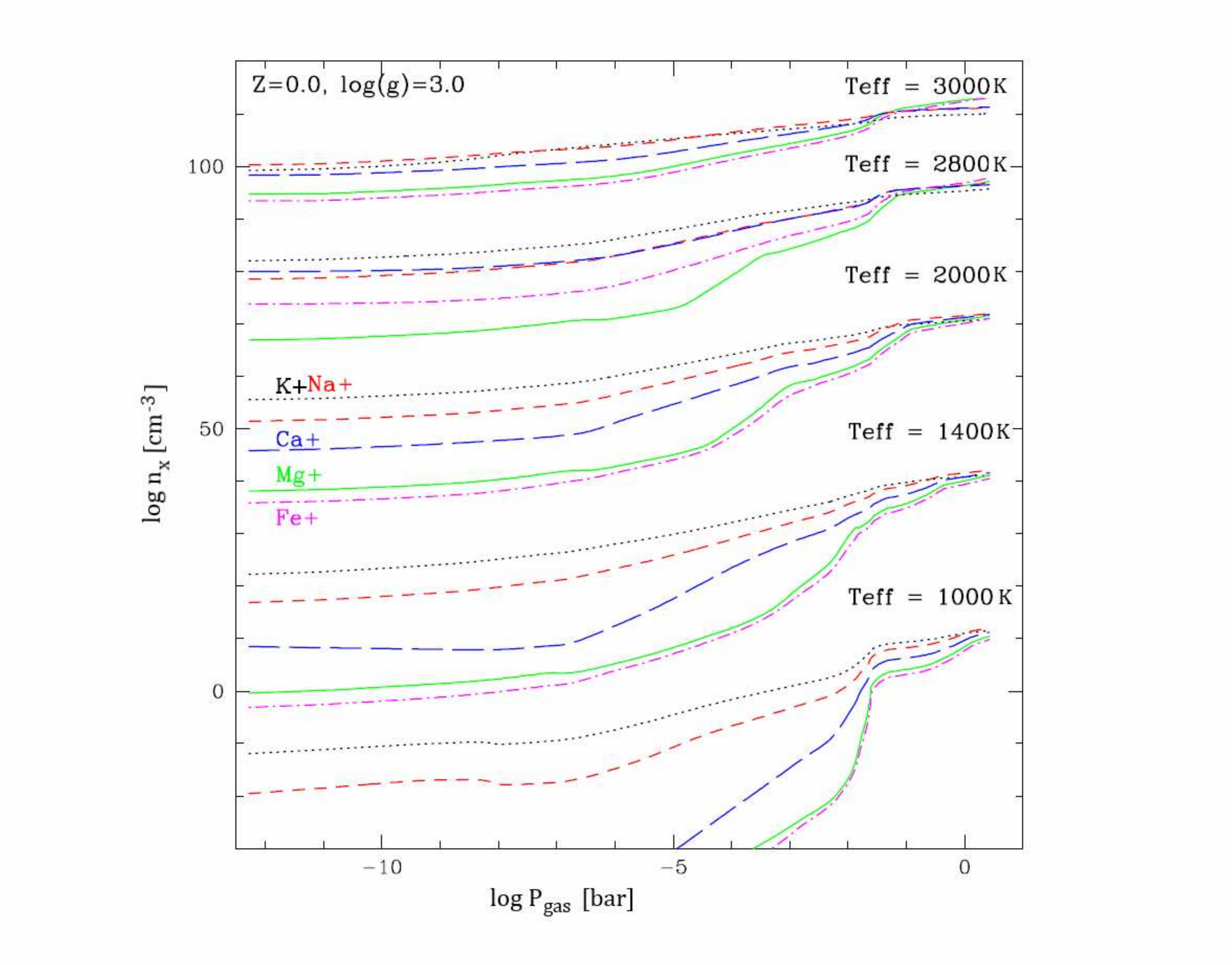}\\*[-0.5cm]
\caption{The dominating thermal electron donors for a subset of
  effective temperatures for log(g)=3,0 and solar element abundances;
  Na$^{+}$, K$^{+}$ and Ca$^{+}$ in the low-density atmosphere for all
  T$_{\rm eff}$; Mg$^{+}$ and Fe$^{+}$ in the inner parts for T$_{\rm
    eff}\ge 2800\,$K. {\it An off-set of 3 orders of magnitudes is
    applied between the models for all T$_{\rm eff}>1000$K to allow a
    better comparison.}}
\label{chem1}
\end{figure*}

 \begin{itemize}
\item {\bf Group 1:} changing {\bf T$_{\rm {\bf eff}}$}~ {\bf
  (Fig.~\ref{fig:fe}, top)}\\ ($\log(g)$=3.0, {\small [M/H]}$=
  0.0$) \vspace{+0.1cm}\\ A solar-metallicity M-dwarf with T$_{\rm
    eff}=3000$~K achieves f$_{\rm e}> 10^{-7}$ in almost the entire
  atmosphere. For cooler atmospheres with T$_{\rm eff}\le2800$~K,
  $f_{\rm e}>10^{-7}$ is only reached for $p_{\rm gas}>10^{-4}$
  bar. The atmospheric fraction that reached $f_{\rm e}>10^{-7}$
  increases with increasing T$_{\rm eff}$.
  \vspace{+0.1cm}

\item {\bf Group 2}: changing {\bf log(g)}~{\bf (Fig.~\ref{fig:fe},
  middle)}\\ (T$_{\rm eff}= 1000$~K, $2000$~K, $2800$~K , {\small
  [M/H]}$= 0.0$) \vspace{+0.1cm} \\ Values of varying surface gravity
  $\log(g)$= 3.0, 4.0, 5.0 are studied here.  Models with T$_{\rm
  eff}=2800$~K reach $f_{\rm e}>10^{-7}$ for $p_{\rm
  gas}>10^{-4}$ bar; models with T$_{\rm eff}=2000$~K for $p_{\rm
  gas}>10^{-2}$ bar. A small part of the upper atmospheric reaches
  $f_{\rm e}>10^{-7}$ for T$_{\rm eff}$=2800\,K and
  $\log(g)$=4.0, 5.0 due to the increasing contribution of Ca+ with
  outwards decreasing $p_{\rm gas}$ (compare Fig.~\ref{chem1}).  Models with T$_{\rm eff}=1000$~K have only a
  small fraction of the atmospheric gas that reaches the $f_{\rm
    e}>10^{-7}$ threshold. This occurs in the deepest layers of
  the atmosphere $1< p_{\rm gas}<10^{2}$ bar. The atmospheric fraction that reached $f_{\rm e}>10^{-7}$
  increases with decreasing $\log(g)$ at high $p_{\rm gas}$ and with increasing $\log(g)$ at low $p_{\rm gas}$.  \vspace{+0.1cm}

\item {\bf Group 3}: changing {\bf{\small [M/H]}$= 0.0$ }{\bf
  (Fig.~\ref{fig:fe}, bottom)}\\ (T$_{\rm eff}= 1000$~K, $2000$~K,
  $2800$~K, $\log(g)$) \vspace{+0.1cm} \\ Values of varying
  metallicity ${\small [M/H]}$= - 0.6, -0.3, 0.0, +0.3 are
  analysed. Models with T$_{\rm eff}$=2800 K and T$_{\rm eff}=2000$~K
  satisfy $f_{\rm e}>10^{-7}$ for all values of metallicity for
  $p_{\rm gas} >10^{-5}$ bar. A small part of the upper atmospheric
  reaches $f_{\rm e}>10^{-7}$ for 
  T$_{\rm eff}$=2800\,K due to an increasing electron donation from Ca+.  Models with T$_{\rm eff}$=1000\,K have only
  a small fraction of the atmospheric gas that can be ionised for
  $p_{\rm gas}>10^{-1}$ bar.
\end{itemize}
  All models of non-irradiated atmospheres show a degree of ionisation which increases from a minimum ($p_{\rm gas}\sim10^{-8}-10^{-6}$ bar) with increasing local gas pressure values towards the deeper layers of the
  atmosphere. Atmosphere models with T$_{\rm eff}\le$2800\,K, $\log(g)$= 3.0, [M/H]= 0.0 can reach
  f$_{\rm e} > 10^{-7}$ only for high $p_{\rm gas}$ (inner parts of
  the atmosphere). Only one model atmosphere achieves $f_{\rm
    e}>10^{-7}$ throughout nearly the entire atmosphere (T$_{\rm
    eff}$= 3000\,K, $\log(g)$= 3.0, [M/H]= 0.0). 
For atmospheres of late M-dwarfs, brown
  dwarfs and giant gas planets for T$_{\rm eff}=1000\,\ldots\,3000$K
  for varying log(g) and metallicity, we observe that:\\*[-0.6cm]
    \begin{itemize}
\item[-] The hottest model has the highest thermal degree of
  ionisation, $f_{\rm e}$.
\item[-] The lowest value of surface gravity causes an increase of
  $f_{\rm e}$ at high $p_{\rm gas}$, however, the highest value of
  surface gravity causes an increase of $f_{\rm e}$ at low $p_{\rm
    gas}$. Both trends are for a given T$_{\rm eff}$ and [M/H].
\item[-] The highest metallicity values cause an increase of $f_{\rm
  e}$ at high $p_{\rm gas}$ compared to the lowest metallicity
  models. For T$_{\rm eff}$=2800\,K and T$_{\rm eff}$=1000\,K the
  lowest value of the metallicity causes an increase of $f_{\rm e}$ at
  low $p_{\rm gas}$. Both trends are for a given T$_{\rm eff}$ and
  $\log(g)$.
 \end{itemize}
  
 Section~\ref{ions} investigates which atoms or molecules are the most
 important electron donors in these cold atmospheres, and hence,
 responsible for the values of the degree of thermal ionisation, $f_{\rm
   e}$. We note that the degree of thermal ionisation will be
 influenced by the formation of clouds if the dominating electron
 donors are amongst the most abundant condensing species. Ca does not fall into this category.

\subsubsection{Dominating electromagnetic interaction}\label{sss:fp}
 
  The criterion $\omega_{\rm pe }\gg \nu_{\rm ne}$ is used to derive
  where in an ultra-cool atmosphere the long-range, electromagnetic,
  collective interactions of many charged particles dominates over
  short-range binary interactions in a ionised gas of a certain degree
  of ionisation.  Figure~\ref{fig:fp} shows the results of this
  criterion for the three groups of model atmosphere structures
  (Fig.~\ref{fig:Tpgas}).

 \begin{itemize}
 \item {\bf Group 1:} changing {\bf T$_{\rm {\bf eff}}$}~ {\bf
   (Fig.~\ref{fig:fp}, top)}\\ ($\log(g)$=3.0, {\small [M/H]}$=
   0.0$) \vspace{+0.1cm}\\ As T$_{\rm eff}$ increases, the range of
   the $p_{\rm gas}$ where $\omega_{\rm pe }\gg \nu_{\rm ne }$
   increase too. For models with T$_{\rm eff}\ge2200$~K the entire
   atmosphere satisfies this criterion; for T$_{\rm eff}=2000$~K
   almost the entire atmosphere; for models with T$_{\rm
     eff}=1800-1400$~K in the uppermost and for the innermost parts of
   the atmosphere and for T$_{\rm eff}=1200$~K only for 10$^{-2}<$
   p$_{\rm gas}<10^{-1}$ bar. The model with T$_{\rm eff}=$1000\,K is
   too cool to fulfill this criterion.\vspace{+0.1cm}

\item {\bf Group 2}: changing {\bf log(g)}~{\bf (Fig.~\ref{fig:fp},
  middle)}\\ (T$_{\rm eff}= 1000$~K, $2000$~K, $2800$~K , {\small
  [M/H]}$= 0.0$) \vspace{+0.1cm} \\ Models with T$_{\rm eff}=2800$~K
  satisfy this criterion throughout the whole atmosphere except for
  $\log$=5.0 at the highest pressures. For T$_{\rm eff}=2000$~K
  and $\log(g)$=3.0 almost the entire atmosphere fulfills this
  criterion; T$_{\rm eff}=2000$~K and $\log(g)$=4.0 only in the
  uppermost and for the innermost parts of the atmosphere; for T$_{\rm
    eff}=2000$~K and $\log(g)$= 5.0 only for p$_{\rm gas}<10^{-6}$
  bar. Models with T$_{\rm eff}=1000$~K do not satisfy the
  criterion. \vspace{+0.1cm}

\item {\bf Group 3}: changing {\bf{\small [M/H]}$= 0.0$ }{\bf
  (Fig.~\ref{fig:fp}, bottom)}\\ (T$_{\rm eff}=1000$~K, $2000$~K,
  $2800$~K, $\log(g)$) \vspace{+0.1cm} \\ Models with T$_{\rm
  eff}=2800$~K and all value of metallicity and T$_{\rm eff}=2000$~K
  with [M/H]= +0.3 satisfy this criterion in the whole atmosphere. For
  T$_{\rm eff}=2000$~K and [M/H]= 0.0 almost the entire atmosphere
  fulfills this criterion. For T$_{\rm eff}=2000$~K and [M/H]= -0.3,
  -0.6 only in the uppermost and for the innermost parts of the
  atmosphere, $10^{-7}>p_{\rm gas}>10^{-3}$; models with T$_{\rm
    eff}=1000$~K and [M/H]= + 0.3 only for $10^{-2}< p_{\rm
    gas}<10^{0}$. Models with T$_{\rm eff}=1000$~K and [M/H]= 0.0,
  -0.3, -0.6 and do not fulfill this criterion for any atmospheric gas
  pressure. 
\end{itemize}

Figure~\ref{fig:fp} demonstrates that the collective, long-range electromagnetic interactions dominate over short-range binary interactions in atmospheres of low degrees of ionisations, i.e. for $2800\ge\,$T$_{\rm eff}\ge2000$~K. As T$_{\rm eff}$ and the metallicity increase, $\omega_{\rm pe}\gg
  \nu_{\rm ne}$ is easier fulfilled at high $p_{\rm gas}$, however, as T$_{\rm eff}$ increases and the metallicity decreases, $\omega_{\rm pe}\gg
  \nu_{\rm ne}$ is easier fulfilled at low $p_{\rm gas}$ for T$_{\rm eff}=2800$~K, $1000$~K. This effect is counteracted by an increase in log(g). The lowest value of $\log(g)$ causes a
  decrease of $\omega_{\rm pe }/\nu_{\rm ne}$  in the uppermost parts of the atmosphere and an increase
  in the innermost parts. 
Consequently, long-range, electromagnetic,
  collective interactions of many charged particles do not require a
  complete ionisation of the atmospheric gas, and a moderate gas
  ionisation is sufficient.

\begin{figure}
{\includegraphics[angle=0,width=0.45\textwidth]{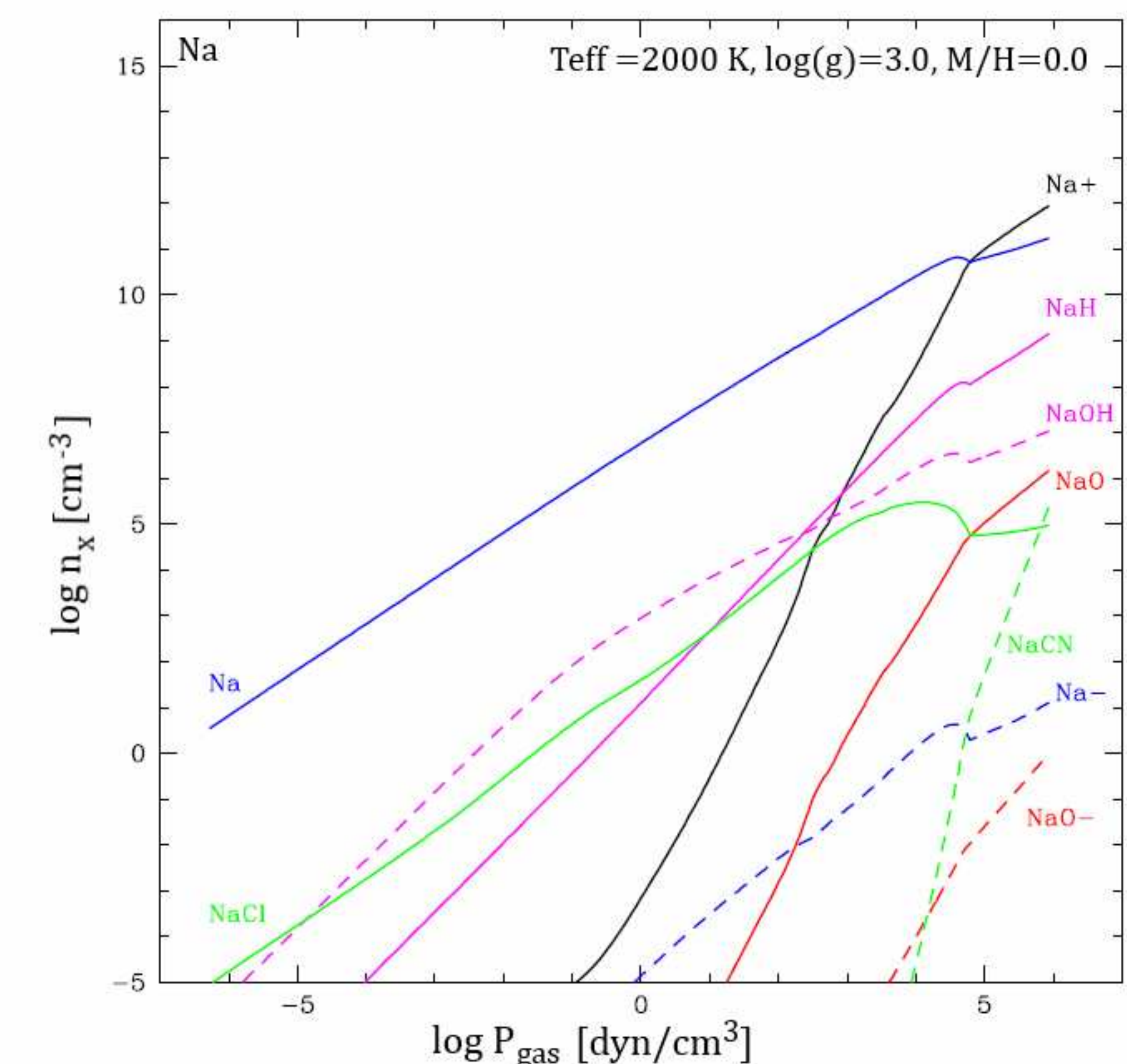}}\\*[-0.0cm]
{\includegraphics[angle=0,width=0.45\textwidth]{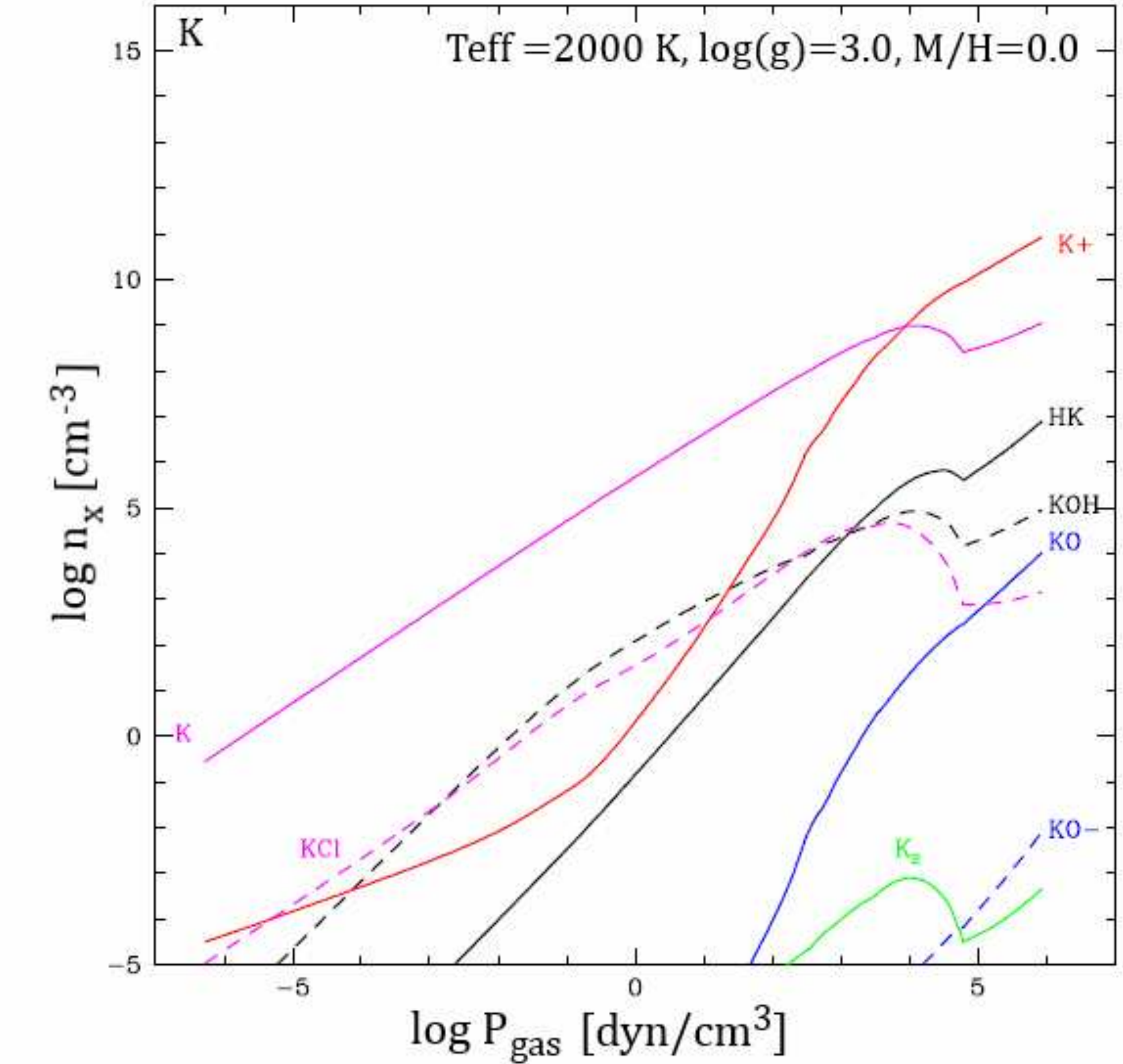}}\\*[-0.0cm]
{\includegraphics[angle=0,width=0.45\textwidth]{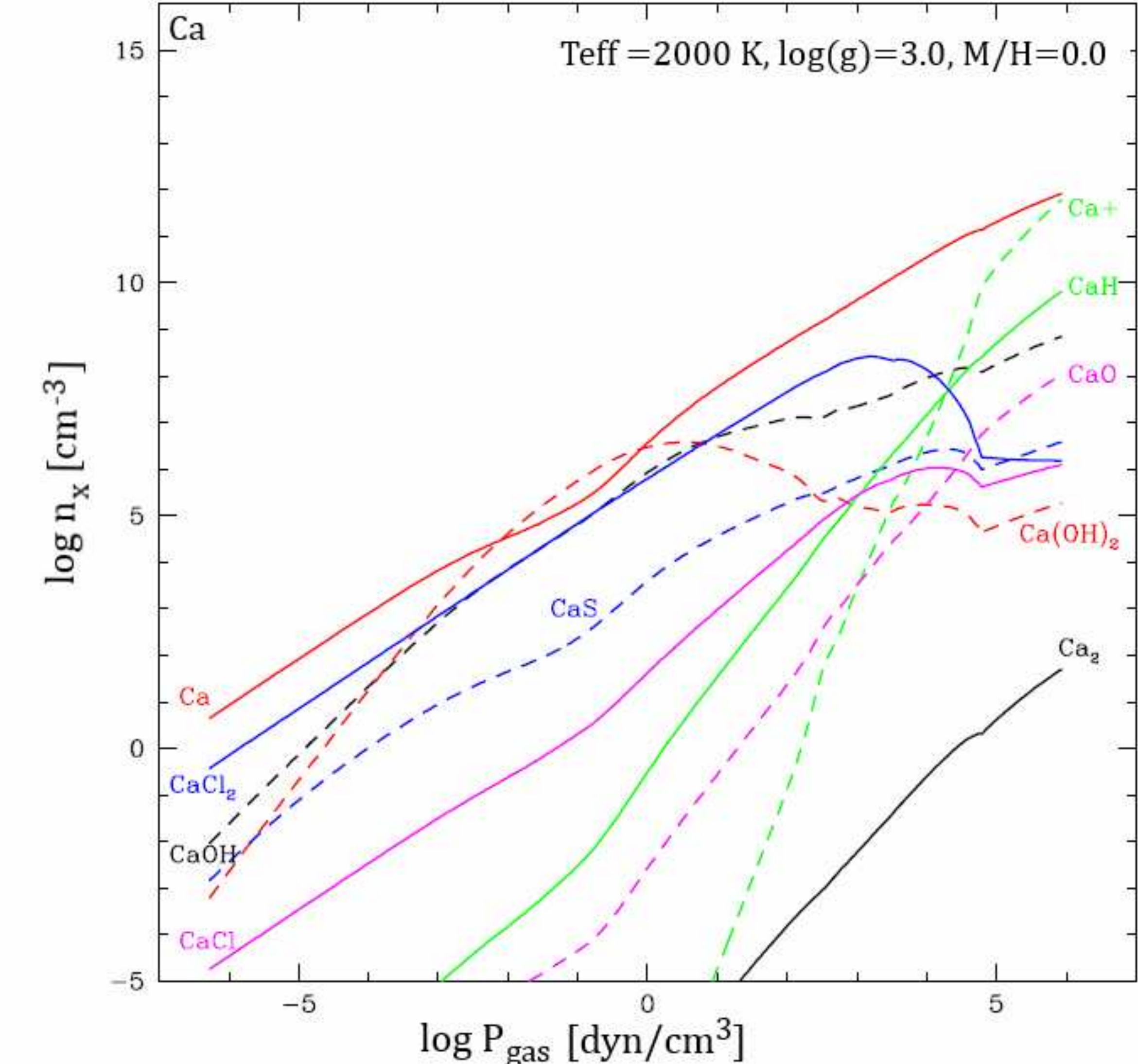}}\\*[-0.5cm]
\caption{Distribution of electron donor's abundance over atoms, molecules and ions for a warm model of T$_{\rm eff}=2000$K, log(g)=3.0 and solar element composition. The upper left corner contains the element considered.}
\label{fig:chemcomp_el1}
\end{figure}

\begin{figure}
{\includegraphics[angle=0,width=0.45\textwidth]{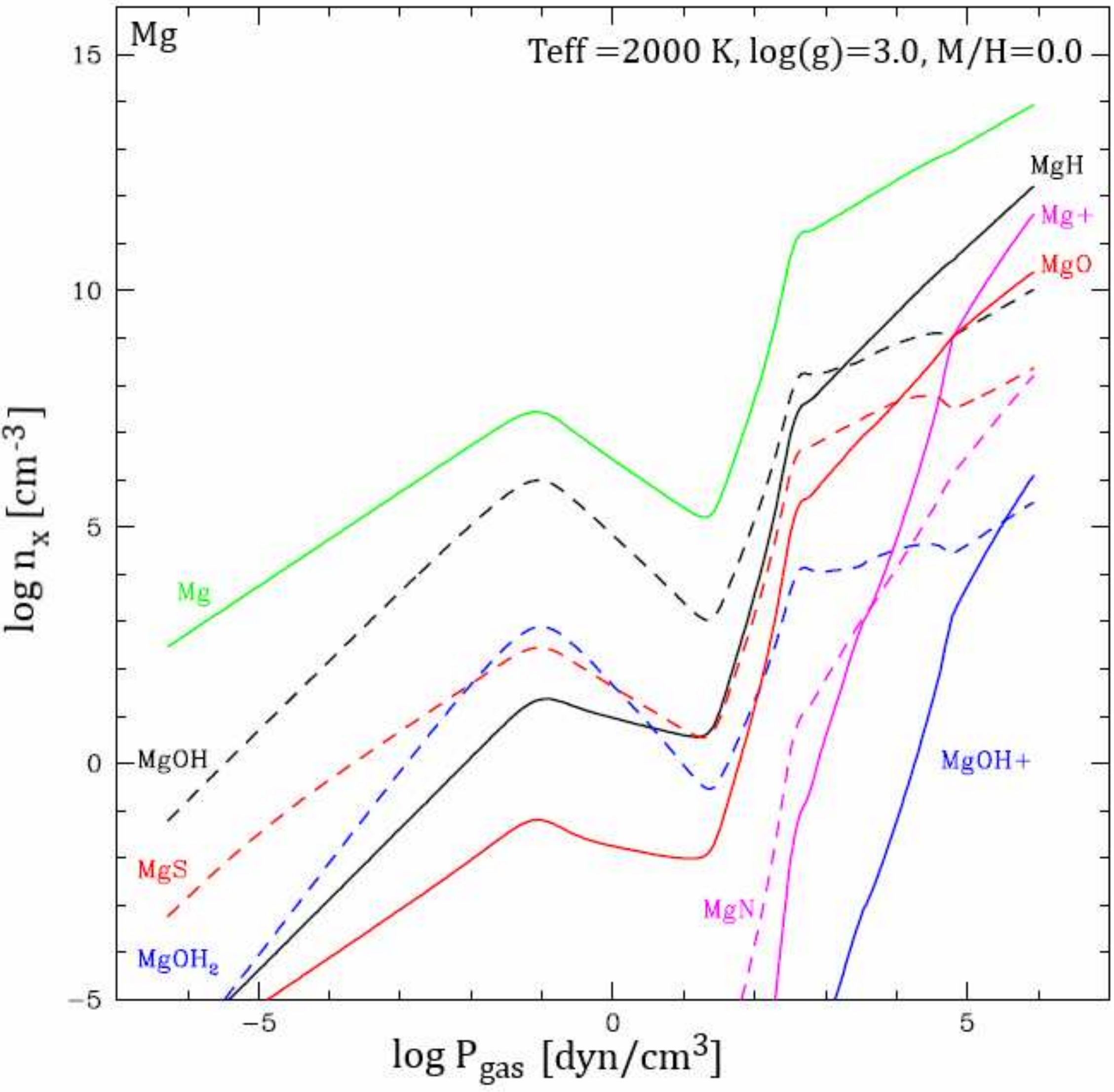}}\\*[-0.0cm]
{\includegraphics[angle=0,width=0.45\textwidth]{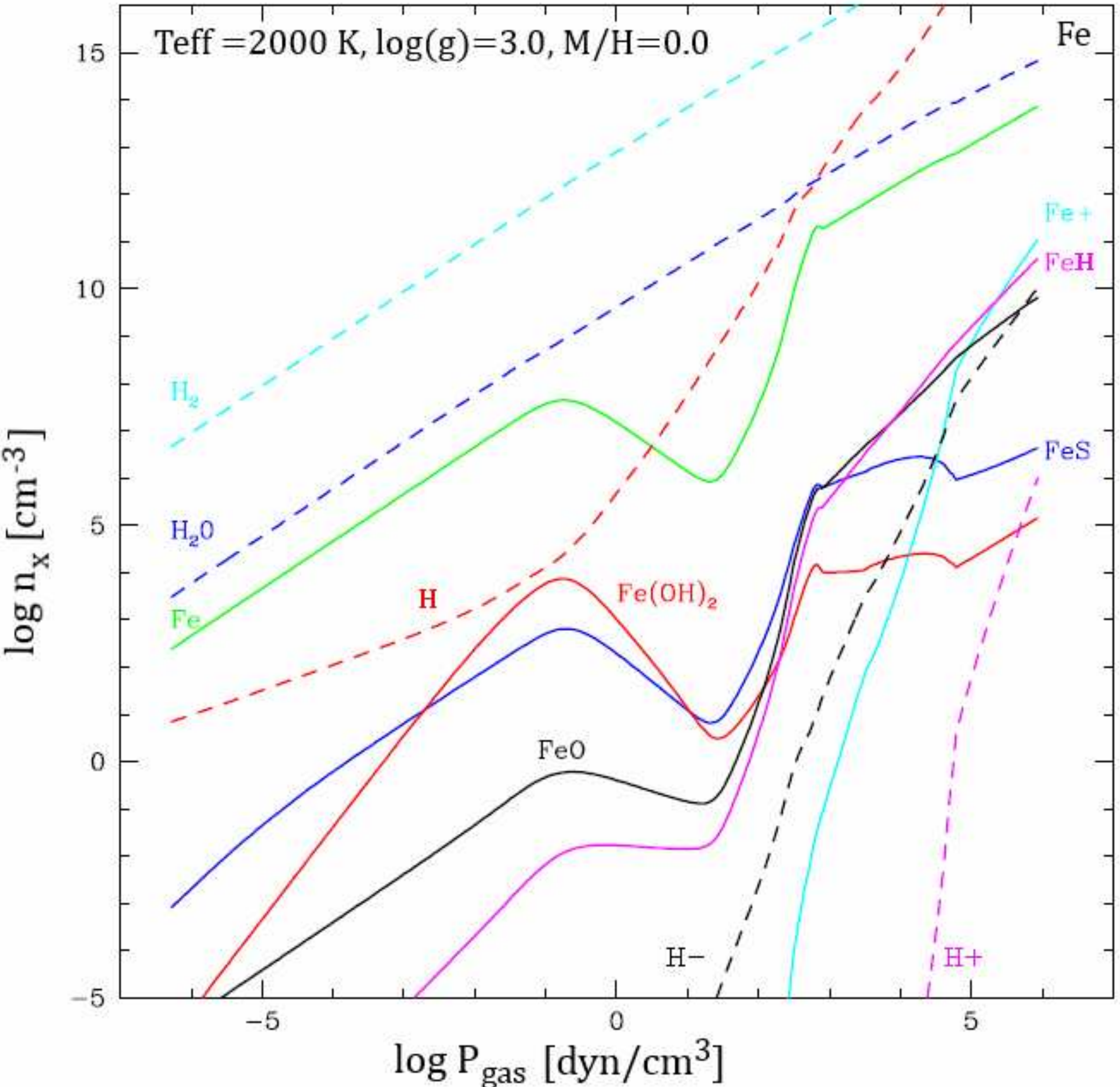}}\\*[-0.3cm]
\caption{Same like Fig.~\ref{fig:chemcomp_el1}. Both, Mg and Fe are
  influenced by dust formation which strongly decreases both elements,
  resulting into the localize large kink. The upper left corner
  contains the element considered. H-binding species are shown
  for comparison.}
\label{fig:chemcomp_el2}
\end{figure}

 \subsubsection{Electrostatically effected atmospheric length scales}\label{sss:lD}
 
The Debye length, $\lambda_{\rm D}$, is compared to a typical
atmospheric length scale of the order of the pressure scale height,
L=10$^{3}$ m \citep{Helling2011}. Figure~\ref{fig:lD} shows how the Debye length changes
depending on the local atmospheric gas pressure, and where $\lambda_{\rm D}\ll L$ is fulfilled.
 \begin{itemize}
 \item {\bf Group 1}: changing {\bf T$_{\rm {\bf eff}}$}~ {\bf
   (Fig.~\ref{fig:fp}, top)}\\ ($\log(g)$=3.0, {\small [M/H]}$=
   0.0$) \vspace{+0.1cm}\\ The criterion $\lambda_{\rm D}\ll L$ is
   fulfilled throughout the whole atmosphere for T$_{\rm
     eff}\ge1800$~K; only for T$_{\rm eff}=1600$~K a small atmospheric
   gas volume for $p_{\rm gas}<10^{-10}$ bar cannot reach this
   criterion. For models with T$_{\rm eff}\le1400$~K this criterion is
   fulfilled for $p_{\rm gas}>10^{-6}$ bar. As T$_{\rm eff}$
   increases, the range of $p_{\rm gas}$ where $\lambda_{\rm D}\ll L$
   increases.\vspace{+0.1cm}
 \item {\bf Group 2}: changing {\bf log(g)}~{\bf (Fig.~\ref{fig:fp},
   middle)}\\ (T$_{\rm eff}= 1000$~K, $2000$~K, $2800$~K , {\small
   [M/H]}$= 0.0$) \vspace{+0.1cm} \\ Models with T$_{\rm eff}=2800$~K and $2000$~K
   satisfy this criterion for all surface gravity values and
   throughout the whole atmosphere. For T$_{\rm eff}=1000$~K and all surface gravity values
   then $\lambda_{\rm D}\ll L$ is fulfilled only for $p_{\rm
     gas}>10^{-4}$ bar.\vspace{+0.1cm}
 \item {\bf Group 3}: changing {\bf{\small [M/H]}$= 0.0$ }{\bf
   (Fig.~\ref{fig:fp}, bottom)}\\ (T$_{\rm eff}=1000$~K, $2000$~K,
   $2800$~K, $\log(g)$) \vspace{+0.1cm} \\ Models with T$_{\rm
   eff}=2800$~K and $2000$~K satisfy this criterion for all metallicities and
   $p_{\rm gas}$. Models with T$_{\rm eff}=1000$~K satisfy this
   criterion in the inner atmospheric regions only where $p_{\rm
     gas}>10^{-4}$ bar.
 \end{itemize} 
 
Figure~\ref{fig:lD} demonstrates that the Debye length is generally
very large in the upper atmospheric regions throughout the whole
regime of ultra-cool objects, i.e. late M-dwarfs, brown dwarfs and
giant gas planets. In the upper atmosphere the electron density is low
causing an increasing Debye length; whereas deeper in the atmosphere
the electron number density is high and so the Debye length is
relatively lower. According to \cite{Cravens1997}, at the top of the
Earth's ionosphere $\lambda_{\rm D}\approx 1$cm for $T_{\rm e}\approx
1000$K and $n_{\rm e}\approx 10^{11}$cm$^{-3}$, compared to a vertical
extant of $\approx 300$km of the ionosphere. For the solar wind at
$\approx$ 1AU, $\lambda_{\rm D}\approx 700$cm ($T_{\rm e}\approx
10^5$K, $n_{\rm e}\approx 10^{7}$cm$^{-3}$) .  A comparison of
different values of the Debye sphere for different astrophysical
environments is presented in Table\,\ref{tab2} and in
Fig.\,\ref{fig:lDvsne}. \cite{Duru2008} investigate the electron
density in the upper ionosphere of Mars, \cite{Trotignon2001} the
interaction of the Martian's atmosphere with the solar wind. Both
consider the presence of the dust in the plasma
environment. \cite{Yaroshenko2011} model a plasma composed of
electrons, water group ions and protons with the presence of
photoemission due to the UV radiation. \cite{Kremer2006} work with a
pure electron plasma.

\begin{table}
\caption{Debye lengths for different astrophysical environments. Figure\,\ref{fig:lDvsne} provides a comparison to results of these papers.}
\centering
\scriptsize\addtolength{\tabcolsep}{-3.8pt}
\hspace*{-0.3cm}\begin{tabular}{|c|c|c|c|c|c|} 
\hline\hline
      Object &$T_{\rm e}$ [K] & $n_{\rm e}$ [cm$^{-3}$]					  & $\lambda_{D} [cm]$	 & References	 \\ [0.5ex]
 \hline\hline
 \rule{0pt}{4ex} Martian's & & & &\cite{Duru2008}  \\
                        
                        Ionosphere         & 5000    & 10$^{-3}$ & 1.5$\times$10$^{2}$ & (Mar's atmosphere)  \\
                        &\\
\hline
\rule{0pt}{4ex}   & 347 & 9.78& 1.3&\\
                          Martian's  &3131& 4.12$\times$10$^{-1}$&19& \cite{Trotignon2001}\\ 
			  Ionosphere &12222 &5.82$\times$10$^{-2}$&10$^{2}$& (Mar's atmosphere)\\
                           &86665&3.47$\times$10$^{-3}$&1.1$\times$10$^{3}$\\
                           &\\
\hline
\rule{0pt}{4ex}     & 1.16$\times$10$^{4}$ & 50& 10$^{2}$&\\
                          Saturn  &3$\times$10$^{4}$& 30&$2\times$10$^{2}$& \cite{Yaroshenko2011}\\ 
			  Orbit &7$\times$10$^{4}$ &10&5.6$\times$10$^{2}$ & (electron, water group ions \\
                           Insertion&1.7$\times$10$^{5}$&2&1.9$\times$10$^{3}$ &and protons plasma )\\
                            &4$\times$10$^{5}$&0.1&5$\times$10$^{3}$\\
                            &\\
\hline
\rule{0pt}{4ex}     Laboratory & & & &\cite{Kremer2006}  \\
                            Experiments      & 46418    &7.5$\times$10$^{6}$  & 1.7 &(pure electrons plasma)\\                           
      [1ex] \hline
\end{tabular}
\label{tab2}
\end{table}

  Section \ref{ss:Nd} provides supplementary material about N$_{\rm
    D}$, the average number of charges in the Debye sphere. The values for  N$_{\rm
    D}$ are $\gg10^5$ in the rarefied upper part of the atmospheres ($p_{\rm gas}<10^{-4}$ bar) for all M-dwarf, brown dwarf and giant gas planet model atmospheres investigated here. Values
  for the above quoted Debye length for the Earth ionosphere and solar
  wind are N$_{\rm D}^{\rm ionosphere}=10^5$ and N$_{\rm D}^{\rm sol wind}=10^{10}$.      The parameter N$_{\rm D}$ indicates that the gas is dominated by many long-range interactions of charged particles, rather than the short-range binary interactions. 
  If any charged particle is close enough to
  others then, it interacts in a collective behaviour, not only with
  the closest one. This collective behaviour distinguishes 
   a plasma from a kinetic gas. Collective behaviour occurs if $N_{\rm D}\gg
  1$. Figure~\ref{fig:Nd} shows that $N_{\rm D}\gg 1$ is satisfied
  throughout the whole atmosphere for model atmosphere structures
  presented in Table~\ref{tab}. A substantially higher N$_{\rm D}$ is
  required for cooler atmospheres than in atmospheres for higher
  T$_{\rm eff}$ if thermal ionisation is considered only (i.e.
  $T_{\rm e}= T_{\rm gas, LTE}$).

\begin{figure*}
\centering
\hspace*{-1.cm}\includegraphics[width=1\textwidth]{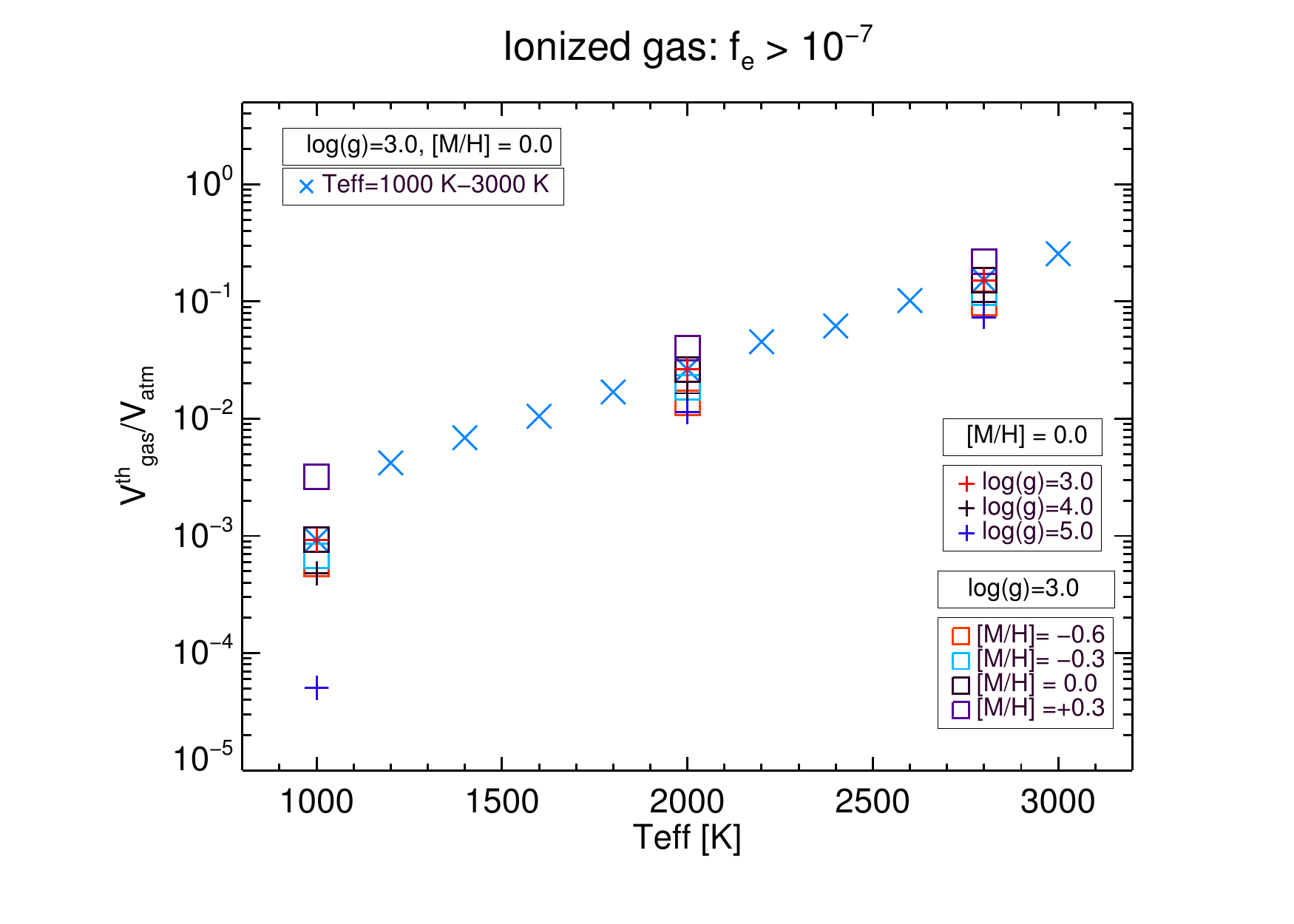}
\vspace{-1cm}\caption{The fraction of the thermally ionized atmospheres volume, $V^{\rm th}_{\rm gas}/V_{\rm atm}$, for $f_{\rm e}>10^{-7}$ and for M-dwarf, brown dwarf and giant gas planet atmospheres.}
\label{fig:vfe}
\end{figure*}

\begin{figure*}
\centering
\hspace*{-1.cm}\includegraphics[width=1\textwidth]{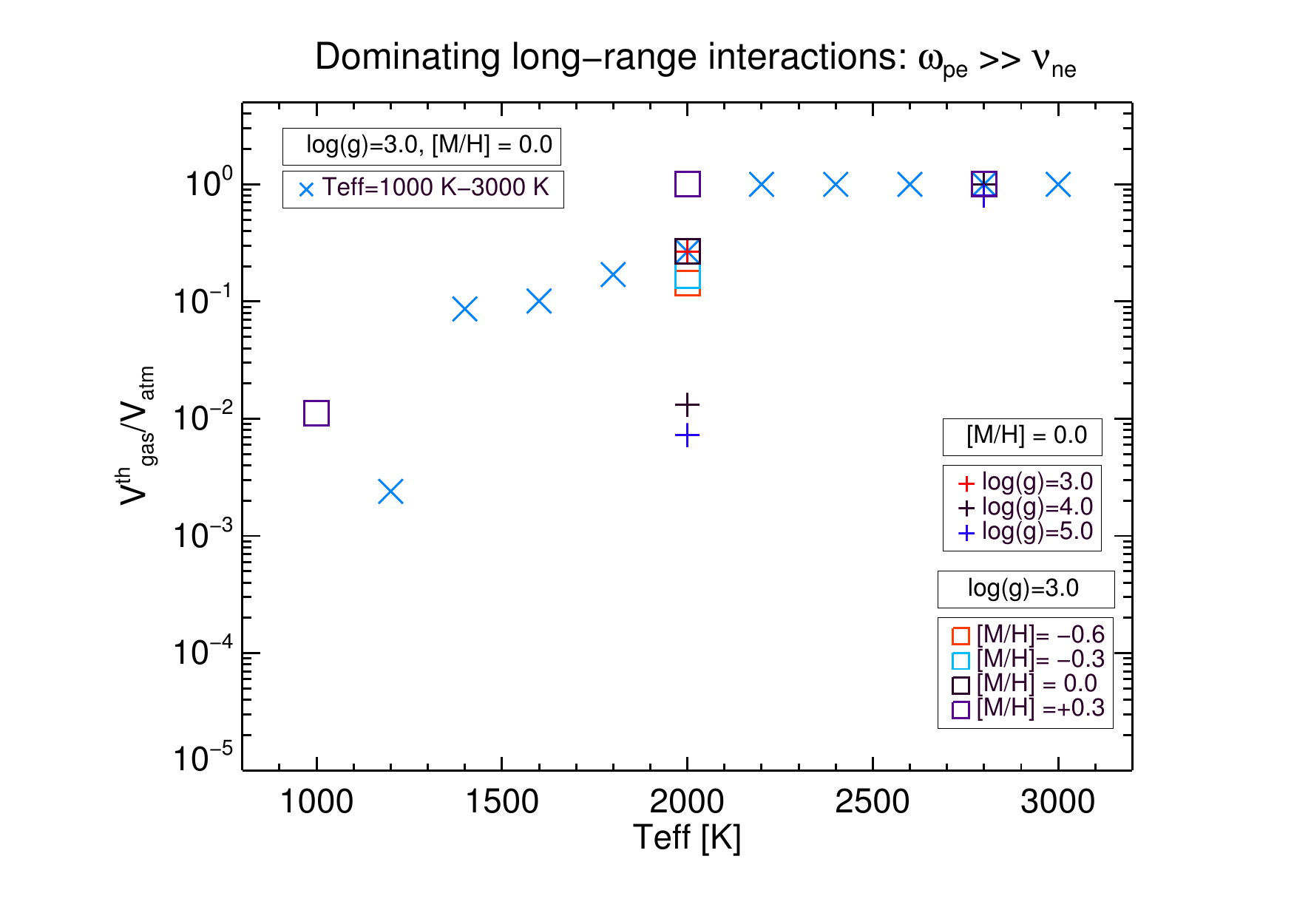}
\vspace{-1cm}\caption{The fraction of the thermally ionized atmospheres volume, $V^{\rm th}_{\rm gas}/V_{\rm atm}$ for $\omega_{\rm pe}\gg\nu_{\rm ne}$ for M-dwarf, brown dwarf and giant gas planet atmospheres.
}
\label{fig:vwpe}
\end{figure*}

 \subsection{Comparing different plasma criteria}\label{s:comp}
 Radio and X-ray observations from low-mass objects suggest that their
 atmospheres contain enough free charges to constitute a magnetized
 plasma (\citealt{Hallinan2008}). From our evaluation of the thermal
 degree of ionisation (Sect.~\ref{sss:fe}), we chose a threshold for the
 degree of ionisation of $f_{\rm e}>10^{-7}$ above which an
 atmospheric gas can be sufficiently ionised that it may exhibit
 plasma behaviour. In a plasma,    electron-electron
 interactions dominate over collisions between electrons and neutral
 particles, $\omega_{\rm pe}\gg\nu_{\rm ne}$ (Sect.~\ref{sss:fp}). 
 Additionally, for a plasma to be considered magnetized, the magnetic
 field must be sufficiently strong that it significantly influences
 the electron and ion dynamics, which we address in Sect.~\ref{s:magnetic}.

 We now cast the results in terms of atmospheric volumes to allow
 a comparison between  the results for different parameters.  Figure~\ref{fig:vfe} and
 ~\ref{fig:vwpe} summarise our findings in terms of the volume
 fraction, $V^{\rm th}_{\rm gas}/V_{\rm atm}$, with $V^{\rm th}_{\rm
   gas}$ the thermally ionised volume of the atmosphere and $V_{\rm
   atm}$ the total atmospheric volume.  $V^{\rm th}_{\rm gas}$ is
 derived by calculating the fraction of the atmospheric volume for
 which $f_{\rm e}>10^{-7}$ (Fig.~\ref{fig:vfe}).
 Figure~\ref{fig:vwpe} visualises the atmospheric volume fraction
 where $\omega_{\rm pe}\gg\nu_{\rm ne}$ is fulfilled. The atmosphere volume that reached $f_{\rm
   e}>10^{-7}$ and satisfied $\omega_{\rm pe}\gg\nu_{\rm ne}$ is
 affected by the global parameters as follows:
     \begin{description}
   \item - If  T$_{\rm eff}$ increases, then the thermally ionised atmospheric volume fraction increases for a given $\log(g)$ and [M/H]. 
   \item - The $V^{\rm th}_{\rm gas}/V_{\rm atm}$, that reaches $f_{\rm e}>10^{-7}$ and $\omega_{\rm pe}\gg\nu_{\rm ne}$, increases if $\log(g)$ decreases for a given T$_{\rm eff}$ and [M/H].
      \item - Higher values of the metallicity, [M/H], cause a larger
        fraction of the atmosphere volume to have a sufficiently
        ionised gas that large-scale electromagnetic interactions
        dominate over electron-neutral collisions for a given T$_{\rm eff}$ and $\log(g)$.
  \end{description}
   The late M-dwarfs have the largest atmosphere volume fraction, $V^{\rm
       th}_{\rm gas}/V_{\rm atm}$, that reached $f_{\rm
       e}>10^{-7}$ that is for model atmosphere structures with
     T$_{\rm eff}= 2600-3000~$K, $\log(g)$=3.0, [M/H]= 0.0 and T$_{\rm
       eff}= 2800~$K , $\log(g)$=3.0, [M/H]=+0.3. For late M-dwarfs
     and brown dwarfs, the atmospheric gas satisfies $f_{\rm
       e}>10^{-7}$ only for half of their atmosphere. For planetary
     objects the fraction of the volume that reaches $f_{\rm
       e}>10^{-7}$ becomes increasingly small except for those
     that have the highest value of metallicity and the lowest value
     of surface gravity.
 
 Models with 2200\,K$\le$ T$_{\rm eff}\le 3000\,K$, $\log(g)$=3.0, [M/H]= 0.0 have the largest
 atmosphere volume fraction that reached $\omega_{\rm pe}\gg\nu_{\rm
   ne}$ for a given $\log(g)$ and [M/H]. Models with T$_{\rm
   eff}=$2800\, K, $\log(g)$=4.0, [M/H]= 0.0; T$_{\rm
   eff}=$2800\, K, $\log(g)$=3.0, [M/H]= -0.6, -0.3, 0.0, +0.3 and
 T$_{\rm eff}= 2000~$K, $\log(g)$=3.0, [M/H]= +0.3 have the largest
 atmosphere volume fraction that reached $\omega_{\rm pe}\gg\nu_{\rm
   ne}$ as well.   In young brown dwarfs, the atmospheric gas volume that reaches
   $\omega_{\rm pe}\gg\nu_{\rm ne}$ is more than 50\%.  The atmospheric gas volume that reaches
   $\omega_{\rm pe}\gg\nu_{\rm ne}$ for planetary objects is smaller than
   for the rest of the objects, i.e. for T$_{\rm eff}= 1000~$K, $\log(g)$=3.0,
   [M/H]= +0.3. Our results show that $V^{\rm th}_{\rm gas}/V_{\rm atm}$ ($\omega_{\rm
  pe}\gg\nu_{\rm ne}$) is larger than $V^{\rm th}_{\rm gas}/V_{\rm
  atm}$($f_{\rm e}>10^{-7}$) for 1000\,K$ \le $T$_{\rm
  eff}\le$3000\,K. A general observation is that despite a relatively
low degree of ionisation, large-scale electromagnetic interactions can
dominate a considerably larger atmospheric volume than a $f_{\rm e}$ evaluation would suggest for all ultra-cool objects in our
sample.

\section{Most abundant thermal ions in late M-dwarf, brown dwarf and giant gas planet atmospheres}\label{ions}

We investigate the atmospheric gas-phase composition regarding the
most abundant local gas ions to demonstrate which are the dominating
electron donors across the star-planet regime based on our
non-irradiated {\sc Drift-Phoenix} model atmosphere grid. This
investigation allows us to understand which gas-phase species are
responsible for increasing the number of free electrons in ultra-cool
atmospheres and consequently, responsible of satisfying the plasma
parameters given in Eq.\,\ref{eq:fe}-\ref{eq:lD}. It also allows us to
understand how the chemical composition of the gas is linked with the
dominating electron donors in the gas-phase. In the case of
cloud-forming atmospheres, the abundance of the element depleted by
cloud formation has an effect on electron donors at the location
  of the cloud.  This section serve also as reference for future
investigations on ionisation processes and their effect on the gas
composition like in \citep{Rimmer2013}.

\subsection{Dominating ions across the late M-dwarf, brown dwarf and planetary regime}\label{ss:chem1}
Figure~\ref{chem1} demonstrates for a subset of effective temperatures
T$_{\rm eff}$= 1000 ... 3000 K (log(g)=3,0 solar element abundances)
that K$^{+}$, Na$^{+}$, Ca$^{+}$, Mg$^{+}$ and Fe$^{+}$ are the
dominating thermal electron donors. These ionic species are
the most significant contributors to the electron number density from
thermal ionisation.  Species that have sufficiently low first
ionization potentials and sufficiently high atmospheric number
densities will contribute most effectively to the thermal degree of
ionization.  Therefore, K$^{+}$, Na$^{+}$ and Ca$^{+}$ provide the
majority of thermal electrons. Figure~\ref{chem1} demonstrates that
K$^{+}$ is the dominating thermal electron donor where p$_{\rm gas}<
10^{-2}$ bar for all atmospheres except in an M-dwarf atmosphere of
T$_{\rm eff}$= 3000 K. The second dominating electron donor is
Na$^{+}$ from T$_{\rm eff}$= 2000 K. Na$^{+}$ dominates for T$_{\rm
  eff}$= 3000 K. For increasing gas pressure, $p_{\rm gas}> 10^{-2}$
bar, Na$^{+}$ and K$^{+}$ are the dominating thermal electron donor
for T$_{\rm eff}$= 2000\,K, 1400\,K, 1000\,K. For example, Mg$^{+}$
provides most of the electrons in the T$_{\rm eff}$= 3000 K model for
$p_{\rm gas}>10^{-2}$ bar. The detailed results for all model groups
are summarise in Tables~\ref{tab:1} and \ref{tab:2} in the
Appendix~\ref{chem}.

Figures~\ref{fig:chemcomp_el1} and ~\ref{fig:chemcomp_el2} show the
distribution of the most important electron donating elements over
atoms, molecules and ions. For example, Na$^{+}$ is the dominating
Na-species at high temperature, but the atomic Na followed by NaH and
NaOH contain most of the element Na at lower temperatures.  Ca, Mg and
Fe are involved in the formation of many molecules. In the case of Ca,
CaCl$_{2}$, CaOH and Ca(OH)$_{2}$ reach the abundances not much lower
or  even higher than the atomic Ca. Our investigations
show that it is not sufficient to determine the local degree of
ionisation based on one prescribed electron donor species. Such an
approach has been chosen in various complex simulations like MHD
simulations for protoplanetary disks (e.g. \citealt{Sano2000}) or
atmospheric circulation models
(e.g. \citealt{Perna2010}).

 The exact amount with which K, Na, Ca, Mg and Fe contribute to
  the local degree of thermal ionisation will also depend on the
  amount of each element that is chemically locked in cloud particles. Element depletion
  by cloud formation is fully taken into account for Mg and Fe (see
  Sect.~\ref{s:DF}) in the {\sc Drift-Phoenix} atmosphere
  simulations. The more extended work by \cite{Helling2008} (Fig. 6)
  suggests that the effect of cloud formation on the Ca$^+$ abundance is
  negligible. \cite{Morley2012} (Fig. 3) show that Na$_2$S[s] is
  thermally stable for $T_{\rm gas}<1100$K and KCl[s] for $T_{\rm
    gas}<900$K, hence Na$^+$ and K$^+$ would be effected by cloud formation in a similar
  temperature window. The solar element abundances for Na (6.17), K
  (5.08) and Cl (5.5) are lower than for Ca (6.31) and it may
  therefore be reasonable to expect a similarly negligible effect of
  cloud formation on the abundances of Na$^+$ and K$^+$.

\subsection{Summary on electrostatic parameters}
Our reference study suggests that:
\begin{description}
\item - M-dwarfs have almost the entire atmospheric gas ionised, brown
  dwarfs present only an ionised gas for $p_{\rm gas}>10^{-4}$ bar and
  giant gas planets present only a small fraction of an ionised gas
  for values of $p_{\rm gas}>10^{-1}$ bar.
\item - Collective, long-range electromagnetic interactions of electrons dominate over short-range binary interactions with neutrals particles increase as T$_{\rm
    eff}$ increases  in the atmospheres of M-dwarfs and brown
  dwarfs. Giant gas planets are to cool to fulfill this criterion for
  large ranges of $p_{\rm gas}$; only for T$_{\rm eff}$= 1200\,K,
  log(g)=3,0, [M/H]=0.0 and T$_{\rm eff}$= 1000\,K, log(g)=3,0, [M/H]=
  +0.3 a small fraction of the atmospheric gas at most deeper parts of
  their atmosphere.
\item - $\lambda_{\rm D}\ll L$ is fulfilled for M-dwarfs throughout
  their atmospheres. For brown dwarfs only for $p_{\rm gas}>10^{-8}$
  bar and for giant gas planets for $p_{\rm gas}>10^{-3}$ bar.
 \end{description}
K$^{+}$, Na$^{+}$, Ca$^{+}$, Mg$^{+}$ and Fe$^{+}$ are the dominating
thermal electron donors, however, K$^{+}$, Na$^{+}$ and Ca$^{+}$ provide the
majority of electrons for T$_{\rm eff}$= 1000...3000 K for log(g)=3,0
and solar element abundances. In particular, the degree of ionisation
is low in the upper atmospheres where the abundances of those
ions are low. As their abundances increase, $f_{\rm e}$ increases as
well. Long-range electromagnetic interactions dominating over
collisions (Eq.\,\ref{eq:wpe}) and a zero electrostatic
forces inside the plasma (Eq.\,\ref{eq:lD}) require a sufficient
number of free charged particles. Any process (cloud formation, CR
impact) that impacts the element abundance of the dominating electron
donors will affect to the electric state of the atmosphere, and the
potential coupling to a large-scale magnetic field.
 
 \section{Magnetized plasma parameters across the star-planet regime}\label{s:magnetic}
  
In the previous sections we discussed that for a plasma to exists, the gas needs to be ionised to a certain  degree (f$_{\rm e}>10^{-7}$).  The plasma frequency was used to investigate where in the atmosphere, electromagnetic interactions dominate over kinetic collisions between
electrons and neutrals.  

The Debye length provides insight about the length scales on which 
electrostatic interactions influencing the atmospheric gas.  We
demonstrated that the atmospheric volume fraction, $V^{\rm th}_{\rm
  gas}/V_{\rm atm}$, suspected to show plasma behaviour, varies
largely through the M-dwarf to planetary regime. 

A plasma is considered magnetised when the motion and dynamics of the
charged particles are influenced by an ambient magnetic field.  This
requires that the magnetic field is of sufficient magnitude that the
charged particles can on average participate in at least one Larmor
orbit before colliding with a neutral atom or dust particle.
Otherwise, frequent collisions with the ambient neutrals will dominate
the dynamical evolution of the plasma particles and the influence of
the magnetic field will be negligible. In some instances, because of
the differing mass between electrons and ions, the electrons can be
magnetised while the ions are not. For magneto-fluid descriptions of
plasmas (such as magnetohydrodynamics) both the electrons and ions
need to be magnetised.  Radio flares (\citealt{Route2012}), X-ray
flares (\citealt{Berger2010}) and quiescent radio emission
(\citealt{Williams2013}) have been observed in brown dwarfs,
\cite{Schmidt2015} conclude that 45\% of their active L-dwarfs are
also variable. This fraction of L-dwarfs, which is lower compared to the 60\% of
active M-dwarfs that were found to be variable in their sample, lead
\cite{Schmidt2015} to speculate about a brown dwarf chromosphere as
origin for the observed H$\alpha$ emission and variability.  These
observations suggest that there should be a strong magnetic field and
considerable coupling between the magnetic field and the atmospheric
gas, either to directly accelerate free electrons or to allow plasma
waves to travel into the low-density upper atmosphere and deposit
their energy causing a chromosphere to develop.  It is interesting to
note here that old brown dwarfs (log(g)=5.0) have enough time
available to build up a chromosphere even if the acoustic heating
rates should be low due to a insufficient magnetic coupling of the atmosphere. For
young brown dwarfs, rapid rotation may favour chromospheric heating by
a better coupling to a stronger magnetic field.
 
\begin{figure}
\centering
\hspace*{-0.9cm}\includegraphics[angle=0,width=0.6\textwidth]{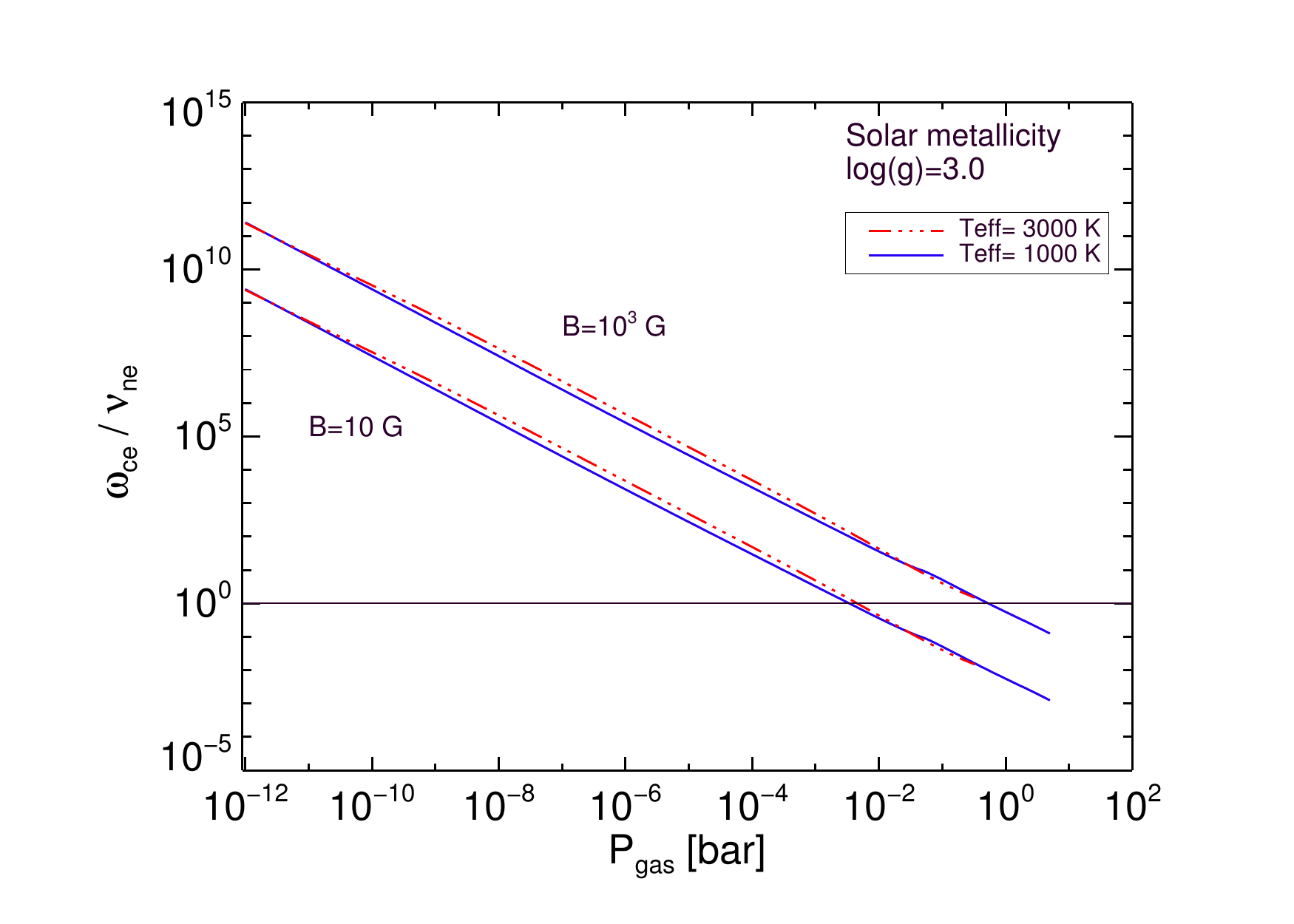}\\*[-0.8cm]
\hspace*{-0.9cm}\includegraphics[angle=0,width=0.6\textwidth]{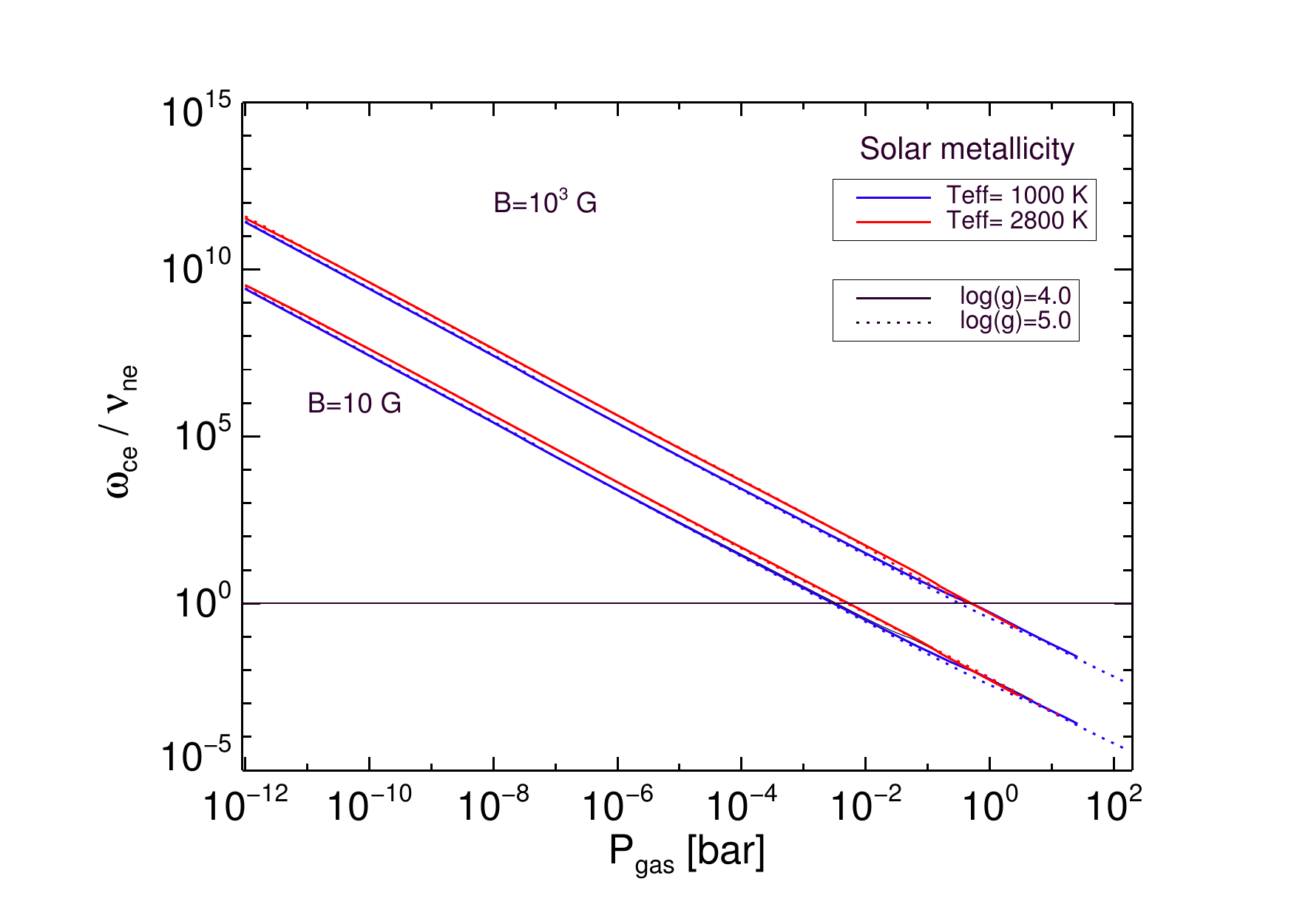}\\*[-0.8cm]
\hspace*{-0.9cm}\includegraphics[angle=0,width=0.6\textwidth]{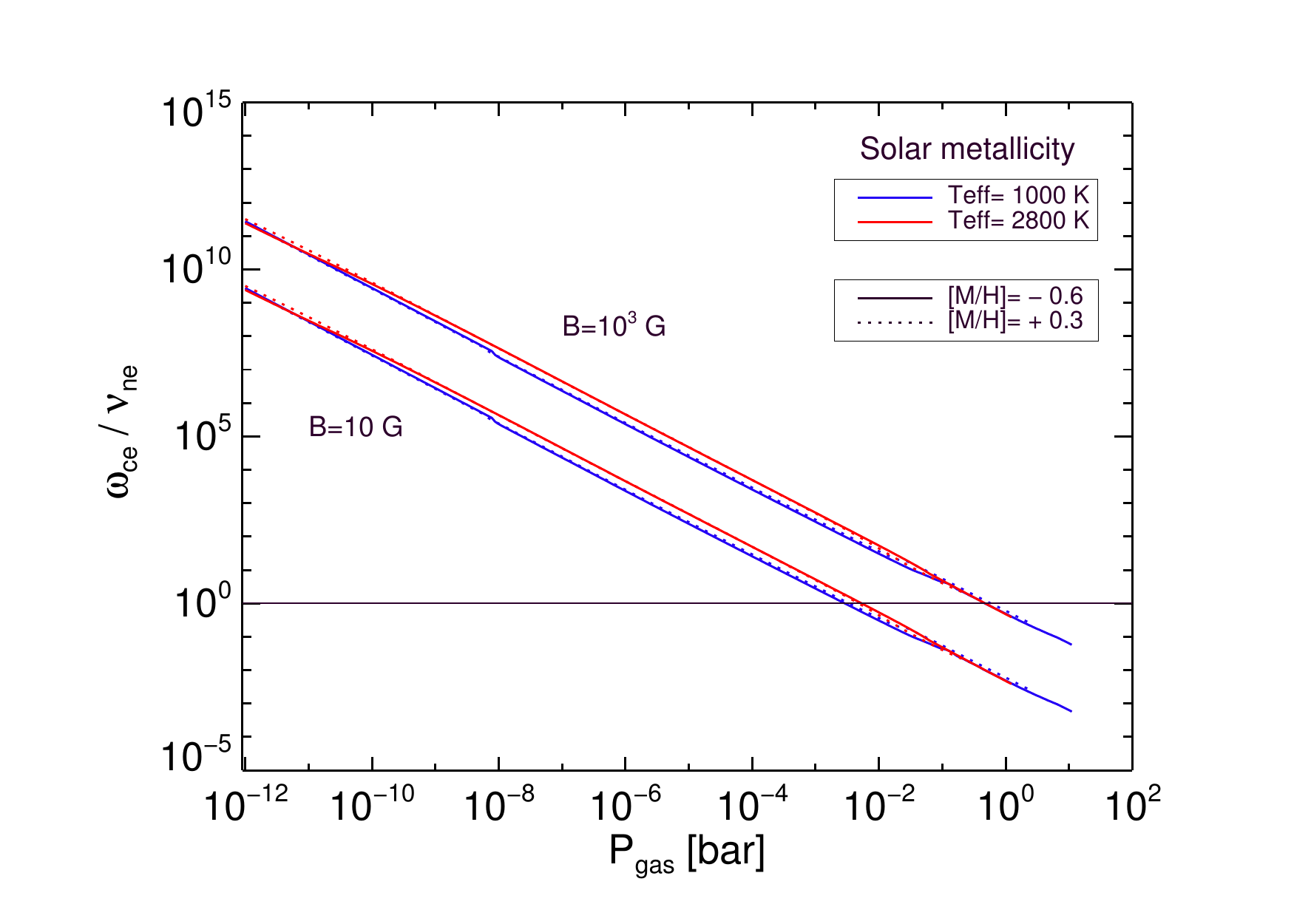}\\*[0.05cm]
\caption{Ratio of cyclotron frequency of the electrons and the frequency of collisions between neutral particles and electrons. 
{\bf Top:} Group 1.
{\bf Middle:} Group 2.
 {\bf Bottom:} Group 3. }
 \label{fig:fc}
\end{figure}

\subsection {Cyclotron frequency versus collisional frequency $\bm \omega_{c}\gg\nu_{\rm coll}$}\label{ss:fc}

The cyclotron frequency is the angular velocity with which charged
particles gyrate around the magnetic field line \citep{Boyd2003},
$\omega_{\rm c,s}=v_{\perp,\rm s}/r_{\rm L,s}=q_{\rm s}B/m_{\rm s}$ in
$[\rm rad\,s^{-1}]$, where $m_{\rm s}$ $[\rm kg]$, $q_{\rm s}$ $[\rm
  C]$, B [T] and $v_{\perp,\rm s}$ $[\rm m\,s^{-1}]$ are the mass of
species $s$, with charge $q_{\rm s}$ and speed perpendicular to the
magnetic field $v_{\perp,\rm s}$ respectively; and $|\overrightarrow {
  B }|= B$ is the magnitude of the external magnetic
flux density present in the medium.  For a charged particle's motion
to be dictated by a magnetic field, the particle needs to complete on
average one gyration before a collision with a neutral atom.
Formally, a magnetised plasma requires
\begin{equation}\label{eq:fc}
\omega_{\rm c,s}\gg\nu_{\rm ns},
\end{equation}
where $\nu_{\rm ns}$ $[\rm s^{-1}]$ is the collision frequency for
neutral particles with charged species $s$.  From Eq.~\ref{eq:fc} we
obtain the minimum value for the external magnetic field flux,
$|\overrightarrow { B }|= B$, that is needed to satisfied this
criteria.  Applying the definitions for $\omega_{\rm c,s}$ and
$\nu_{\rm ns}$, we derive the critical magnetic flux density that is
required for the dynamics of the charged particle to be influenced by
the background magnetic field:
 \begin{eqnarray}
 \frac {eB}{m_{\rm s }}&\gg& \sigma _{\rm gas,e }n_{\rm  gas}v_{\rm s }, \\
\Rightarrow B_{\rm s}&\gg&\frac {m_{\rm s}}{e} \sigma_{\rm gas,e }n_{\rm gas }{ \left( \frac {k_{\rm B}T_{\rm s}}{m_{\rm s }} \right)  }^{ 1/2 },\label{eq:B}
 \end{eqnarray}
 where the collision, or scattering, cross section is
  $\sigma_{\rm gas}= \pi\cdot r_{\rm gas}^2$. The
 atmospheric gas in late M-dwarfs, brown dwarfs and most likely also in
 giant gas planets is composed mostly of molecular hydrogen, H$_2$. The collision cross section is approximated as $\sigma_{\rm gas}\approx\sigma_{\rm H_{2}}\approx\pi\cdot {r_{\rm
     H_{2}}}^{2}= 5.81\cdot10^{-20}$ m$^{2}$ (r$_{\rm
   H_{2}}=1.36\cdot10^{-10}$ m).
 
\begin{figure}
\centering
\hspace*{-0.5cm}{\includegraphics[angle=0,width=0.55\textwidth]{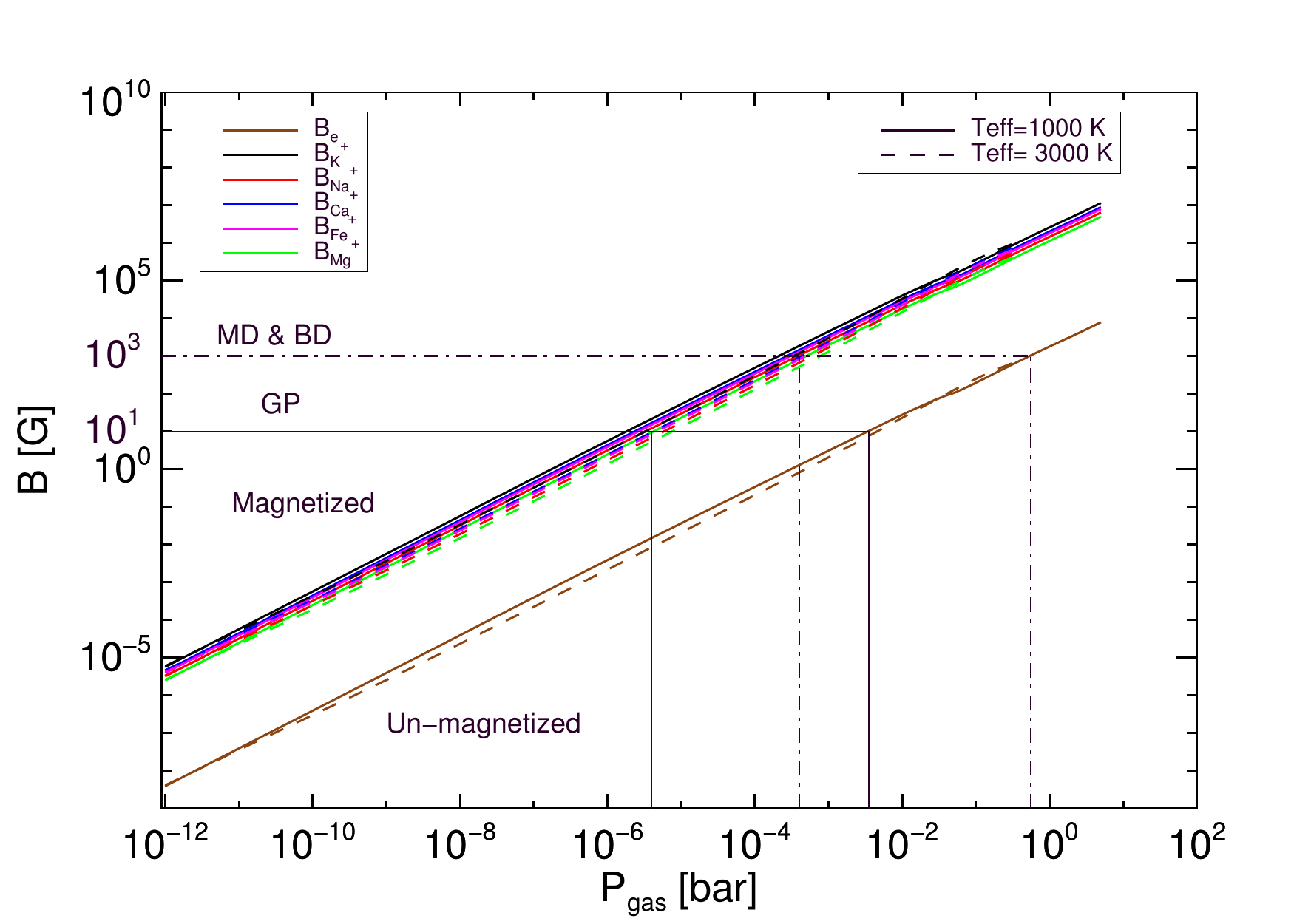}}
\caption{Magnetic flux density required for
  electrons, B$_{\rm e}$ (lower set of lines), and ions, B$_{\rm i}$ (upper set of lines), to be magnetically
  coupled to a background magnetic field in the object (B$=10$G --
  giant gas planets (GP), B$=10^{3}$ G -- M-dwarfs (MD), brown dwarfs
  (BD); black horizontal/vertical lines).  If $B>B_{\rm i}$ (or
  $B>B_{\rm e}$), $\omega_{\rm c,s}\gg\nu_{\rm ns}$ is fulfiled and
  the gas is magnetised by the external magnetic field B. }\label{fig:Be}
\end{figure}

Taking the electrons and ions as the particles that are influenced by an external magnetic flux density, Eq.~\ref{eq:B}  becomes
\begin{eqnarray}
 B_{\rm e}&\gg&\frac {m_{\rm e}}{e} \sigma_{\rm gas }n_{\rm gas }{ \left( \frac {k_{\rm B}T_{\rm e}}{m_{\rm e }} \right)  }^{ 1/2 },\label{eq:Be}\\
 B_{\rm i}&\gg&\frac {m_{\rm i}}{e} \sigma_{\rm gas }n_{\rm gas }{ \left( \frac {k_{\rm B}T_{\rm i}}{m_{\rm i }} \right)  }^{ 1/2 }.\label{eq:Bi}
 \end{eqnarray}
Grouping the constants, we rewrite Eqs.~\ref{eq:Be} and~\ref{eq:Bi} as $B_{\rm e}\propto n_{\rm gas}(m_{\rm e}T_{\rm e})^{ 1/2 }$, with $B_{\rm e}$ as the minimum threshold for the magnetic flux density to ensure that the electrons are magnetised and $B_{\rm i}\propto n_{\rm gas}(m_{\rm i}T_{\rm i})^{ 1/2 }$, with $B_{\rm i}$ as the minimum threshold for magnetic flux density
required to ensure that an ion, $i$, is magnetised in Fig.~\ref{fig:Be}. The ion masses,
$m_{\rm i}$, are taken to be for K$^{+}$, Na$^{+}$ Ca$^{+}$, Fe$^{+}$, and Mg$^{+}$ according
to Sect.~\ref{ions} assuming local thermal equilibrium, T$_{\rm
  gas}\approx$T$_{\rm i}\approx$T$_{\rm e}$.

Figure~\ref{fig:fc} shows that for a $p_{\rm gas}<10^{0}$ bar and
B$=10^{3}$ G and for a $p_{\rm gas}<10^{-2}$ bar and B$=10$ G ,
$\omega_{\rm ce}\gg\nu_{\rm ne}$ (horizontal black line) is reached
for all model atmosphere structures. There is almost no dependence on
T$_{\rm eff}$, $\log(g)$ and the metallicity.  Largest values of
$\omega_{\rm ce}/\nu_{\rm ne}$ are reached for B$=10^{3}$ G
representative for M-dwarfs or brown dwarfs.  Figure~\ref{fig:Be}
shows for which atmospheric gas pressures, $p_{\rm gas}$, electrons
and ions can be magnetised in the atmospheres of M-dwarfs, brown
dwarfs and giant gas planets. For M-dwarfs and brown dwarfs an
background magnetic field flux density of B$=10^{3}$ G is large enough
to magnetise the charged particles: for electrons at $p_{\rm gas}<1$
bar and for ions at $p_{\rm gas}<10^{-3}$ bar. For giant gas planets,
the magnetised part of the atmosphere decreases because of a smaller
background field (B$\le10$ G) compared to M-dwarfs and brown dwarfs.
For electrons this occurs at $p_{\rm gas}<10^{-2}$ bar and for ions at
$p_{\rm gas}< 10^{-5}$ bar. Figure~\ref{fig:vwce} summarises the
results on magnetic coupling in term of the affected atmospheric gas
volume, $V^{\rm th}_{\rm gas}/V_{\rm atm}$, that reach $\omega_{\rm
  c,e}\gg\nu_{\rm ne}$:
     \begin{description}
   \item - If  T$_{\rm eff}$ increases, the magnetically coupled volume of an atmosphere increases for a given $\log(g)$ and [M/H]. 
   \item -  If $\log(g)$ decreases, then the magnetically coupled volume  increases  for a given T$_{\rm eff}$ and [M/H]. 
     \item - Higher values of the metallicity, [M/H], cause an increase of $V^{\rm th}_{\rm gas}/V_{\rm atm}$ for a given T$_{\rm eff}$ and $\log(g)$.  \end{description}
 For a fixed value of magnetic flux density, M-dwarfs and brown dwarf atmospheres have the largest  magnetically coupled volume. Unsurprisingly, a smaller fraction of a giant gas planets atmosphere is magnetically coupled when thermal ionisation is considered as the only source of gas ionisation. However, this fraction can reach 80\% also in a planetary atmosphere.
The fraction of the atmospheric gas volume, $V^{\rm th}_{\rm gas}/V_{\rm atm}$, that reaches $f_{\rm e}>10^{-7}$ (Fig.\,\ref{fig:vfe}) and $\omega_{\rm pe}\gg\nu_{\rm ne}$ (Fig.\,\ref{fig:vwpe}) increases for the same set of global parameters T$_{\rm eff}$, $\log(g)$, [M/H] like the atmospheric gas volume that reaches  $\omega_{\rm ce}\gg\nu_{\rm ne}$. Figure\,\ref{fig:vwce} demonstrates also that a larger atmospheric gas volume can be expected to be magnetically coupling than the thermal degree of ionisation had initially suggested in Fig.~\ref{fig:vfe}.
This finding is  particularly relevant with respect to the effect that the  magnetic field geometry might have on the detection of the H$\alpha$ activity signatures and on the radio emission: \cite{Donati2008} show that partially-convective  M dwarfs host non-axisymmetric large-scale magnetic fields with a strong toroidal component, while fully convective M dwarfs have stronger large-scale field dominated
by a mainly axisymmetric poloidal component. A partial ionisation of the magnetically coupled gas does influence the magnetic flux density.  


\subsection {\bf Magnetic Reynolds Number,  R$_{m}$}\label{ss:Rm}

The previous sections outlined the framework quantifying when an ionised gas in a substellar atmosphere behaves like a magnetized plasma.  The magnetic Reynolds number is an easy to utilise measure of a potentially magnetically coupled ionised gas.  Within the context of MHD, the magnetic Reynolds number ($R_{m}$) is the ratio of the convective and diffusive terms from the magnetic field induction equation. It quantifies whether the MHD plasma is in the ideal or resistive regimes.  When the magnetic Reynolds number is very large (i.e. in the limit of large length scales), the MHD plasma is in the ideal MHD regime and the convective term has the dominant influence. In this regime the motion of the plasma fluid is determined by the magnetic field and vice verse. In the resistive MHD regime, the diffusive term is important and dissipative processes as Ohmic dissipation (\citealt{Huang2012, Perna2010}) become significant. The magnetic Reynolds number is defined through the induction equation
\begin{equation}
\frac{\partial\vec{B}}{\partial t}=\nabla\times(\vec{u}\times\vec{B})+\eta\nabla^{2}B
\end{equation}
where $|\overrightarrow { B }|= B$ $[T]$ is the magnetic flux density, $u$ is the flow velocity (formed by electrons and ions), and $\sigma$ [S\,m$^{-1}$] is the electric conductivity. The magnetic diffusivity, $\eta$, is linked to the conductivity by $\eta=1/\sigma$ in $[\rm m^{2}\,s^{-1}]$ and represents the diffusion of the magnetic field, a measure of  the effect of collisions between the electrons and the neutral particles on the magnetic field.  The collisions between the neutral particles and the charged particles (electrons or ions) have an influence on the diffusivity of the magnetic field. If the effect of the collisions is sufficient to displace them away from the magnetic lines, the coupling between the magnetic field and the fluid may be not effective. Therefore, the diffusion of the magnetic field depends on the frequency of the collisions between neutral particles and charged particles.

  The magnetic Reynolds number, R$_{\rm m}$ can be defined as the ratio between the relative strength between the diffusive term and the advective term of the induction equation. It can be used as a measure of the magnetic coupling calling the plasma coupled to the magnetic field if $R_{\rm m}\ge1$, with

\begin{figure}
\centering
\hspace*{-0.5cm}\includegraphics[angle=0,width=0.55\textwidth]{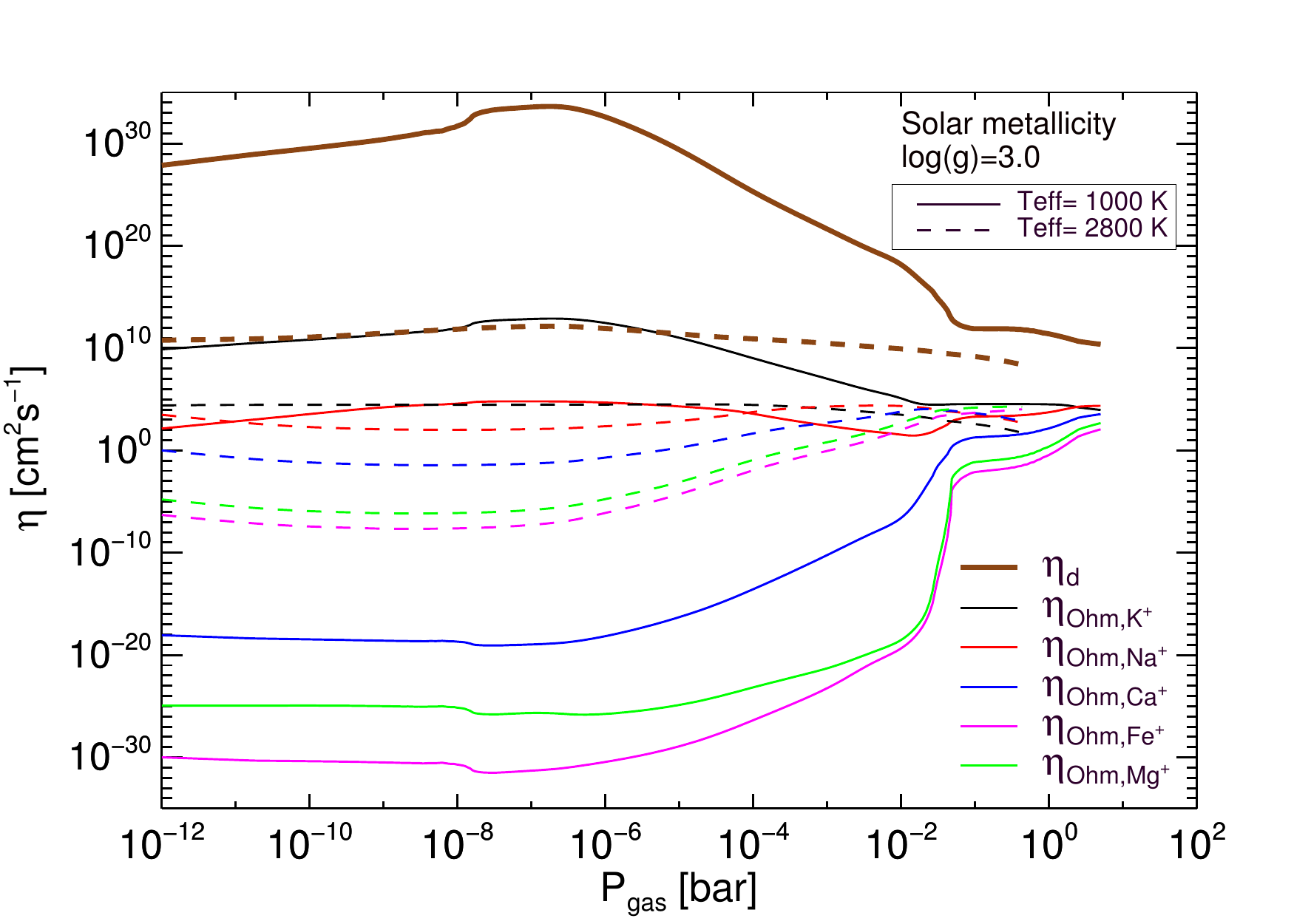}
\caption{Decoupled diffusion coefficient, $\eta_{\rm d}={c^{2}\nu_{\rm
      en}}/{\omega_{\rm pe}^{2}}$ and the Ohmic diffusion coefficient,
  $\eta_{\rm ohm}={c^{2}\nu_{\rm ei}}/{\omega_{\rm pe}^{2}}$ for the
  dominating thermal electron donors for T$_{\rm eff}=1000, 2800\,$K,
  for log(g)=3,0 and solar element abundances; K$^{+}$, Na$^{+}$,
  Ca$^{+}$, Fe$^{+}$ and Mg$^{+}$ (Fig.\,\ref{chem1}). The Ohmic
  diffusion coefficient is smaller than the decoupled diffusion
  coefficient. This result suggest that the binary interactions
  between the ions and electrons ($\eta_{\rm Ohm}$) are not
  significant compared to the binary interactions between electrons
  and neutral particles ($\eta_{\rm d}$) in the case of thermal
  ionisation. {\it The lines in both temperature sets appear from
    top to bottom in the order $\eta_{\rm d}$, $\eta_{\rm Ohm,K^+}$,
    $\eta_{Ohm,Na^+}$,  $\eta_{Ohm,Ca^+}$,  $\eta_{Ohm,Mg^+}$,  $\eta_{Ohm,Fe^+}$.}} \label{fig:eta}
\end{figure}

\begin{center}
\begin{equation}
R_{m}=\frac{\mid\nabla\times(\vec{u}\times\vec{B})\mid}{\eta\mid\nabla^{2}B\mid}\label{eq:Rm}.
\end{equation}
\end{center}
Applying a dimensional analyses, Eq.~\ref{eq:Rm} reduces to
\begin{center}
\begin{equation}\label{eq:Rmf}
R_{m}\approx\frac{vB/L}{\eta B/L^{2}}=\frac{vL}{\eta}\,,
\end{equation}
\end{center}
where L [cm] is a typical length scale of the plasma over which
$|\overrightarrow { B }|= B$ varies through a hydrodynamic motion of a
velocity $|\overrightarrow { v }|$, can be approximated by L=10$^{3}$
[m] (Helling et al. 2011).  The diffusion coefficient, $\eta$, can be
approximated by $\eta\approx\eta_{\rm d}$
(Fig.~\ref{fig:eta}). \footnote{The diffusion coefficient used in
  Eq.\,\ref{eq:Rmf} is given by $\eta=\eta_{\rm d}+\eta_{\rm ohm}$
  being $\eta_{\rm d}={c^{2}\nu_{\rm ne}}/{\omega_{\rm pe}^{2}}$ as
  the decoupled diffusion coefficient and $\eta_{\rm
    ohm}={c^{2}\nu_{\rm ei}}/{\omega_{\rm pe}^{2}}$ as the Ohmic
  diffusion coefficient and $\eta_{\rm d}\gg \eta_{\rm ohm}$. Both
  measure the degree of the dominance of the collisions between
  electrons-neutral particles and electrons-ions respectively over
  long-range electromagnetic collective interactions. It is easy to
  relate $\eta_{\rm d}$ with $\omega_{\rm pe}/\nu_{\rm ne}$ (see
  Fig.\,\ref{fig:fp}). If the latter increase, the diffusion
  coefficient decrease and the magnetic field may be generated and
  transported by fluid motions allowing the magnetic energy to be
  released into upper layers of the atmosphere as radio, X-ray and
  H$_{\alpha}$ emissions. }
 
\begin{figure}
\centering
\hspace*{-0.9cm}\includegraphics[angle=0,width=0.6\textwidth]{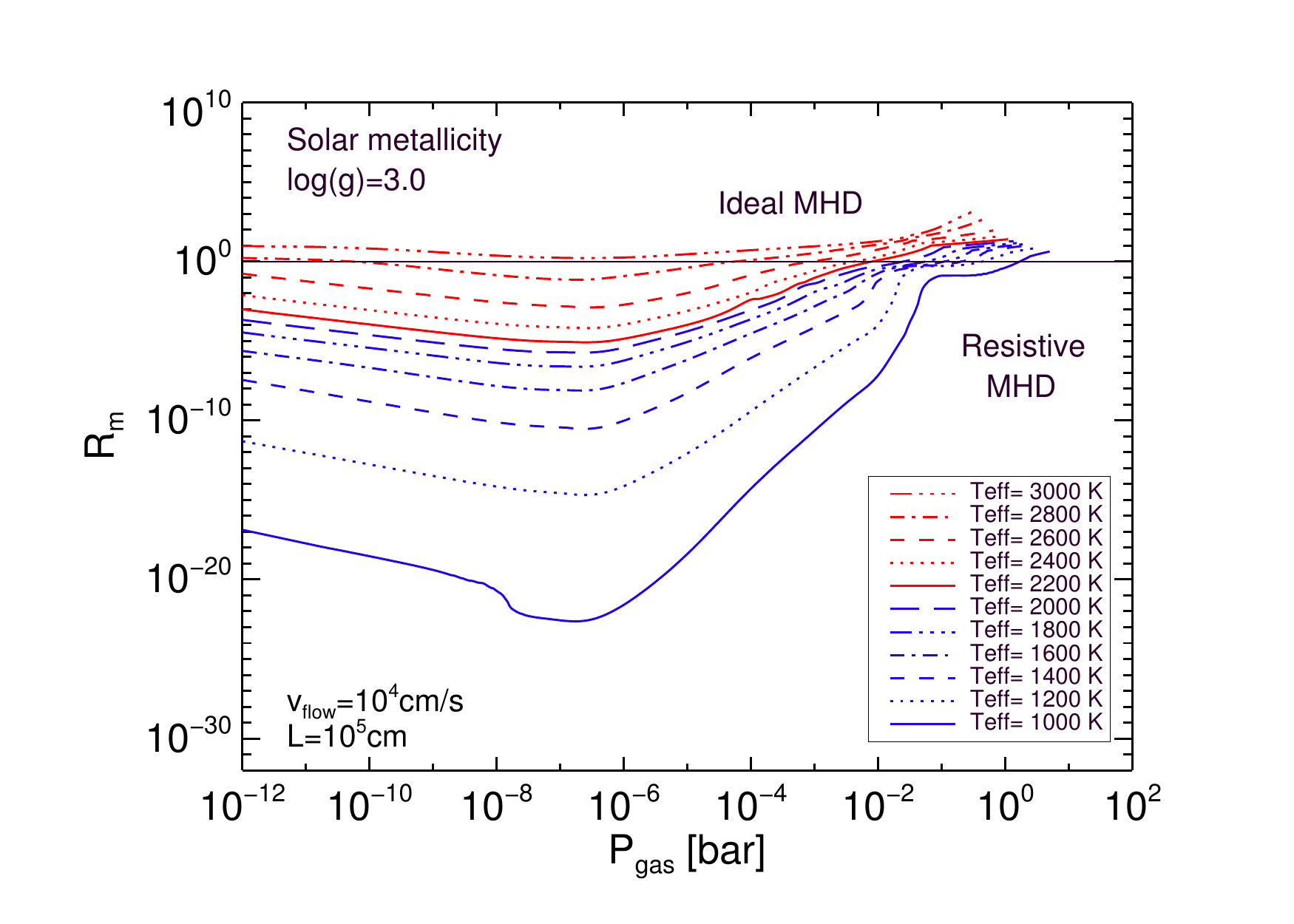}\\*[-0.8cm]
\hspace*{-0.9cm}\includegraphics[angle=0,width=0.6\textwidth]{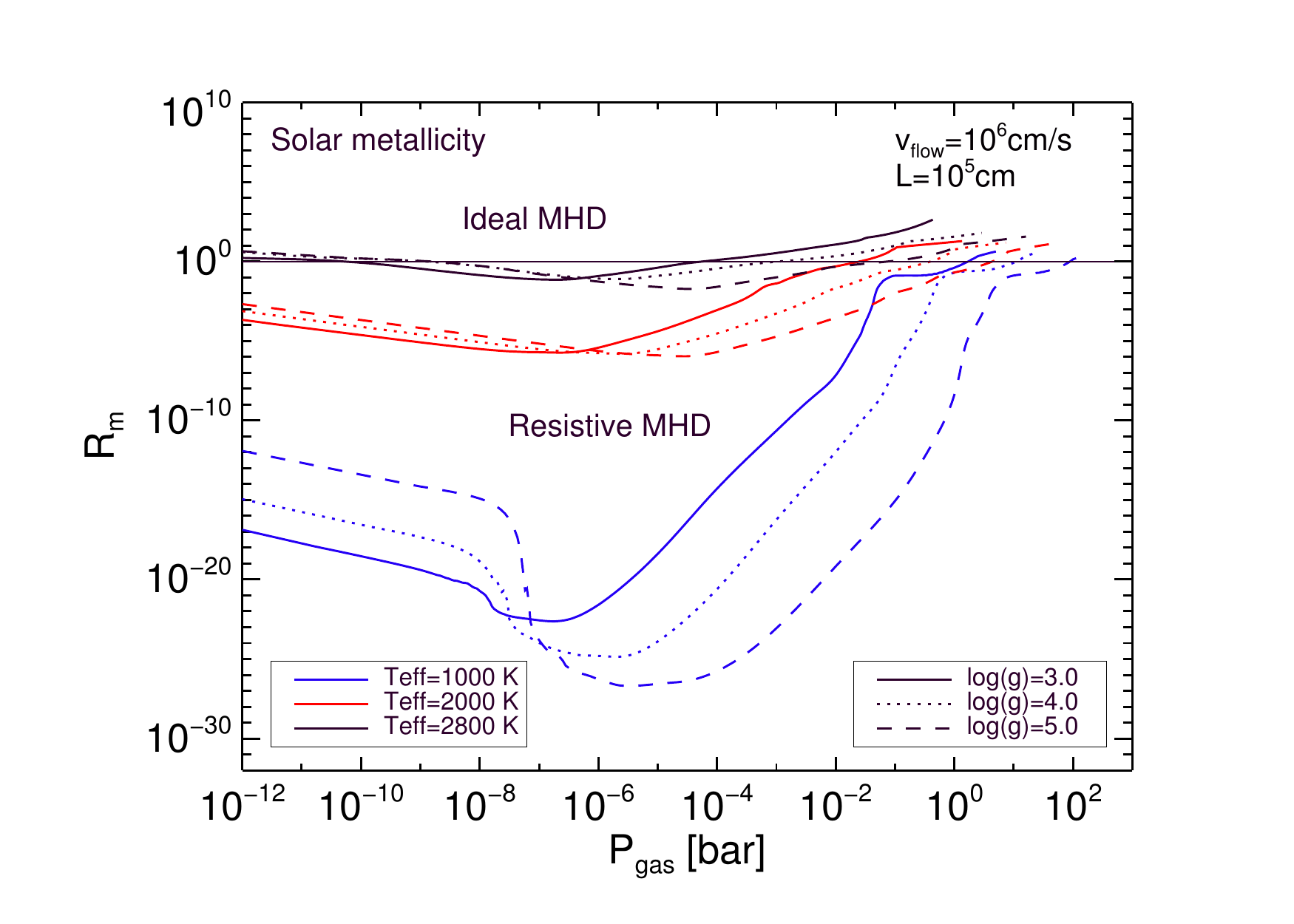}\\*[-0.8cm]
\hspace*{-0.9cm}\includegraphics[angle=0,width=0.6\textwidth]{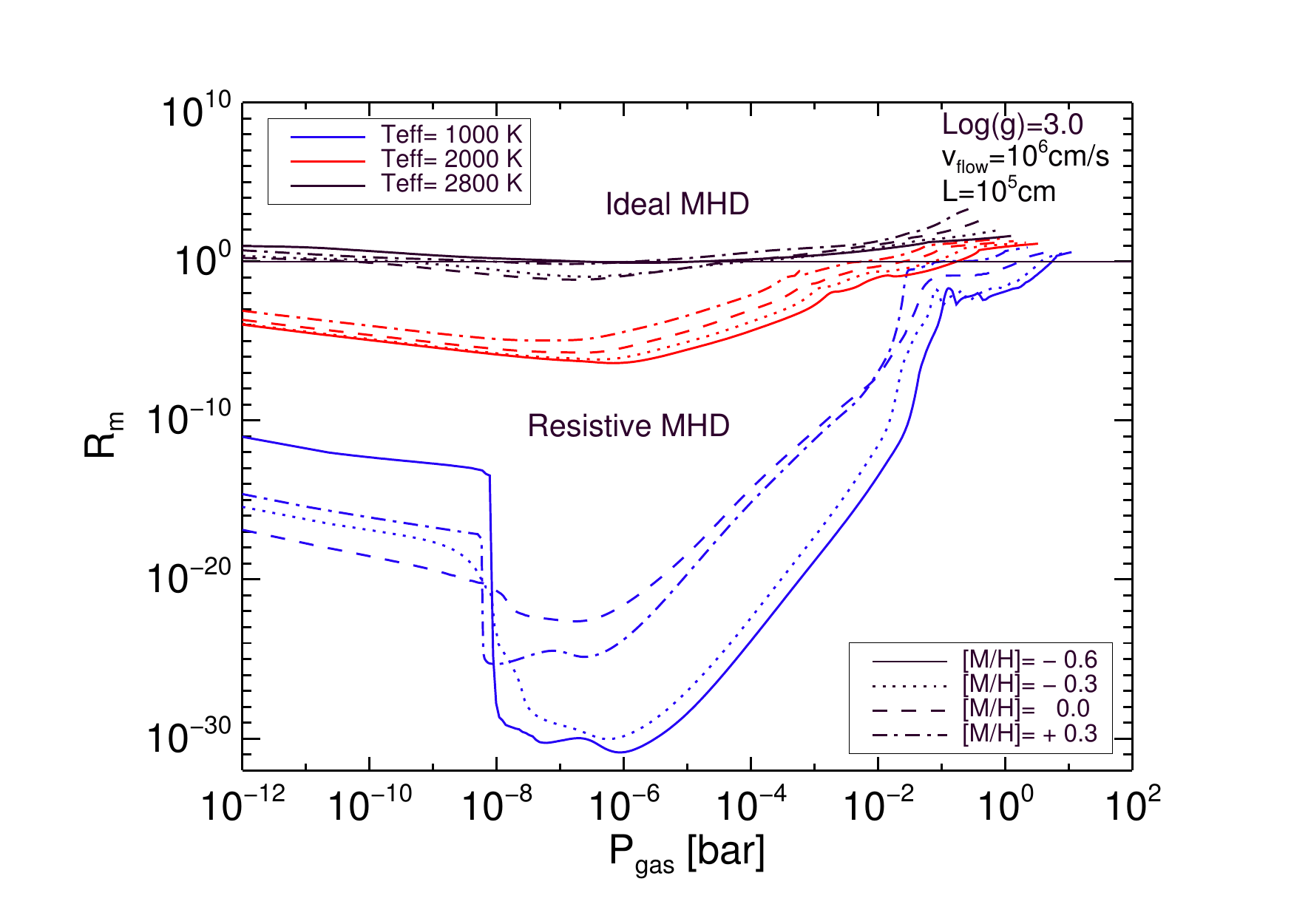}\\*[0.05cm]
\caption{Magnetic Reynolds number R$_{\rm m}$ for the three different model atmosphere structure groups described in Sec~\ref{s:DF}. R$_{\rm m}$ is calculated for a flow speed of $v_{\rm flow}=10^{6}$ cm\,s$^{-1}$. If the flow speed increases, then R$_{\rm m}$ increases. 
{\bf Top:} Group 1.
{\bf Middle:} Group 2.
 {\bf Bottom:} Group 3. }
 \label{fig:Rm2}
\end{figure}
\begin{figure*}
\centering
\hspace*{-1.cm}\includegraphics[width=1\textwidth]{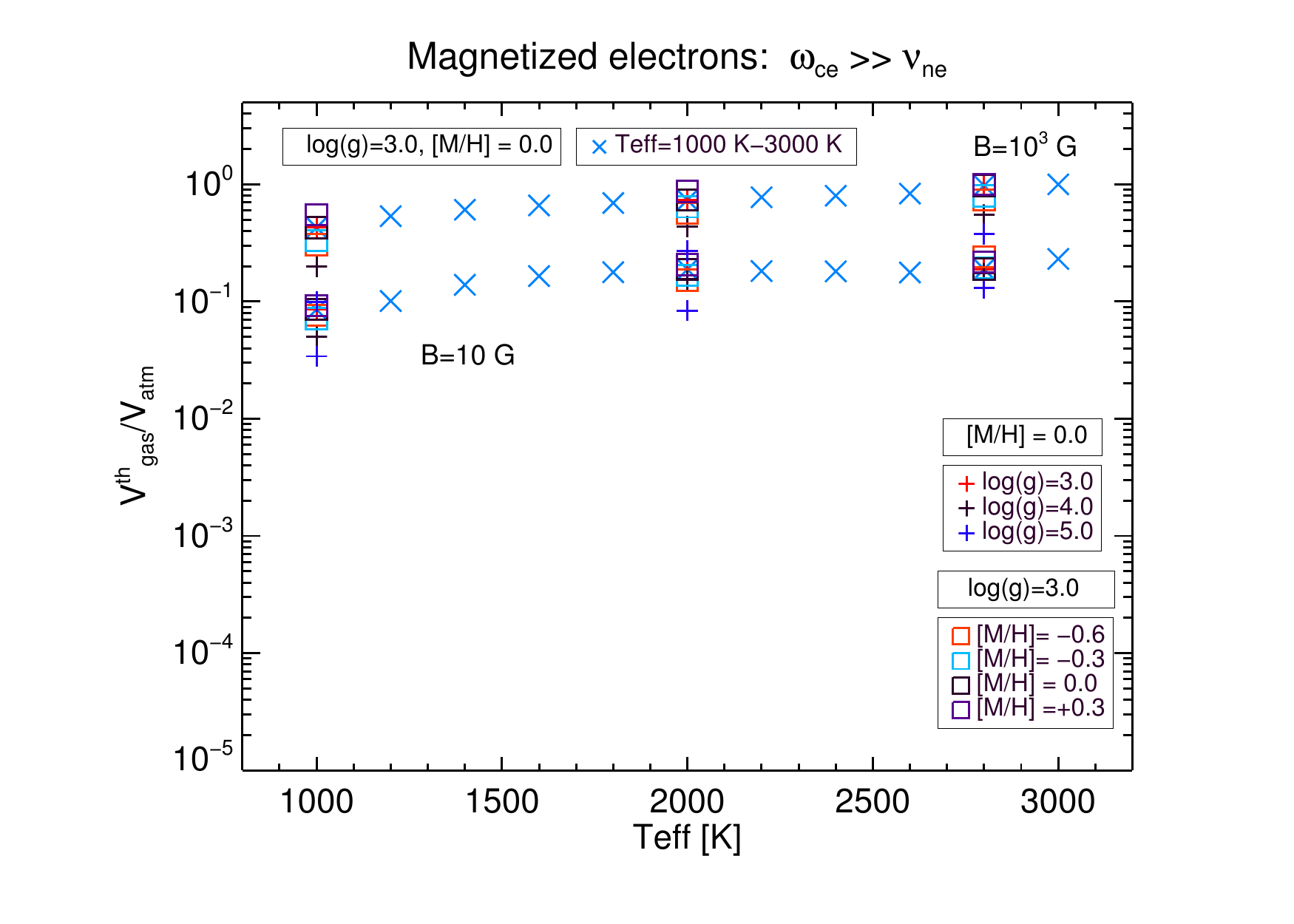}
\vspace{-0.5cm}\caption{The fraction of the atmospheric volume, $V^{\rm th}_{\rm gas}/V_{\rm atm}$, where $\omega_{\rm ce}\gg\nu_{\rm ne}$ for M-dwarf, brown dwarf and giant gas planet atmospheres. M-dwarfs and Brown dwarfs are represented by B$=10^{3}$ G and giant gas planets for B$=10$ G.}
\label{fig:vwce}
\end{figure*}
Therefore, the expression for the magnetic Reynolds number is rewritten as

\begin{equation}
R_{\rm m}\approx 10^{-4}\cdot v_{\rm flow}\left(\frac{n_{\rm e}}{n_{\rm gas}T_{\rm e}^{1/2}}\right).
\end{equation}

The values for the flow velocity chosen are v$_{\rm flow}$= 10$^{4}$ cm\,s$^{-1}$ and v$_{\rm flow}$= 10$^{6}$ cm\,s$^{-1}$ guided by values of circulation models (\citealt{Cooper2005}; \citealt{Menou2009}; \citealt{Rauscher2013}; \citealt{Heng2014}; \citealt{Rauscher2014}). 
Figure~\ref{fig:Rm2} represents the magnetic Reynolds number for the {\sc Drift-Phoenix}  model atmosphere structures which are comparable to the earlier results by \cite{Mohanty2002} that were based on {\sc Dusty-} and {\sc Cond-Phoenix}:
\begin{description}
\item - Reynolds number increases as T$_{\rm eff}$ increases because of the increasing of n$_{\rm e}$/n$_{\rm gas}$ in globally and locally hotter atmospheres (see Sec.~\ref{sss:fe}). The highest $R_{\rm m}$ is reached for T$_{\rm eff}$= 3000 K, $\log(g)$= 3.0, [M/H]= 0.0, v$_{\rm flow}$= 10$^{6}$ cm\,s$^{-1}$. 
\item-  Lower gravities cause an increased $R_{\rm m}$ for a given T$_{\rm eff}$ and [M/H] in the inner, high-density part of the atmosphere only. However, higher gravities cause an increased $R_{\rm m}$ for a given T$_{\rm eff}$ and [M/H] in the outer, low-density atmosphere.
\item -  Higher values of metallically cause a increased of $R_{\rm m}$ at high values of $p_{\rm gas}$  for a given T$_{\rm eff}$ and $\log(g)$. Low-metallicity atmospheres with low effective temperatures, i.e planet or T- and Y-dwarf atmospheres, have an increasing $R_{\rm m}$ in the outer, low density.  This correlates with a drastically changing gas-phase chemistry as shown in Fig.~\ref{chem1}.
\end{description}
Our results suggest that ideal MHD, where a fully ionised gas is assumed, is best suited for models atmospheres with T$_{\rm eff}\ge 3000~$K which includes M-dwarfs and young brown dwarfs. For cooler brown dwarfs and planetary regime objects only a small fraction of their atmosphere can be considered in ideal MHD.

\section{Discussion}\label{discussion}

\subsection{Chromospheres on  ultra-cool objects}

Observations in radio, soft X-ray and H$_{\alpha}$ wavelengths from
low-mass objects infer that their atmospheres are populated with
magnetized plasmas. 
Radio and
X-ray emission from ultra-cool dwarfs have been well established by
different authors (e.g. \citet{Berger2002}, \citealt{Route2012},
\citealt{Burgasser2013}).  \citet{Sorahana2014} and
\citet{Schmidt2015} suggest the presence of chromospheres in brown
dwarfs.  \citet{Sorahana2014} suggest that weakened H$_2$O
(2.7$\mu$m), CH$_4$ (3.3$\mu$m) and CO (4.6$\mu$m) absorption in
combination with moderate H$\alpha$ emission could be linked to
chromospheric activity. They represent a potential chromospheric
heating by an increased, constant temperature (with $p_{\rm gas}=nkT$)
in the upper atmosphere of their UCM 1D model atmosphere which allows
a considerably better data fit of their observation.
\citet{Schmidt2015} use a comparable approach by replacing the outer
atmospheric temperature of BT-settle model atmospheres with a
chromospheric temperature structure where the start of the
chromosphere, a chromospheric break and the start of the transition
region are used as parameters.  \cite{Metchev2015} discuss the likely
correlation of magnetic spots and high-amplitude photometric
variability in brown dwarfs with low surface gravity
values. \citet{Williams2014} demonstrate that ultra-cool objects do
not follow the classical G\"udel-Benz relationship where the radio
luminosity increases proportional to the X-ray luminosity in F\,$-$\,M
stars \citep{Gudel1993}. The deviation of ultra-cool stars in the
G\"udel-Benz relationship beyond than approximately M5 may suggest a
change in the dynamo mechanism that produces the magnetic field in
such ultra-cool objects \citep{Cook2014}. Another interpretation of
radio emission is the concurrence of an auroral region
(\citealt{Nichols2012}).  \cite{Spiers2014} describe a theoretical
approach for cyclotron radio emission from Earth's auroral region
providing a physical description for the widely used loss cone
parameterisation (e.g. \citealt{Osten2015}).  \cite{Spiers2014} show
that the radiation results from a backward-wave cyclotron-maser
emission process. The radio emission is generated by electrons
following a horseshoe velocity distribution, instead of a cone, that
travel the magnetic field lines downward. The backwards traveling
waves cause the upward refraction of the radiation which will be
further enhanced by density inhomogeneities.

If the atmospheric gas is well coupled with the background magnetic field in brown dwarfs and planets, the kinetic energy carried by large-scale convective motions may be transported to the top of the atmosphere and released. \citet{Tanaka2014} suggest that energy from the convective part of the atmosphere might be transported through the upper atmosphere by magneto-convection processes and suggest the formation of a chromosphere by Alfv\'en wave heating. \citet{Mohanty2002} carried out a study of magnetic field diffusivity to explain why the chromospheric $H_{\alpha}$ activity in brown dwarfs is low in spite of being rapid rotators. 
   \citet{Mohanty2002} based their work on a grid of model atmospheres (mid-M and L dwarfs) in a parameter range T$_{\rm eff}= 3000\,$K..1500$\,$K, $\log(g)$=5.0 and [M/H]=0.0. They explained why in these range of T$_{\rm eff}$ the observation of chromospheric levels activity are lower than early M-dwarfs, considering mid-M and L dwarfs as rapid rotators. 
   In our work, we extend our model atmosphere grid until T$_{\rm eff}= 1000\,$K and we include atmosphere structures with different values of $\log(g)$ and [M/H]  (Table\,\ref{tab}). 
  A linear field diffusion equation, MHD regime and LTE  are used in both works. 
  Results obtained for R$_{\rm m}$ as a measure of the ideal or resistive MHD atmosphere (see Fig.\,\ref{fig:Rm2}) could be incomplete and therefore misleading. According to Eq.\,\ref{eq:Rmf}, R$_{\rm m}\propto1/\eta$. Our results demonstrate that the regions where R$_{\rm m}>1$ satisfy also f$_{\rm e}>10^{-7}$. Atmospheres of ultra-cool objects could be ionised and treated as an ideal MHD gas only at deep layers. Furthermore, $\omega_{\rm ce}/\nu_{\rm ne}\gg1$ measures the coupled between the magnetic field and the atmospheric gas and it  depends,  mostly, on the strength of the magnetic field (see Fig.\,\ref{fig:vwce}). Therefore, it is possible to find large volumes of the atmospheric gas that are magnetised, but smaller magnetised volumes that are strongly ionised.

 Our results further suggest not only that higher effective
 temperatures (in agreement with \cite{Mohanty2002}) and higher
 metallicity atmospheres are the best candidates for forming a
 magnetised atmospheric plasma in support of radio, X-ray and
 H$_{\alpha}$ observations in ultra-cool objects. Also low surface
 gravity atmospheres fall in this category which supports the
 interpretation by \cite{Metchev2015} that high-amplitude photometric
 variability in L3-L5.5 dwarfs can also be related to magnetic spot
 appearance.  While M-dwarfs have been shown to be fully magnetised,
 L-dwarfs and later brown dwarfs have smaller atmospheric volume that
 can be magnetised in an external magnetic field. This findings relate
 to the activity-vs-SpecT results in \citet{Schmidt2015} (e.g. their
 Fig. 6). The threshold of T$_{\rm eff}=2300$\,K
   given in their Fig.6 is the \cite{Mohanty2002} threshold that
   points out the limit from which models R$_{\rm m}>1$ using
   v$=10^{4}$ cms$^{-1}$ and 10$^{-2}\le\tau_{J}\le$10$^{2}$
   (convection zone) being $\tau_{J}$ the optical length in J
   band. Our paper suggests that this threshold move towards T$_{\rm
     eff}=1400$\,K for the same value of flow velocity and same
   region. This result suggests that atmospheres cooler than T$_{\rm
     eff}=2300$\,K may be susceptible to be magnetized. Another
   criterion to consider a gas magnetised is $\omega_{\rm
     ce}\gg\nu_{\rm ne}$ (see Fig.\,\ref{fig:fc}). For all models
   considered in this study (Table\,\ref{tab})) the atmospheric gas
   fulfill $\omega_{\rm ce}\gg\nu_{\rm ne}$ for $p_{\rm
     gas}<1$ bar. Combining both criteria $\omega_{\rm ce}\gg\nu_{\rm
     ne}$ and R$_{\rm m}>1$, a large fraction of possible active
   objects are found for T$_{\rm eff}$=3000\,K-1400\,K, $\log(g)$=3.0,
   [M/H]=0.0 (Group1).

 \subsection{Ionisation through non-thermal processes}
 
Our work focuses on determining global parameters to provide the suitable local atmospheric conditions for a magnetised plasma to be present.  The results suggest that ultra-cool atmospheres are susceptible to plasma and magnetic processes even if only thermal ionisation processes are considered and the influence of dust beyond element depletion is neglected. An atmospheric plasma regime and magnetised gas were found in the where M-dwarf atmosphere, which cooler brown dwarfs and planetary objects require high $p_{\rm gas}$, hence, their magnetised volume is smaller than in M-dwarf atmospheres.

Additional potentially non-thermal ionisation processes will enhance the degree of ionisation and increase the local volume affected by a magnetic field. Local enhancement  can result from dust-dust collisions in large cloud areas \citep{Helling2011b}  and Alfv\'en ionisation if the local, hydrodynamic wind speed is high enough \citep{Stark2013}.
Electric storms that develop inside an atmosphere effects the extent of an ionosphere causing a link between the local ionisation processes and the global effects (\citealt{Luque2014}).  Irradiation from a host star for close-in exoplanets or in  white dwarfs - brown dwarf binaries  \citep{Casewell2013} will increase the local thermal ionisation globally. Galactic cosmic rays increase the number of free charge particles in single brown dwarfs. Cosmic rays are effective at ionising the upper atmospheric parts, however, the exact amount  is hard to quantify without extensive chemistry simulations \citep{Rimmer2013}.  

 \section{Conclusions}\label{s:con}
 
 We present a reference study for late M-dwarfs, brown dwarfs and giant gas planet to identify which ultra-cool objects are
  most susceptible to atmospheric gas-phase plasma processes.  Only thermal ionisation is considered for this reference study and the influence of dust beyond element depletion is neglected.  The effect of additional processes like cosmic ray ionisation, irradiation, Alfv{\'e}n ionisation, lighting can be evaluated against the reference results in this paper.  
  
 Ultra-cool atmospheres with high T$_{\rm eff}$, high [M/H] and low $\log(g)$ have large fraction of  atmospheric volume where plasma processes occur, and are therefore the best candidates for radio, X-ray and H$_{\alpha}$ emissions. M-dwarfs have a considerable degree of ionisation throughout the whole atmosphere, the degree of thermal ionisation for a L-dwarf is  low but high enough to seed other local ionisation processes like  Alfv{\'e}n ionisation or lightning discharges. Electromagnetic interaction dominates over electron-neutral interactions also in regions of a very low degree of ionisation in most model atmospheres in our sample. The relevant length scales effected by electromagnetic interactions in the gas phase are larger in low-density regions of any atmosphere. The minimum threshold for the magnetic flux density required for electrons and ions to be magnetised is smaller than typical values of the global magnetic field strengths of a brown dwarf and a giant gas planet. A considerably lower magnetic flux density is required for magnetic coupling of the atmosphere in the rarefied upper atmosphere than in the dense inner atmosphere. Na$^{+}$, K$^{+}$and Ca$^{+}$ are the dominating electron donors in low-density atmospheres (low log(g), solar metallicity) independent of Teff. Mg$^{+}$ and Fe$^{+}$dominate the thermal ionisation in the inner parts of M-dwarf atmospheres. Molecules remain unimportant for thermal ionisation. Chemical processes (e.g. cloud formation, cosmic ray ionisation) that affect the abundances of Na, K, Mg, Ca and Fe will have a direct impact on the state of ionisation in ultra-cool atmospheres.

Our results suggest that it is not unreasonable to expect ultra-cool atmospheres (M-dwarfs and brown dwarfs) to emit H$_{\alpha}$ or even in radio wavelength as in particular the rarefied upper parts of the atmospheres fulfill plasma criteria easily despite having low degrees of ionisation.  Our results therefore suggest that an ionosphere may emerge also in brown dwarf and giant gas planet atmospheres, and that the built-up of a chromosphere is likely. Both effects will contribute to atmospheric weather features and to space weather occurrences  in extrasolar, planet-like objects.  Ultra-cool atmospheres could also drive auroral emission without the need for a companion's wind or an outgassing moon.

\section*{Acknowledgments}

\bigskip
We highlight financial support of the European Community under the FP7
by the ERC starting grant 257431. We thank G.~Lee, G.~Hod{\'o}san, I.~Vorgul
and I.~Leonhardt for valuable discussions of the manuscript. Most
literature search was performed using the ADS. We acknowledge our
local computer support highly.


\footnotesize{
\bibliographystyle{mn2e}
\bibliography{bib4}{}

}

\appendix

\section{Most abundant thermal ions in M-dwarf, brown dwarf and giant gas planet atmospheres}\label{chem}

\paragraph*{Varying T$_{\rm eff}$ and $\log(g)$:}

Table~\ref{tab:1} includes model atmospheres where [M/H] is kept at a constant value of 0.0 and T$_{\rm eff}$ is varied alongside $\log(g)$ (see Sec.\,\ref{ions}).
The ions shown in the table are the most abundant; it does not include ions that are very prominent but not the most abundant.
Note: When there are two ions listed (for example, Al+/Na+) the second ion occurs at the higher pressure.

Higher values of log(g) correspond to higher overall values of pressure, due to the increased surface gravity.

It is worth noting that there are other ions which are highly prominent within these models, however the following tables only show which ones are the most prominent.

\paragraph*{Varying T$_{\rm eff}$ and [M/H]}

Table~\ref{tab:2} shows model atmospheres where $\log(g)$ has been kept at a constant value of 3.0 and T$_{\rm eff}$ is varied alongside [M/H].

\clearpage
\begin{table*}
\caption{First, second and third ion represent the most abundant positive ion in the high, middle and lower pressure regions respectively. First ion corresponds to the higher pressure (and therefore higher abundances) and visa verse.}
\begin{tabular}{|l|lll|lll|lll|}
\hline
\begin{tabular}[c]{@{}l@{}}Teff\\ log(g)\end{tabular} &  & 3.0 &  &  & 4.0 & & 5.0 & & \\ \hline
 \rule{0pt}{4ex}     1000   & K$^{+}$& K$^{+}$    & K$^{+}$ & K$^{+}$ & K$^{+}$   & K$^{+}$ & K$^{+}$ & K$^{+}$ & K$^{+}$   \\
						   & Na$^{+}$ &  &  & Na$^{+}$ &  &  & Na$^{+}$ &  & \\ 
\rule{0pt}{4ex}      1200   & K$^{+}$ &  K$^{+}$   & K$^{+}$    &  &  &  &&& \\ 
							& Na$^{+}$ & & &&& &&&\\  
\rule{0pt}{4ex}   1400   & K$^{+}$ & K$^{+}$   & K$^{+}$     &  &  &  &&&    \\
							& Na$^{+}$ &&&&& &&&\\  
 \rule{0pt}{4ex}   1600   & Na$^{+}$ & K$^{+}$    & K$^{+}$     &  &  &  &&&    \\
\rule{0pt}{4ex}   1800   & Na$^{+}$ & K$^{+}$    & K$^{+}$     &  &  &  &&&    \\
 \rule{0pt}{4ex}   2000   &Na$^{+}$  & K$^{+}$    & K$^{+}$ & Na$^{+}$ &  K$^{+}$   & K$^{+}$ & Na$^{+}$ & K$^{+}$ & K$^{+}$  \\
							&Ca$^{+}$ &&&&&&  && \\  
 \rule{0pt}{4ex}   2200   & Na$^{+}$ &  K$^{+}$   & K$^{+}$     &  &  &   &&&    \\
 \rule{0pt}{4ex}   2400   & Na$^{+}$ & K$^{+}$    & K$^{+}$     &  &  &  &&&    \\
							& Mg$^{+}$ & & &&& &&&\\  
 \rule{0pt}{4ex}   2600   & Mg$^{+}$ &  K$^{+}$   & K$^{+}$     &  &  &  &&&     \\
							& & Na$^{+}$ &&&&&&& \\ 
 \rule{0pt}{4ex}   2800   & Mg$^{+}$ &  K$^{+}$   & K$^{+}$ &Na$^{+}$ &  K$^{+}$   & K$^{+}$ & Na$^{+}$ & K$^{+}$& K$^{+}$ \\
							&  & Na$^{+}$ &  &Mg$^{+}$& Na$^{+}$ & &Mg$^{+}$ & &\\
 \rule{0pt}{4ex}   3000   & Na$^{+}$ &  K$^{+}$   & Na$^{+}$     &  &  &  &&&     \\
							& Mg$^{+}$ & Na$^{+}$ & K$^{+}$ &&& &&&\\ 
							& H$^{+}$ & &&&&&&&\\
                            \hline
\end{tabular}
\label{tab:1}
\end{table*}

\clearpage
\begin{table*}
\caption{First, second and third ion represent the most abundant positive ion in the high, middle and lower pressure regions respectively. First ion corresponds to the higher pressure (and therefore higher abundances) and visa verse.} \vspace{0.1 cm}
\begin{tabular}{|l|lll|lll|lll|lll|}
\hline
\hline
\begin{tabular}[c]{@{}l@{}}M/H\\ Teff\end{tabular} &  & -0.6 &  &  & -0.3 &  &  & 0.0 &  & +0.3 &  &  \\ \hline
 \rule{0pt}{4ex}1000& K$^{+}$& K$^{+}$    & K$^{+}$ & K$^{+}$ & K$^{+}$     & K$^{+}$ & K$^{+}$& K$^{+}$    & K$^{+}$ & K$^{+}$& K$^{+}$    & K$^{+}$  \\
                                                    & Na$^{+}$ &  &  & Na$^{+}$ &      &  & Na$^{+}$ &   & &Na$^{+}$ & &  \\
 \rule{0pt}{4ex}1200&  &      &  &  &        & & K$^{+}$ &  K$^{+}$   & K$^{+}$ &  &  &  \\
                                                   &  &      &  &  &      &  & Na$^{+}$ &    &  &  &  &  \\
 \rule{0pt}{4ex}1400&  &      &  &       &  &   & K$^{+}$ & K$^{+}$   & K$^{+}$     &   & & \\
                                                   &  &      &  &  &      &  & Na$^{+}$ &&& &  &  \\
 \rule{0pt}{4ex}1500&  &      &  &  &      &  &  &     &   & K$^{+}$ & K$^{+}$ & K$^{+}$ \\
                                                   &  &      &  &  &      &  &  &   &  & Na$^{+}$ &  &  \\
 \rule{0pt}{4ex}1600&  &      &  &  &      &  & Na$^{+}$ & K$^{+}$    & K$^{+}$     &   &  &  \\
 \rule{0pt}{4ex}1800&  &      &  &  &      &  & Na$^{+}$ &  K$^{+}$   & K$^{+}$     &  &  &  \\
 \rule{0pt}{4ex}2000& Na$^{+}$ &  K$^{+}$   & K$^{+}$     & Na$^{+}$ &  K$^{+}$   & K$^{+}$     & Na$^{+}$ &  K$^{+}$   & K$^{+}$     & Na$^{+}$ &  K$^{+}$   & K$^{+}$ \\
 \rule{0pt}{4ex}2200&  &      &  &  &      &  & Na$^{+}$ &  K$^{+}$   & K$^{+}$     &  &  &  \\
 \rule{0pt}{4ex}2400&  &      &  &  &      &  & Na$^{+}$ &  K$^{+}$   & K$^{+}$     &  &  &  \\
 \rule{0pt}{4ex}2600&  &      &  &  &      &  & Mg$^{+}$ &  K$^{+}$   & K$^{+}$     &  &  &  \\
                                                   &  &      &  &  &      &  &  & Na$^{+}$    &  &  &  &  \\
 \rule{0pt}{4ex}2800& Na$^{+}$ & K$^{+}$     & Na$^{+}$ & Na$^{+}$ & K$^{+}$   & K$^{+}$ & Na$^{+}$ & K$^{+}$    & K$^{+}$ & Na$^{+}$ & K$^{+}$ & K$^{+}$ \\
                                                   &Ca$^{+}$  &  Na$^{+}$    & K$^{+}$ & Ca$^{+}$ &  Na$^{+}$    &  &Mg$^{+}$  & Na$^{+}$    &  & Mg$^{+}$ & Na$^{+}$ & \\
                                                   & &      &  &  K$^{+}$&      &  &  &     &  &  &  & \\
 \rule{0pt}{4ex}3000&  &      &  &  &      &  & Na$^{+}$  & K$^{+}$   & Na$^{+}$ &  &  &  \\
                                                   &  &      &  &  &      &  & Mg$^{+}$ &   Na$^{+}$  & K$^{+}$ &  &  &  \\
                                                   &  &      &  &  &      &  & H$^{+}$ &     &  &  &  & \\
                                                   \hline
\end{tabular}
\label{tab:2}
\end{table*}

\section{Plasma parameter: Number of particles in a Debye sphere: $\bm{N_{\rm D}\gg1$} }\label{ss:Nd}
  \begin{figure}
\centering
\hspace*{-0.7cm}\includegraphics[angle=0,width=0.6\textwidth]{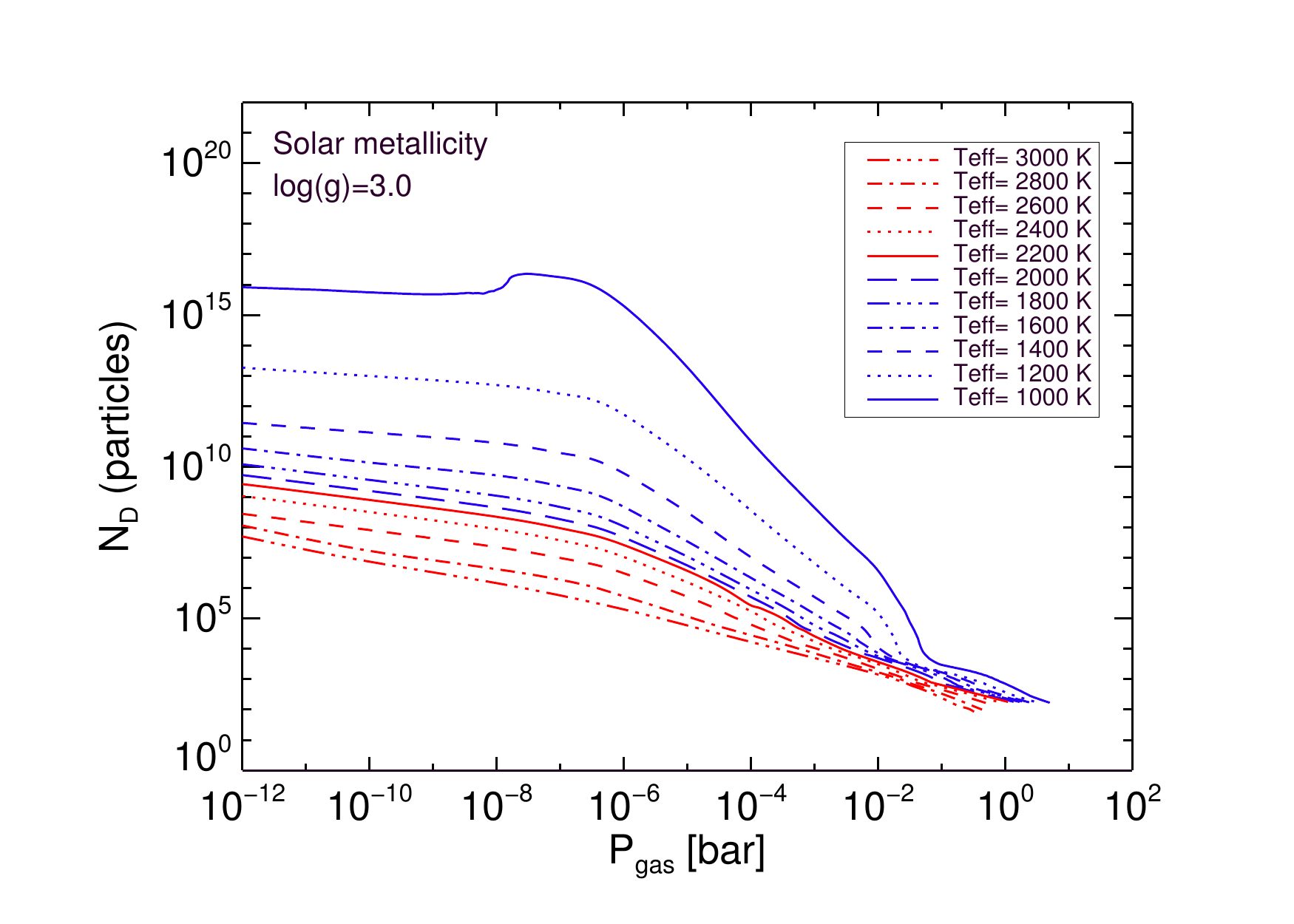}\\*[-0.8cm]
\hspace*{-0.7cm}\includegraphics[angle=0,width=0.6\textwidth]{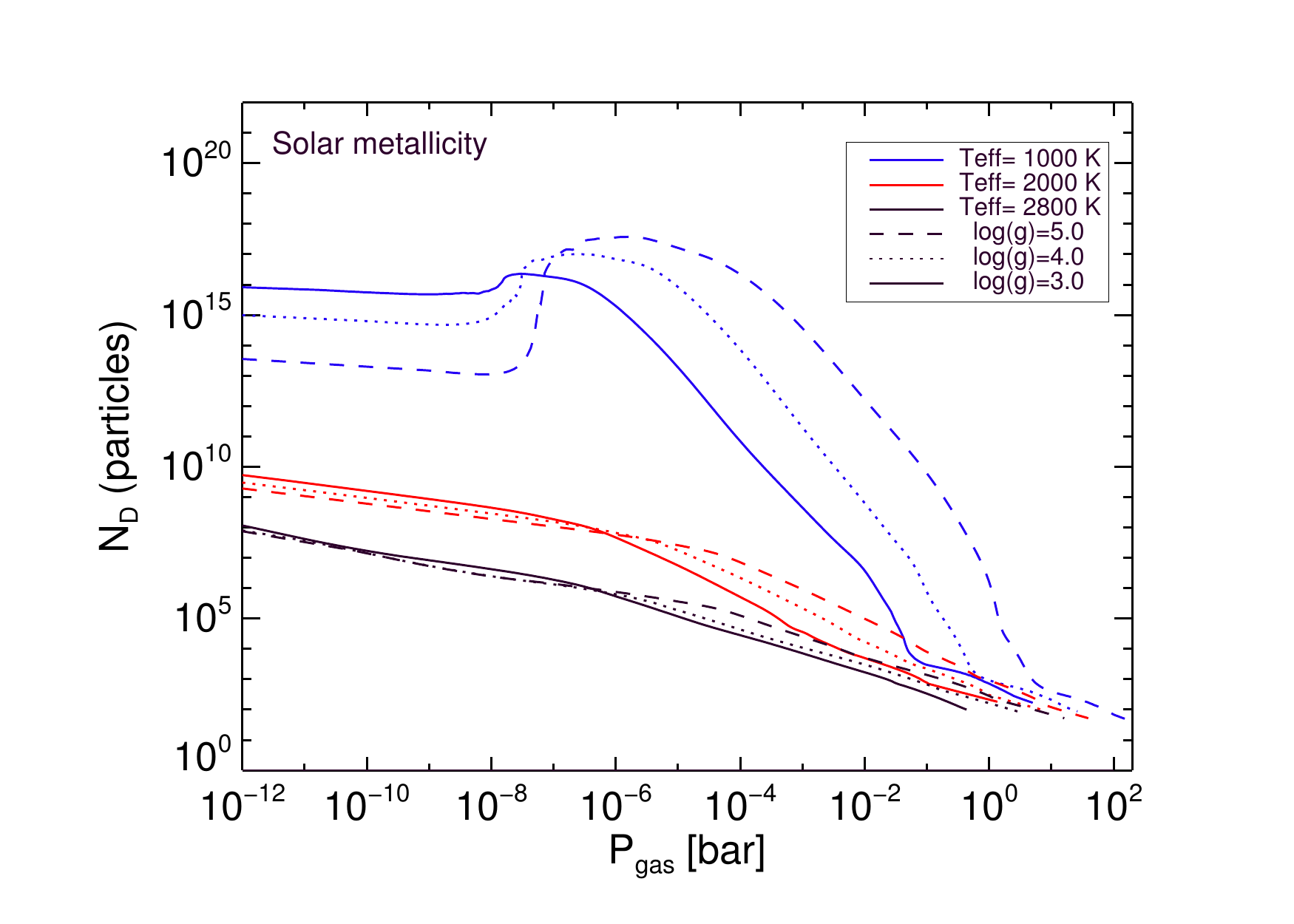}\\*[-0.8cm]
\hspace*{-0.7cm}\includegraphics[angle=0,width=0.6\textwidth]{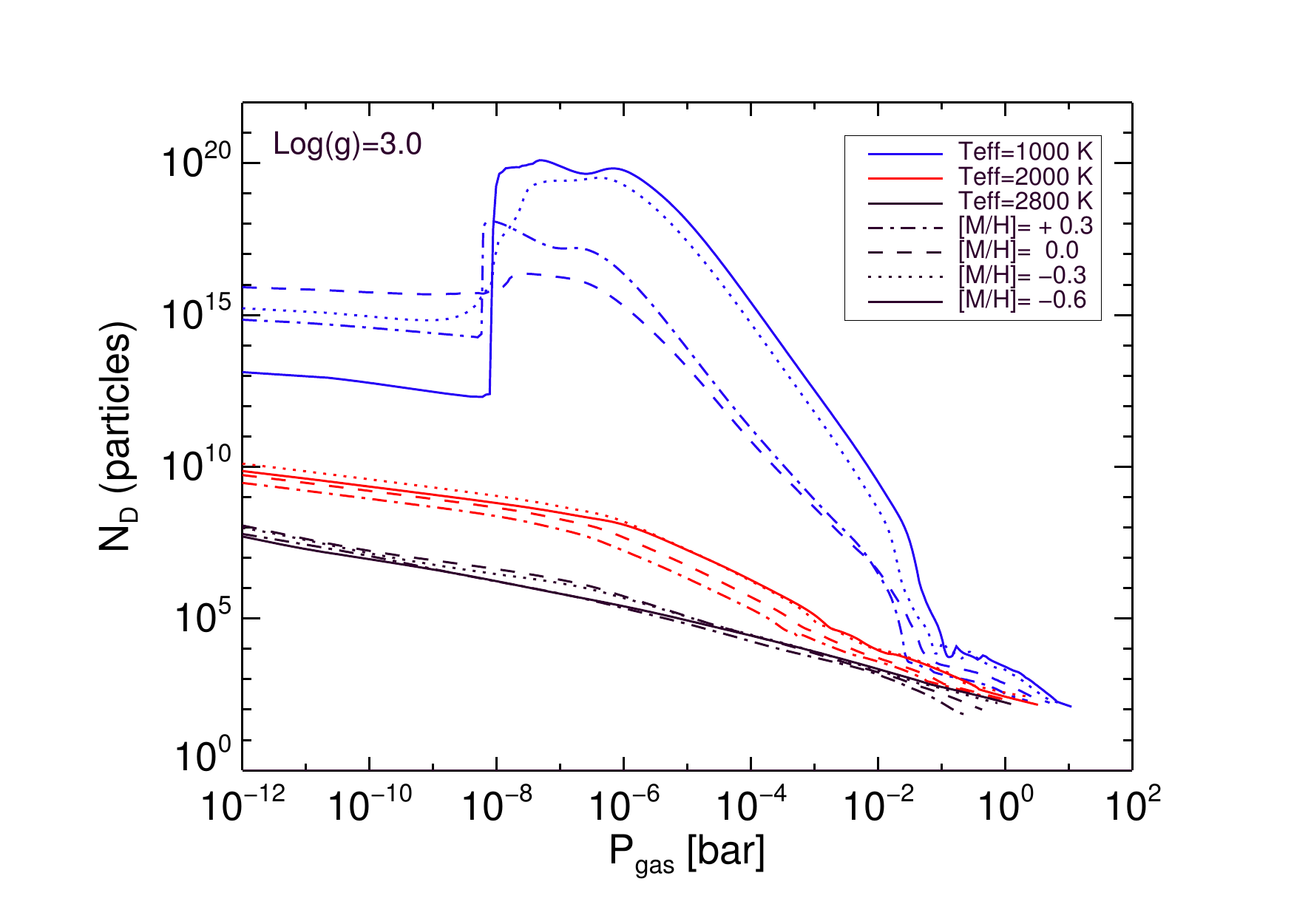}
\\*[-0.6cm]
\vspace{+0.1cm}\caption{Number of particles inside of Debye sphere measure the efficiency of the screening given by the Debye sphere in the plasma. N$_{\rm D}\gg1$ results in the collective interactions dominate over short-range collisions in the gas.
{\bf Top:} Group 1.
{\bf Middle:} Group 2.
 {\bf Bottom:} Group 3. 
}
\label{fig:Nd}
\end{figure}


A plasma has the capacity to screen a single charged particle placed
at any point. That means that any single charged particle attracts
oppositely charged particles producing a screening and repels those
who have the same charge. A net space is produced in the neighbourhood
of any single charged particle, reducing the electric field generated
by it.  The effective range of the net force between particles is
restricted to the order of the Debye length (see
Sec.\,\ref{s:plasma}). As a consequence, a test particle in the Debye
sphere interacts only with particles that lie within this
sphere. $N_{\rm D}$ measured the efficiency of this screening and
allows us to calculate how many gas particles are required to
participate. Hence, only particle inside this screening areas
($\lambda_{\rm D}$) can be considered as electrostatically active.

All particles have a thermal velocity due to the temperature of the plasma. 
The deflected angle due to the electrostatic interactions is bigger if the number of particles around of the screened particle in the Debye sphere is small. The movement of the screened particle will not be smooth, unlike in the situation when the number of particles in the Debye sphere is sufficiently large to reduce it. That is why the Debye length increases as the number of particles in the screened sphere decreases. This is demonstrated in Fig.~\ref{fig:lD} where all Debye length increase with height in the atmosphere, i.e. with the outward decreasing local gas pressure. 

The change in velocity due to the interactions with the particles produces a non-negligible net electrostatic force inside the Debye sphere.
Therefore, large numbers of particles that are uniformly distributed inside the Debye sphere are required to avoid a large-angle deflection on a test particle. Hence the Debye length will be small in comparison to the length scale of the plasma. In this case, the plasma is dominated by many long-range interactions, rather than the short-range binary collisions of a neutral gas.
A measure of the efficiency of the screening is the plasma parameter $N_{\rm D}$.
  
  The plasma parameter is defined as
  \begin{equation}\label{eq:Nd}
  N_{\rm D}=(4/3)\pi n_{\rm e}\lambda_{\rm D}^{3}, 
   \end{equation}
   the number of particles in a Debye sphere  with radius  $\lambda_{\rm D}$ and centred on a single charge particle that produced the charge imbalance. 
 When there are many plasma particles in a Debye sphere ($N_{\rm D}\gg 1$) and long-range collective interactions are dominant over short-range collisions, the plasma frequency is much larger than the electron-ion collision frequency. 
 Figure~\ref{fig:Nd} shows $N_{\rm D}\gg1$ for all model atmosphere structures. This indicates that thermal electrons interact over large distances in atmosphere of ultra-cool atmospheres.
%


\appendix
\bsp

\label{lastpage}

\end{document}